\documentclass[prb,twocolumn]{revtex4-1}

\usepackage{graphicx}
\usepackage{comment}
\usepackage{amsmath,amssymb,amsthm} 
\usepackage{verbatim}
\usepackage{float}

\setcitestyle{super}

\begin{document}

\title{Isomorphs in model molecular liquids}
\author{Trond S. Ingebrigtsen, Thomas B. Schr{\o}der, and Jeppe C. Dyre}
\affiliation{DNRF Centre ``Glass and Time'', IMFUFA, Department of Sciences, Roskilde University, Postbox 260, DK-4000 Roskilde, Denmark}
\date{\today}

\begin{abstract}
  Isomorphs are curves in the phase diagram along which a number of static and dynamic 
  quantities are invariant in reduced units. A liquid has good isomorphs if and only if it is strongly correlating, i.e., the equilibrium virial/potential energy fluctuations are more than 90\% correlated in 
  the \textit{NVT} ensemble. This paper generalizes isomorphs to liquids composed of rigid molecules and study the isomorphs of two systems of small 
  rigid molecules, the asymmetric dumbbell model and the Lewis$-$Wahnstr{\"o}m \textit{OTP} model. In particular, for both systems we find that the isochoric heat capacity, the excess entropy, 
  the reduced molecular center-of-mass self part of the intermediate scattering function, the reduced molecular center-of-mass radial distribution function to a good approximation are invariant along 
  an isomorph. In agreement with theory, we also find that an instantaneous change of temperature and density from an equilibrated state point to another isomorphic state point leads to no relaxation.
  The isomorphs of the Lewis$-$Wahnstr{\"o}m \textit{OTP} model were found to be more approximative than those of the asymmetric dumbbell model, which is consistent with the \textit{OTP} model being less 
  strongly correlating. For both models we find ''master isomorphs'', i.e., isomorphs have identical shape in the virial/potential energy phase diagram.
\end{abstract}

\maketitle

\section{Introduction}

For supercooled liquids near the glass transition changing slightly the density $\rho$ or temperature $T$ the structural relaxation time $\tau_{\alpha}$ may change several orders of magnitude. In the study of these 
liquids\cite{dscale,dscale1,reviewRoland}
it is found that $\tau_{\alpha}$ does not change when $\rho^{\gamma}/T$ is kept constant, where $\gamma$ is a fixed material-specific exponent. 
This phenomenon is called \textit{density scaling} and has been established for many liquids, excluding associative liquids such as water\cite{reviewRoland}. A related observation 
is \textit{isochronal superposition}\cite{isochone1,isochone2,reviewRoland}, i.e., that supercooled state points with identical $\tau_{\alpha}$ have the same dielectric spectrum. A different and at first sight
unrelated concept is \textit{Rosenfeld's excess entropy scaling}\cite{rosenfeld1,rosenfeld2}. In this procedure a relation is established between hard-to-predict dynamic properties and 
easier-to-predict thermodynamic quantities, here 
the excess entropy via a scaling of the dynamics to so-called reduced units. Initially this was observed for model atomic liquids\cite{rosenfeld1,rosenfeld2}, but later it was extended 
to model molecular liquids\cite{truskwater, truskhydro,truskdumb} and experimental liquids\cite{exviscous}. 

In a recent series of papers\cite{paper1,paper2,paper3,paper4,paper5} a new class of liquids was identified.  
These liquids are characterized by having strong correlation in the \textit{NVT} ensemble between the equilibrium fluctuations of the instantaneous potential energy $U$ and the virial $W$. 
Recall that the instantaneous energy $E$ and  pressure $p$ can be 
written as a sum of a kinetic part and a configurational part: $E = K + U$ and $pV = Nk_{B}T + W$, respectively.
The correlation between $U$ and $W$ is quantified via the linear correlation coefficient $R$ defined as

\begin{equation}\label{R}
  R = \frac{\langle \Delta W \Delta U \rangle}{\sqrt{\langle (\Delta W)^{2} \rangle}\sqrt{ \langle (\Delta U)^{2} \rangle}}.
\end{equation}
The class of strongly correlating liquids is defined by $R \geq 0.90$\cite{paper1}. An inverse power-law (\textit{IPL}) 
system $r^{-n}$ has correlation coefficient R = 1, since $\Delta W = (n/3)\Delta U$, and only \textit{IPL} systems are perfectly correlating. In the study of strongly correlating liquids 
it was discovered that they obey Rosenfeld's excess entropy scaling, isochronal 
superposition, as well as density scaling. These types of scalings can be explained in the framework of so-called isomorphs (definition follows later).

Model systems that have been identified\cite{first,paper1,paper2,coslovich1,coslovich2,hidden} to belong to this class of liquids include the standard single-component Lennard-Jones liquid (\textit{SCLJ}), 
the Kob-Andersen binary \textit{LJ} mixture \cite{ka1, ka2} (\textit{KABLJ}), 
the asymmetric dumbbell model \cite{hidden}, the Lewis-Wahnstr{\"o}m $o$-$terphenyl$ model\cite{otp1, otp2} (\textit{OTP}), 
and several others. Strong $WU$ correlation has been experimentally verified for a molecular van der Waals liquid \cite{gamma} and for supercritical argon\cite{first}. The class of strongly correlating
liquids is believed to include most or all van der Waals and metallic liquids, whereas covalently, hydrogen-bonding or ionic liquids are not strongly correlating\cite{paper1}. 
The latter reflects the fact that competing interactions tend to destroy the strong correlation.

\begin{figure}[H]
  \centering
  \includegraphics[width=70mm]{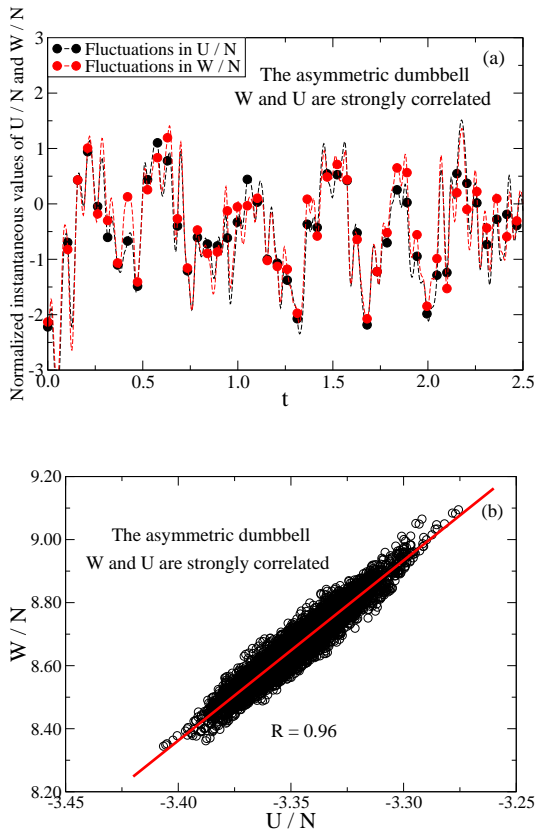}
  \caption{Two different ways of visualizing the strong virial/potential energy correlation for the asymmetric 
    dumbbell model at $\rho$ = $1.863$ and $T$ = $0.465$ (see Sec. \ref{simulation} for details of the model and the units used). (a) The time evolution 
    of $U$ (black) and $W$ (red) per particle normalized to zero mean and unity standard deviation. (b) A scatter plot 
    of the instantaneous values of $W$ and $U$ per particle. The correlation coefficient $R$ is $0.96$.}
  \label{strongR}
\end{figure}

An example of strong $WU$-correlation is given in Fig. \ref{strongR} for the asymmetric dumbbell model\cite{hidden} (details of this model are 
provided in Sec. \ref{simulation}).
Figure \ref{strongR}(a) shows the time evolution of the equilibrium fluctuations of $U$ and $W$ normalized to zero mean and unity standard deviation, 
Fig. \ref{strongR}(b) shows a scatter plot of the corresponding instantaneous values of $U$ and $W$. $U$ and $W$ are clearly strongly correlated in their equilibrium fluctuations.

\begin{figure}[H]
  \centering
  \includegraphics[width=70mm]{dumbbell_WU_instant}
\end{figure}    
References \onlinecite{paper1} and \onlinecite{paper2} identified the cause of strong $WU$-correlation in the \textit{SCLJ} liquid. The \textit{LJ} 
pair potential can be well approximated from about $r=0.95\sigma$ to $r = 1.5\sigma$ (see Pedersen \textit{et al.}\cite{overviewscl}) by a sum of an \textit{IPL}, 
a linear term, and a constant via the so-called
''extended \textit{IPL} potential''\cite{paper2}: $v_{LJ}(r) \approx Ar^{-n} + B + Cr$. At moderate pressures this covers the entire first peak of the radial distribution function, i.e., 
the first coordination shell. 
The constraint of constant volume in the \textit{NVT} ensemble has the effect that when one nearest neighbor distance increases another one decreases; upon summation the contribution from
the linear term to $U$ and $W$ is almost constant. The latter observation has the consequence that some of the scaling properties of pure \textit{IPL} systems are inherited in the \textit{LJ} system in the form 
of isomorphs.

Reference \onlinecite{paper4} introduced a new concept referring to a strongly correlating liquid's phase diagram, namely isomorphic curves or more briefly: isomorphs. 
Two state points with density and temperature ($\rho_{1}$, $T_{1}$) and ($\rho_{2}$, $T_{2}$) are defined to 
be isomorphic\cite{prac} if the following holds: Whenever a configuration of 
state point ($1$) and one of state point ($2$) have the same reduced coordinates ($\rho_{1}^{1/3} \textbf{r}^{(1)}_{i} = \rho_{2}^{1/3} \textbf{r}^{(2)}_{i}$ for all particles $i$), these two 
configurations have proportional Boltzmann factors, i.e.,

\begin{equation} \label{defiso}
  e^{-U(\textbf{r}_ {1}^{(1)}, ..., \textbf{r}_ {N}^{(1)})/k_{B}T_{1}} = C_{12}e^{-U(\textbf{r}_{1}^{(2)}, ...,  \textbf{r}_ {N}^{(2)})/k_{B}T_{2}}.
\end{equation}
Here $C_{12}$ is a constant that depends only on the state points ($1$) and ($2$). An isomorph is defined as a continuous curve of state points that are all
pairwise isomorphic. In other words, Eq. (\ref{defiso}) defines an equivalence relation with the equivalence classes being the isomorphs. Only \textit{IPL} systems  
have exact isomorphs; these are characterized by having $\rho^{\gamma} /T = const$ where $\gamma = n/3$. Reference \onlinecite{paper4} argued and demonstrated by simulations that 
strongly correlating liquids have isomorphs to a good approximation.

From the defining property of an isomorph [Eq. (\ref{defiso})] it follows that the structure in reduced units ($\tilde{\textbf{r}}_{i} \equiv \rho^{1/3}\textbf{r}_{i}$) is invariant along an isomorph, since 
the proportionality constant $C_{12}$ disappears when normalizing the configurational canonical probabilities\cite{paper4}. Thus  
the reduced unit radial distribution function and the excess entropy $S_{ex} = S - S_{id}$ are isomorph invariants, 
where $S_{id}$ is the ideal gas contribution to the entropy at the same temperature and density. 
Isomorph invariance 
is, however, not limited to static quantites, also the mean-square displacement, time auto-correlation functions, and higher-order correlation functions 
 are invariant in reduced units along an isomorph. The reader is referred to Ref. \onlinecite{paper4} for a detailed description of isomorph invariants, and the 
proof that a liquid is strongly correlating if and only if it has good isomorphs. A brief 
overview of strongly correlating liquids and their isomorphs can be found in Pedersen \textit{et al}\cite{overviewscl}.

Reference \onlinecite{paper5} studied isomorphs of atomic single- and multicomponent \textit{LJ} liquids with generalized exponents $m$ and $n$. It  was found that 
for given exponents ($m$, $n$) all isomorphs have the same shape in the $WU$-phase diagram, i.e., a so-called master isomorph exists from which all 
isomorphs can be generated via a simple scaling of the $WU$-coordinates. For instance, the shape of isomorphs in the $WU$-phase diagram of
the \textit{SCLJ} liquid and the \textit{KABLJ} liquid are the same.

References \onlinecite{paper1}-\onlinecite{paper5} focused on understanding strong $WU$-correlation and its implication for atomic systems.  
Schr{\o}der \textit{et al.}\cite{hidden} in $2008$ studied two rigid molecular liquids that are strongly correlating: the asymmetric dumbbell model and 
the Lewis-Wahnstr\"om \textit{OTP} model (see Sec. \ref{simulation}). At that time the 
isomorph concept had not yet been developed and state points with the same $\rho^{\gamma}/T$, as inspired from the \textit{IPL} system, were tested for 
collapse of, for instance, the reduced unit radial distribution function (note that in Refs. \onlinecite{paper1}, \onlinecite{paper2}, and \onlinecite{hidden} $\gamma$ is defined slightly 
different from subsequent papers). The dynamics in reduced units was also found to be a function of $\rho^{\gamma}/T$, to a good approximation, as is the case for \textit{IPL} systems\cite{paper3}.   
Chopra \textit{et al.} \cite{truskdumb} found that the $S_{ex}$ can be written (approximately) as a function of $\rho^{\gamma}/T$ 
for rigid symmetric \textit{LJ} dumbbells with different bond lengths. They also found that the reduced diffusion constant and relaxation time are functions of $S_{ex}$. 
These results suggest that the isomorph concept is relevant also for rigid molecular systems. In this paper we expand on earlier results by 
studying in detail the same systems as Schr{\o}der \textit{et al.}\cite{hidden}. 

The isomorph definition Eq. (\ref{defiso}) must be modified for rigid molecules, since the bond lengths are fixed and cannot follow the overall scaling. A simple modification of Eq. (\ref{defiso}), which 
is consistent with the atomic definition, is to define the mapping amongst configurations in terms of the molecular center of masses, instead of the atomic positions. We thus define 
two state points in the phase diagram of a liquid composed of \textit{rigid} molecules to be isomorphic 
if the following holds: Whenever two configurations of the state points 
have identical reduced center-of-mass coordinates for all molecules,

\begin{equation}\label{moleculeiso}
  \rho_{1}^{1/3} \textbf{r}^{(1)}_{CM,i} = \rho_{2}^{1/3} \textbf{r}^{(2)}_{CM,i}, 
\end{equation}
as well as identical Eulerian angles\cite{goldstein} 
\begin{align}
  \theta_{i}^{(1)} = \theta_{i}^{(2)}, \ \phi_{i}^{(1)} = \phi_{i}^{(2)}, \ \psi_{i}^{(1)} = \psi_{i}^{(2)}, \label{moleculeiso1}
\end{align}
these two configurations have proportional Boltzmann factors, i.e., [where $\textbf{R}$ $\equiv$ ($\textbf{r}_{CM,1}$, $\theta_{1}$, $\phi_{1}$, $\psi_{1}$, ..., 
$\textbf{r}_{CM,N}$, $\theta_{N}$, $\phi_{N}$, $\psi_{N}$)]

\begin{equation} \label{defiso1}
  e^{-U(\textbf{R}^{(1)})/k_{B}T_{1}} = C_{12}e^{-U(\textbf{R}^{(2)})/k_{B}T_{2}}.
\end{equation}
Again, $C_{12}$ is a constant that depends only on the state points ($1$) and ($2$). 
An isomorph is defined as a set of state points that are pairwise isomorphic. It should be noted that, in contrast to what is the case for atomic systems, 
since the bonds do not follow the overall scaling of the system this definition does not 
imply the existence of exact isomorphs for rigid molecules with 
intermolecular \textit{IPL} interactions,.

Taking the logarithm of Eq. (\ref{defiso1}) implies

\begin{equation}\label{linear}
 U(\textbf{R}^{(2)}) = T_{2}/T_{1}\cdot U(\textbf{R}^{(1)}) + k_{B}T_{2}\ln\, C_{12}.
\end{equation}
Equation (\ref{linear}) provides a convenient way of testing to which extent Eq. (\ref{defiso1}) is obeyed for a 
given system. A simulation is performed at one state point ($1$) and the obtained configurations are scaled to a different density $\rho_{2}$, where 
the potential energy is evaluated. The respective potential energies of the two state points are then plotted against each other. 
In the resulting plot a near straight-line \textit{indicates} that there exists an isomorphic state point with density $\rho_{2}$. The temperature $T_{2}$ of the isomorphic state point can be found from the slope of a 
linear regression fit. This procedure is termed the ''direct isomorph check''\cite{paper4}.
If this test is performed for an atomic \textit{IPL} system a correlation coefficient of $R$ = $1$ is obtained, consistent with these systems having exact isomorphs.

\begin{figure}[H]
  \centering
  \includegraphics[width=70mm]{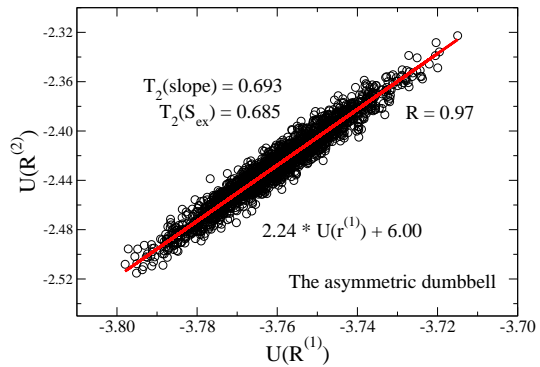}
  \caption{The ''direct isomorph check''\cite{paper4} for the asymmetric dumbbell model. During a simulation 
    at  state point ($\rho_{1}$, $T_{1}$) = (1.737, 0.309) the center of mass of each dumbbell
    is scaled to density $\rho_{2}$ =  1.998, keeping the Eulerian angles fixed. The potential energy is then evaluated from the scaled 
    configurations and plotted against the potential energy of the unscaled configurations. The temperature $T_{2}$(slope) 
    of the isomorphic state point at density $\rho_{2}$ is calculated by multiplying the linear regression slope with $T_{1}$ [Eq. (\ref{linear})].}
  \label{direct}
\end{figure}

As an example we perform a direct isomorph check for the asymmetric dumbbell model in Fig. \ref{direct}. A correlation 
coefficient of $R = 0.97$ is observed for a 15\% density increase. Calculating the temperature of 
the isomorphic state point from the linear regression slope the result differs only 1\% from
the prediction by requiring constant excess entropy (see Sec. \ref{iso1}). 
\newline \newline
In the present paper we show that liquids composed of simple rigid molecules have good isomorphs in their phase diagram as defined in Eqs. 
(\ref{moleculeiso}) - (\ref{defiso1}). Section \ref{isoinv} derives several isomorph invariants in molecular systems composed of rigid molecules.
Section \ref{simulation} describes the simulation 
setup and the investigated model systems. 
Section \ref{iso1} investigates the existence of isomorphs for the asymmetric dumbbell and the Lewis-Wahnstr{\"o}m \textit{OTP} models\cite{otp1, otp2}.
Section \ref{master} investigates the existence of a master isomorph\cite{paper5} for these model systems. Section \ref{sum} summarizes the results and presents an outlook.

\section{Isomorph invariants in liquids composed of rigid molecules}\label{isoinv}

From the single assumption of curves of isomorphic state points in an atomic liquid's phase diagram, Ref. \onlinecite{paper4} derived several invariants along an isomorph. Since we have  
extended this definition in  Eqs. (\ref{moleculeiso}) - (\ref{defiso1}) to molecular systems composed of rigid molecules, it is natural to wonder 
which of these invariants can be extended to molecular systems. The molecular isomorph concept 
is different from the atomic case in that there is no ''ideal'' reference system (the \textit{IPL} system). 
The following sections and simulations, however, show that isomorphs can nevertheless be a useful tool for understanding such liquids. 

In the following we derive several invariants from exact isomorphs.
We start by noting that the generalization of isomorphs to molecular systems define a bijective map amongst configurations of state points 
($1$) and ($2$). The \textit{NVT} configurational probability density for a system of $N$ rigid molecules is given by\cite{gubbins} (where $d\textbf{R} \equiv d\textbf{r}_{CM}^{N}d\tau^{N}$ 
with $\tau$ $\equiv$ ($\theta$, $\phi$, $\psi$) and $d\tau = sin \ \theta \ d\theta \, d\phi \, d\psi$)

\begin{align}
  \hat{P}(\textbf{R}) & = \frac{ e^{- U(\textbf{R})/k_{B}T}}{\int e^{- U(\textbf{R})/k_{B}T}d\textbf{R}}.
\end{align}
In combination with Eq. (\ref{defiso1}) it follows that all mapped configurations of state points ($1$) and ($2$) have identical Boltzmann probabilities, i.e.,

\begin{equation}\label{test}
  \hat{P}(\textbf{R}^{(1)})d\textbf{R}^{(1)} = \hat{P}(\textbf{R}^{(2)})d\textbf{R}^{(2)}.
\end{equation}
For convenience we introduce two configurational distribution functions\cite{paper4,gubbins}

\begin{align}
  P(\textbf{R}) & = (V\Omega)^{N}\hat{P}(\textbf{R}), \\
  \tilde{P}(\tilde{\textbf{r}}_{CM}^{N},\tau^{N}) & =  \frac{ e^{- U(\rho^{-1/3}\tilde{\textbf{r}}_{CM}^{N},\tau^{N})/k_{B}T}}{\int e^{- U(\rho^{-1/3}\tilde{\textbf{r}}_{CM}^{N},\tau^{N})/k_{B}T}d\tilde{\textbf{r}}_{CM}^{N}d\tau^{N}},
\end{align}
where $\Omega$ is the integral over the Eulerian angles for one molecule ($\Omega = 8\pi^{2}$ for a non-linear molecule).
$P$ has been introduced to make $\hat{P}$ dimensionless.
$\tilde{P}(\tilde{\textbf{r}}_{CM}^{N},\tau^{N})d\tilde{\textbf{r}}_{CM}^{N}d\tau^{N}$ is the probability to observe the system represented by a point in the 
volume-element $d\tilde{\textbf{r}}_{CM}^{N}d\tau^{N}$ located at $\{\tilde{\textbf{r}}_{CM}^{N},\tau^{N}\}$. $\tilde{P}$ is invariant along an isomorph and is 
related to $P$ via $\tilde{P}(\tilde{\textbf{r}}_{CM}^{N},\tau^{N}) = (N\Omega)^{-N}P(\textbf{R})$ = $\rho^{-N}\hat{P}(\textbf{R})$. 
We note that the excess entropy $S_{ex} = - (\partial F_{ex}/\partial T)_{N,V}$, where $F_{ex}$ is the excess free energy, can be written as\cite{paper4}

\begin{align}
  S_{ex} & = -k_{B}\int (V\Omega)^{-N}P(\textbf{R}) \ln P(\textbf{R})d\textbf{R},  \\
  & = -k_{B}\int \tilde{P} \ln \tilde{P} \ d\tilde{\textbf{r}}_{CM}^{N}d\tau^{N} - k_{B}N \ln(N\Omega).\label{excess}
\end{align}
From the above observations we can now derive a number of isomorph invariants in liquids composed of rigid molecules.
 
\begin{enumerate}

\item \textit{The molecular center-of-mass structure in reduced units}. For a given configuration of the molecular center-of-mass structure in reduced units, 
  all orientations of the molecules of state points ($1$) and ($2$) by Eq. (\ref{test}) have identical probabilities. The reduced center-of-mass 
  structure is thus invariant along an isomorph. \label{a1}

\item \textit{Any normalized distribution function describing the (relative) orientations of molecules with respect to their center-of-mass}. \label{a2}
  All orientations of the molecules with respect to a given molecular center-of-mass configuration of state point ($1$) are mapped 
  to configurations of state point ($2$) with identical probabilities. It thus follows that the normalized distribution function is invariant along an isomorph.
  
\item \textit{The isochoric heat capacity $C_{V}$}. The excess heat capacity in the \textit{NVT} ensemble is given by $C_{V}^{ex} =  \langle (\Delta U)^{2}\rangle /k_{B}T^{2}$. 
  Defining $X$ = $U/k_{B}T$ we may write $C_{V}^{ex} = k_{B}\langle (\Delta X)^{2} \rangle$. By Eqs. (\ref{linear}) and (\ref{test}) it follows that 
  $C_{V}^{ex}$ is invariant along an isomorph, since the constant $k_{B}T_{2}\ln\, C_{12}$ disappears when subtracting the mean. The ideal gas contribution 
  to $C_{V}$ is independent of state point ($C_{V}^{id} = 6Nk_{B}/2$ for non-linear molecule).

\item \textit{The translational two-body entropy\cite{excessexp,ex,truskdumb} $S_{t}/N$ = $-\rho k_{B}/2 \int [g_{CM}(r) \ln g_{CM}(r) - g_{CM}(r) + 1]d\textbf{r}$}, where
  $g_{CM}(r)$ is the radial distribution function for the center of mass of the molecules.
  The density dependence disappears when switching to reduced units and by Statement \ref{a1}. the molecular 
  center-of-mass structure in reduced units is invariant along an isomorph, and thus also
  the radial distribution function (in reduced units). 

\item \textit{The orientational two-body entropy\cite{excessexp,ex,truskdumb} $S_{o}/N$ = $-\rho k_{B}/(2\Omega^{2}) \int g_{CM}(r) g(\omega^{2}|r) \ln g(\omega^{2}|r) d\omega^{2}d\textbf{r}$}, where
  $\omega^{2}$ denotes a set of angles used to describe the relative orientation of two molecules, and $g(\omega^{2}|r)$ is the conditional distribution
  function for the relative orientation of two molecules separated by a distance $r$. Applying reduced units this invariant follows from 
  Statements \ref{a1}. and \ref{a2}. 

\item \textit{All $N$-body entropy terms}\cite{excessexp,ex}. The excess entropy can be expanded as $S_{ex}$ = $\sum_{i=1}^{\infty}S_{i}$. The two-body expression $S_{2} = S_{t} + S_{o}$ is given above, where as the 
  higher-order terms are more involved. \label{an}

\item \textit{The excess entropy $S_{ex}$}. The excess entropy is given by Eq. (\ref{excess}), and since $\tilde{P}$ is invariant along isomorph, so is the excess entropy. The latter also follows from Statement \ref{an}., 
  since each term is invariant.

\item \textit{The molecular center-of-mass \textit{NVE} and Nos$\acute{e}$-Hoover \textit{NVT} dynamics in reduced units}. 
  The reduced dynamics of the atomic positions on account of the constraints is \textit{not} invariant along an isomorph. Considering instead
  the molecular center-of-mass motion the constraint force disappears and these equations of motion are invariant along an isomorph in reduced units.
  The proof is given in Appendix \ref{appInvdyn} (a brief summary of constrained dynamics is given in Appendix \ref{appConstrained}). \label{a7}

\item \textit{Any average molecular center-of-mass dynamic quantity in reduced units}. This follows immediately from Statement \ref{a7}., 
  since the molecular center-of-mass equations of motion in reduced units are invariant along an isomorph. In particular, this would include a reduced relaxation time $\tilde{\tau}_{\alpha}$.

\end{enumerate}
As detailed above it is necessary to consider the (reduced) center of mass motion and the motion relative to the center of mass separately. Nevertheless, during 
the investigation of isomorphs we will also consider the reduced atomic quantities to examine their ''invariance''.

An additional consequence of isomorphs is that, since by Eq. (\ref{test}) two isomorphic state points have identical canonical probabilities, 
an instantaneous change of temperature and density from an equilibrated state point to another isomorphic state point does not lead to any relaxation. This is called an isomorphic jump\cite{paper4}.

\section{Simulation details}\label{simulation}

\begin{figure}[H]
  \centering
  \includegraphics[width=70mm]{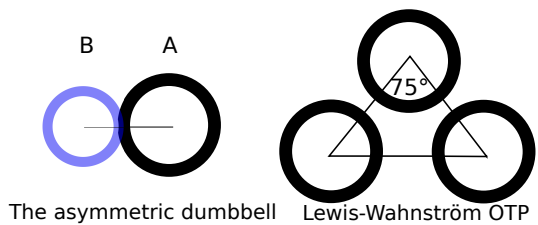}
  \caption{A sketch of the two model systems studied: The asymmetric dumbbell and Lewis-Wahnstr\"om \textit{OTP} models. The asymmetric dumbbell is a simplistic model of 
    toluene with the methyl side-group tightly bonded to the benzene molecule. The Lewis-Wahnstr\"om \textit{OTP} model is an isosceles triangle with an angle of 
    $75^\circ$, different from the $60^{\circ}$ of the real 1,2-diphenylbenzene molecule\cite{otpulf}.}
  \label{cartoon}
\end{figure}

We studied two model systems of rigid molecules (Fig. \ref{cartoon}): the asymmetric dumbbell model ($N$ = 500) and the Lewis-Wahnstr{\"o}m \textit{OTP} model ($N$ = 320).  

For both models the potential energy $U$ and the virial $W$ are given by

\begin{align}
  U & = U_{LJ} \\
  W & = W_{LJ} + W_{CON}.
\end{align}
The first term in the virial is the \textit{LJ} virial $W_{LJ}$, the second term is the contribution to 
the virial due to the constraints (fixed bond lengths), $W_{CON}$. $U_{LJ}$ is a sum over intermolecular pair interactions given by the ($12$, $6$)-\textit{LJ} potential 

\begin{equation}
  u(r_{ij}) = 4\epsilon_{\alpha \beta}\Big[ \Big(\frac{\sigma_{\alpha \beta}}{r_{ij}}\Big)^{12} - \Big(\frac{\sigma_{\alpha \beta}}{r_{ij}}\Big)^{6}\Big].
\end{equation}
The potential energy has no contribution from the fixed bonds. A force smoothing procedure\cite{gromacs} was applied from $r_{s} = 2.45\sigma_{\alpha \beta}$ to $r_{c} = 2.50\sigma_{\alpha \beta}$. 

The bond lengths were held fixed using the Time Symmetrical Central Difference (\textit{TSCD}) algorithm \cite{tox1,tox2}, which is a 
central difference time-discretization of the constrained equations of motion preserving time-reversibility. The simulations were 
performed in the \textit{NVT} ensemble applying the Nos$\acute{e}$-Hoover (\textit{NH}) algorithm\cite{nose,hoover,canotox} using \textit{RUMD}\cite{rumd}, 
a molecular dynamics package optimized for state of the art \textit{GPU}-computing. 

The \textit{NVT} simulations were performed without adjusting the time-constant of the \textit{NH} algorithm (see Appendix \ref{appInvdyn}). This choice is not expected to influence the results  
over the observed density and temperature range, since the dynamics is not 
particular sensitive to the absolute value of the \textit{NH} time-constant\cite{frenkel}. 

Appendix \ref{appConstrained} gives a brief summary of constrained dynamics 
and the effect on the virial (see also Refs. \onlinecite{tox1}, \onlinecite{edberg}, and \onlinecite{ryck}). The specific details of the investigated two models follow below.

\subsection{The asymmetric dumbbell}

The asymmetric dumbbell model consists of a big ($A$) and small ($B$) \textit{LJ} particle, rigidly bonded  with bond distance of $r_{ij} = 0.584$ (here and henceforth units are given 
in \textit{LJ} units referring to the $A$ particle, $\sigma_{AA}$ = 1, $\epsilon_{AA}$ = 1, and $m_{A}$ = 1). The parameters were chosen to roughly mimic Toluene \cite{hidden}. 
The asymmetric dumbbell model can be cooled to a highly viscous state without crystallizing, making it feasible to study slow dynamics. 
The asymmetric dumbbell model has $\sigma_{AB}$ = 0.894, $\sigma_{BB}$ = 0.788, $\epsilon_{AB}$ = 0.342, and $\epsilon_{BB} $ = 0.117 with $m_{B}$ = 0.195.

\subsection{Lewis-Wahnstr{\"o}m \textit{OTP}}

The Lewis-Wahnstr{\"o}m \textit{OTP} model\cite{otp1,otp2} consists of three identical \textit{LJ} particles rigidly bonded together in an isosceles triangle with sides of $r_{ij} = 1.000$ and 
top-angle of $75^\circ$, i.e., different from the $60^\circ$ of the real 1,2-diphenylbenzene molecule\cite{otpulf}. All parameters (including the masses) are unity for the \textit{OTP} model.

\section{Numerical study of isomorphs for the two model systems}\label{iso1}

In order to investigate whether the two model systems have good isomorphs, we first describe how to generate an isomorph in a simulation.
The excess entropy $S_{ex}$ is invariant along an isomorph, and a method for 
generating an isomorph is to generate a curve of constant $S_{ex}$ (see Sec. \ref{isoinv} and also Refs. \onlinecite{paper4} and \onlinecite{paper5}). 
A curve of constant excess entropy can be 
found by using the exact \textit{NVT} ensemble relation\cite{paper4}

\begin{equation}\label{iso}
  \frac{\langle \Delta U \Delta W \rangle}{\langle (\Delta U)^{2} \rangle} = \Big(\frac{\partial \ln T}{\partial \ln \rho}\Big)_{S_{ex}} \equiv \gamma.
\end{equation}
In simulations an isomorph is generated as follows: 1) The left-hand side is calculated from the 
fluctuations at a given state point; 2) A new state point is identified by a dicretization of Eq. (\ref{iso}) by changing the density by 1\%, where the new 
temperature is calculated from $\Delta \ln T = \gamma \Delta \ln \rho$; 3) The procedure 
is repeated and in this way an isomorph is generated in the phase diagram. 

\subsection{Isomorphs of the asymmetric dumbbell model}\label{isoasymsec}

\begin{figure}[H]
  \centering
  \includegraphics[width=70mm]{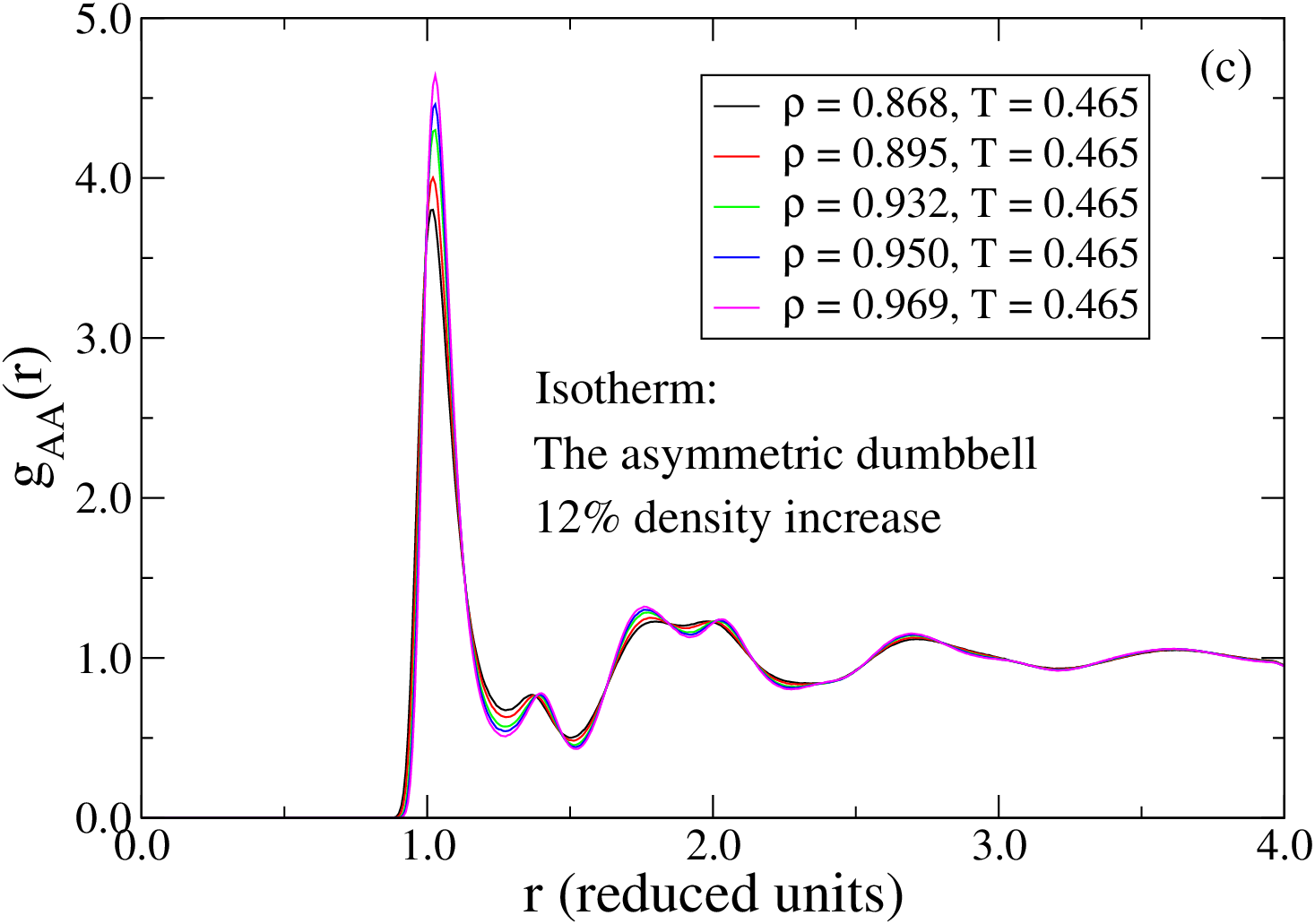}
  \caption{Radial distribution functions for the asymmetric dumbbell model. (a) $AA$ pair-correlation function
    along an isomorph with 19\% density increase before scaling the distance $r$ into reduced units $\tilde{r} = \rho^{1/3}r$. (b) $AA$ pair-correlation function along the same isomorph 
    after scaling the distance $r$ into reduced units. (c) $AA$ pair-correlation function along an isotherm with $12\%$ density increase
    function after scaling of the distance.}
  \label{asymrdfAA}
\end{figure}

\begin{figure}[H]
  \centering
  \includegraphics[width=70mm]{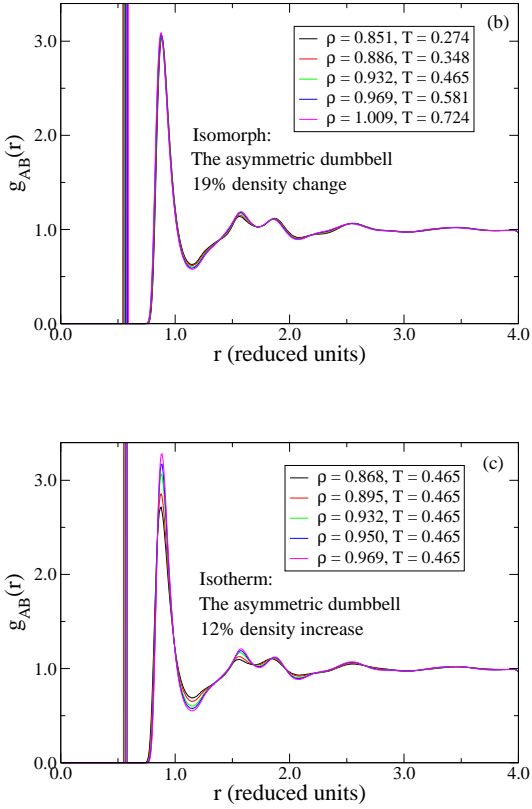}
  \caption{Radial distribution functions for the asymmetric dumbbell model. (a) $AB$ pair-correlation function
    along the isomorph of Fig. \ref{asymrdfAA} before scaling the distance $r$ into reduced units $\tilde{r} = \rho^{1/3}r$. (b) $AB$ pair-correlation function along the same isomorph 
    after scaling the distance $r$ into reduced units. (c) $AB$ pair-correlation function along the isotherm of Fig. \ref{asymrdfAA} after scaling of the distance.}
  \label{asymrdfAB}
\end{figure}

This section investigates the asymmetric dumbbell model. Isomorphs were mapped out following the procedure described above.
Figure \ref{asymrdfAA} shows the $AA$ radial distribution functions along an isomorph with 19\% density increase before ($a$) and after ($b$) scaling the distance
$r$ into reduced units

\begin{equation}
  \tilde{r} = \rho^{1/3}r.
\end{equation}
Also shown for reference in Fig. \ref{asymrdfAA}(c) is the $AA$ radial distribution functions along an isotherm with 12\% density increase.
Although the reduced structure of the atomic positions is not predicted to be invariant along 
an isomorph, Fig. \ref{asymrdfAA}  shows that it nevertheless is a resonable approximation. The reduced structure of the atomic positions is less invariant along the isotherm.
\newline \newline
\begin{figure}[H]
  \centering
  \includegraphics[width=70mm]{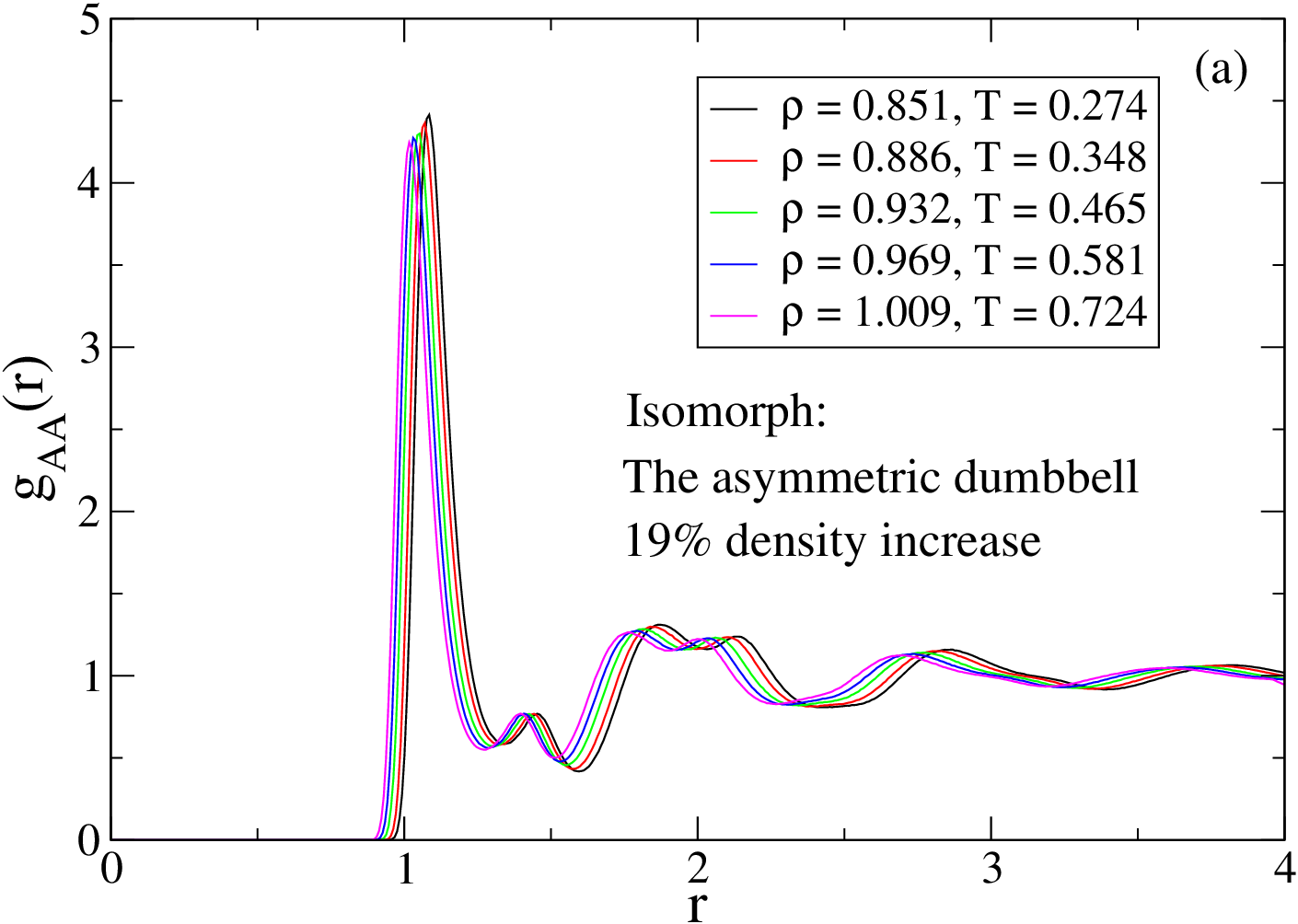}
\end{figure}
\begin{figure}[H]
  \centering
  \includegraphics[width=70mm]{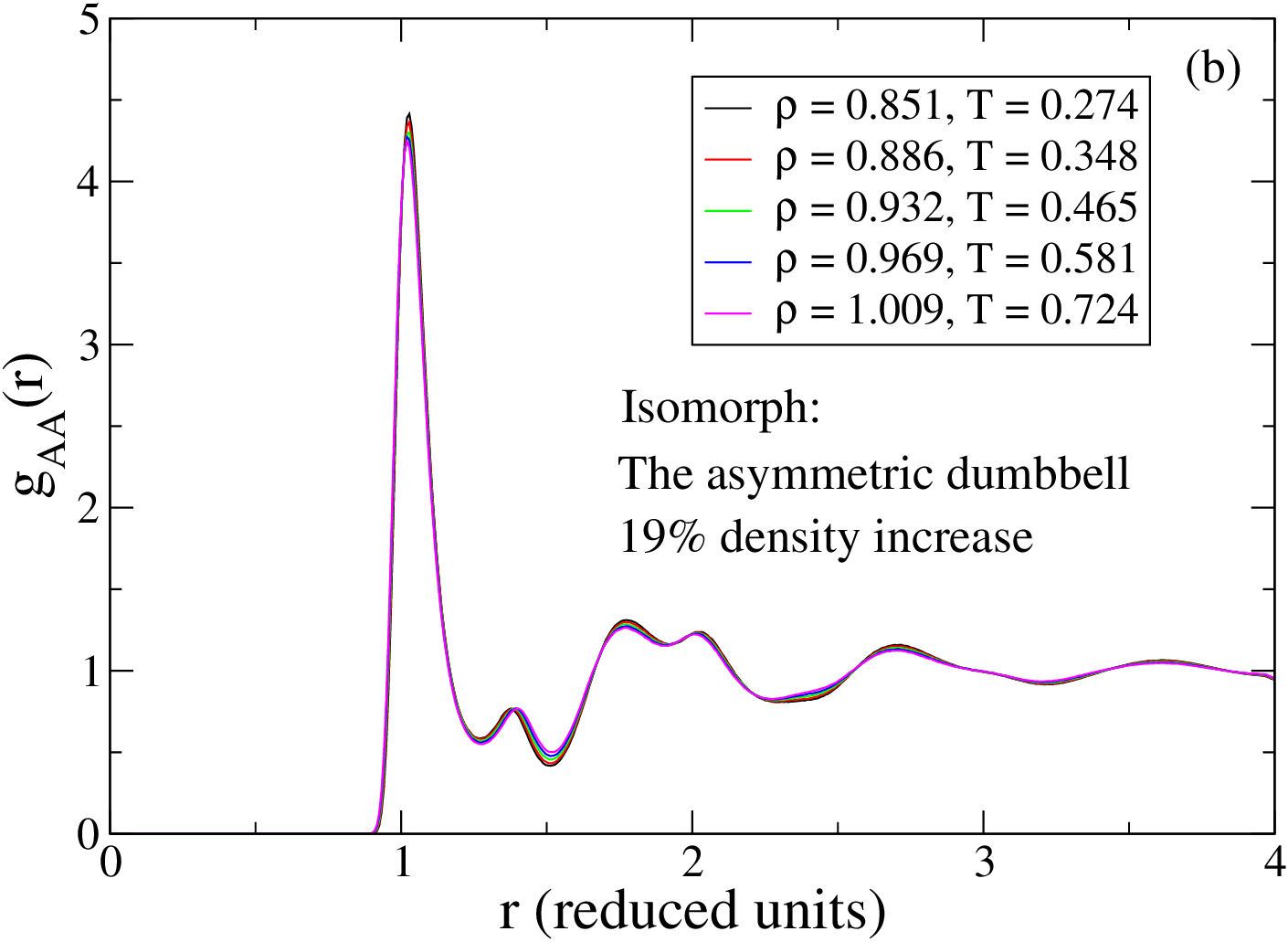}
\end{figure}    
Figure \ref{asymrdfAB} considers the $AB$ radial distribution functions, where 
the constrained bond distance shows up as a sharp peak. The analogous conclusion as with the $AA$ distribution functions is reached, and likewise for the $BB$ 
distribution functions (not shown).

\begin{figure}[H]
  \centering
  \includegraphics[width=70mm]{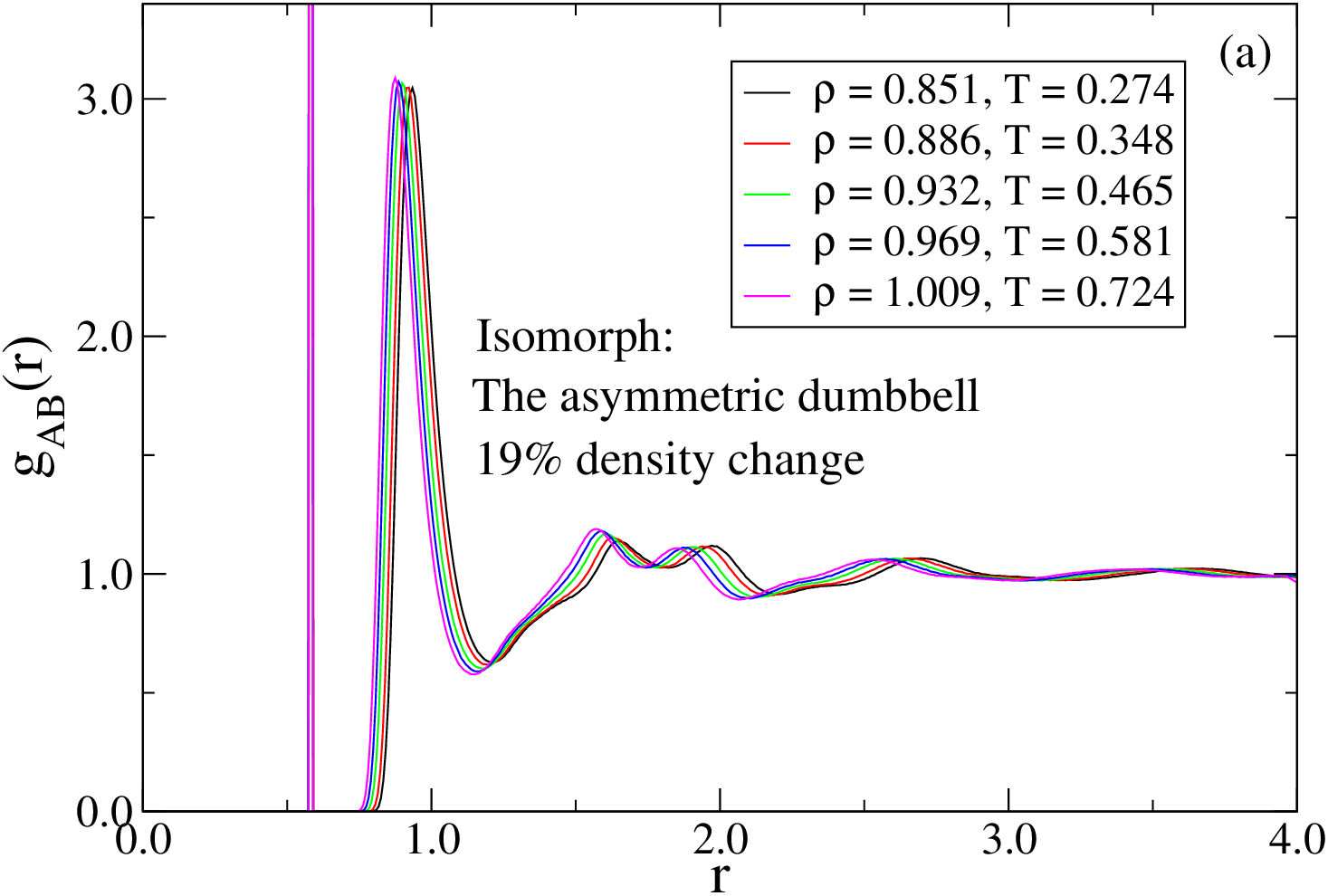}
\end{figure}  
 \begin{figure}[H]
  \centering
  \includegraphics[width=70mm]{dumbbell_isomorph1_rdfAB_after}
\end{figure} 
\begin{figure}[H]
  \centering
  \includegraphics[width=70mm]{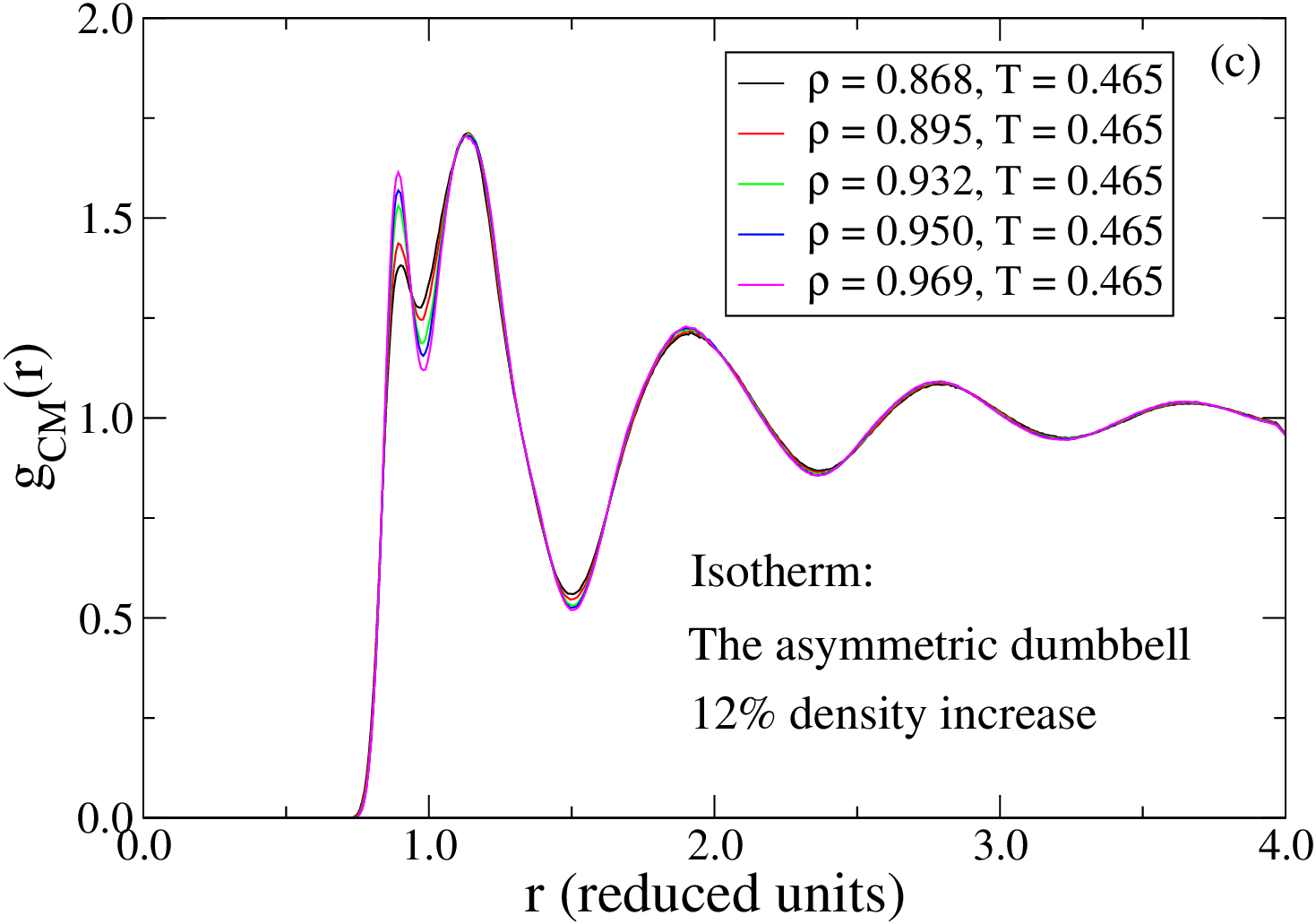}
  \caption{Molecular center-of-mass radial distribution functions for the asymmetric dumbbell model. (a) Pair-correlation function
    along the isomorph of Figs. \ref{asymrdfAA}-\ref{asymrdfAB} before scaling the distance $r$ into reduced units $\tilde{r} = \rho^{1/3}r$. (b) Pair-correlation function along 
    the same isomorph after scaling the distance $r$ into reduced units. (c) Pair-correlation function along the isotherm of Figs. \ref{asymrdfAA}-\ref{asymrdfAB} 
    after scaling of the distance.}
  \label{asymrdfCM}
\end{figure}

Next, we consider in Fig. \ref{asymrdfCM} the molecular center-of-mass radial distribution functions along the isomorph and isotherm of Figs. \ref{asymrdfAA}-\ref{asymrdfAB}. 
This quantity is predicted to be invariant along an isomorph (see Sec. \ref{isoinv}). The molecular center-of-mass structure is to a good approximation invariant in reduced units along the isomorph, while this 
is less so along the isotherm as can be seen from the first peak.
\newline \newline
\begin{figure}[H]
  \centering
  \includegraphics[width=70mm]{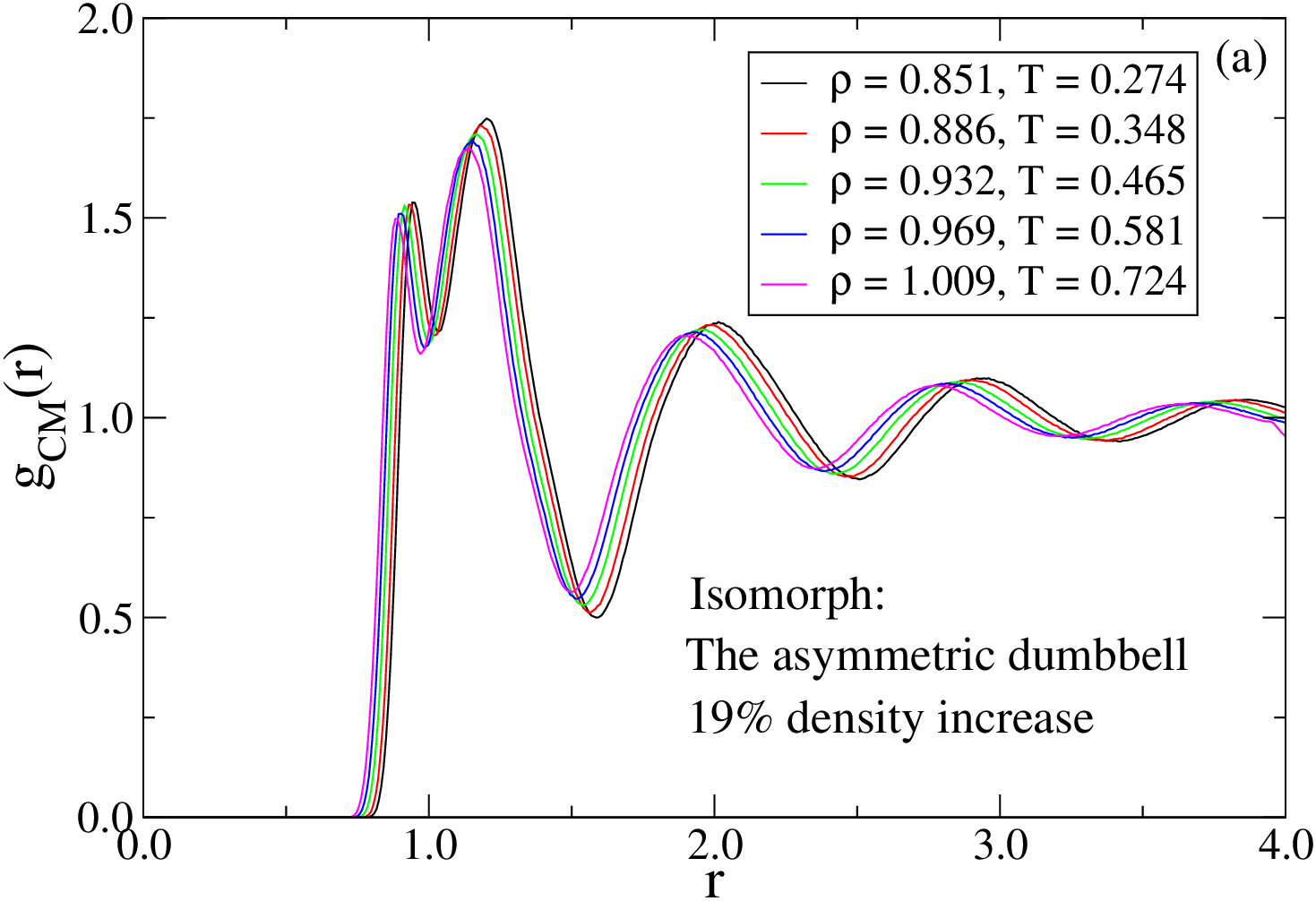}
\end{figure}
\begin{figure}[H]
  \centering
  \includegraphics[width=70mm]{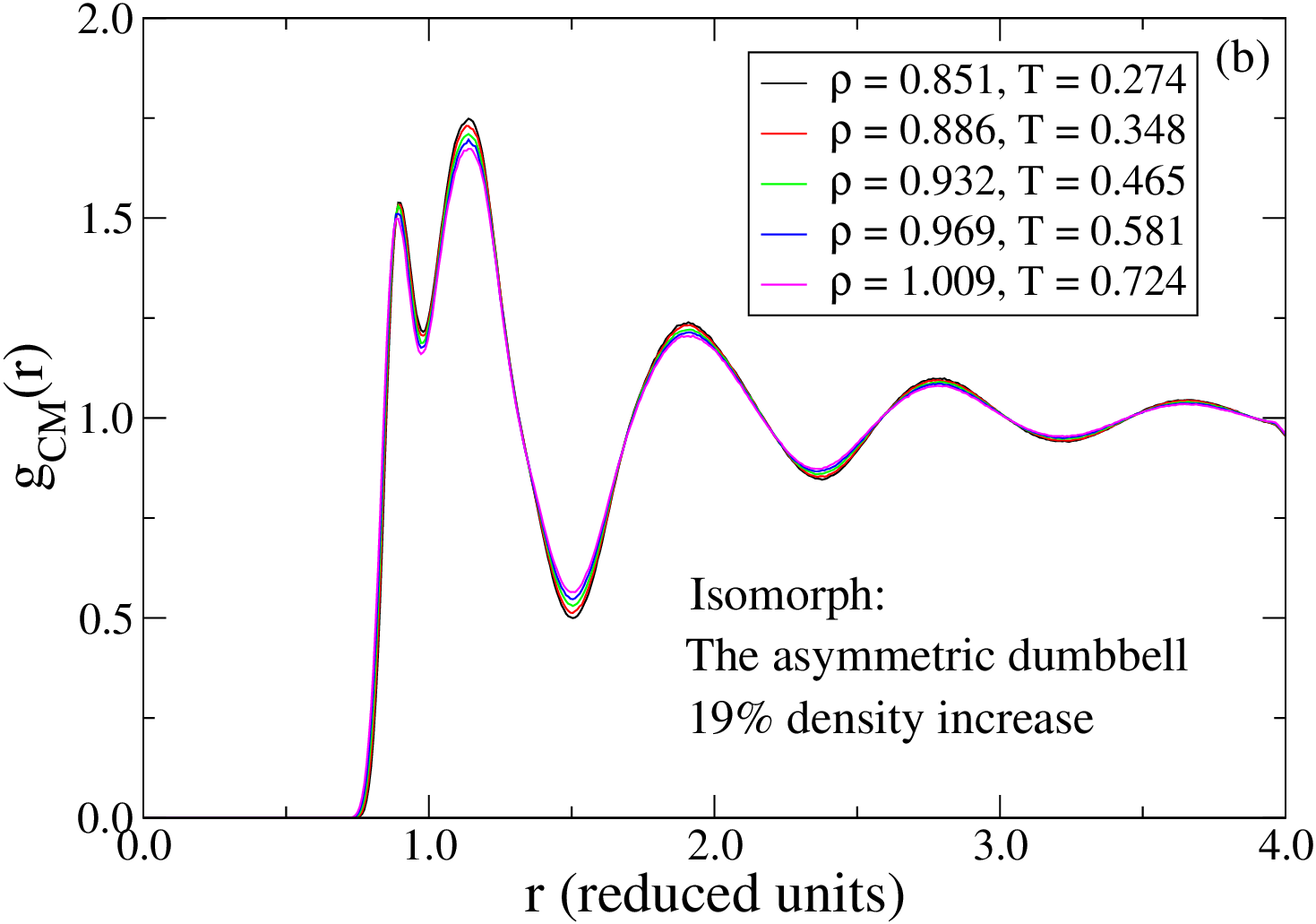}
\end{figure}    
We consider in Fig. \ref{asymfs} the dynamics in terms of the reduced $A$-particle incoherent intermediate scattering function.
The reduced dynamics of the atoms is not predicted to be invariant along an isomorph (see Appendix \ref{appInvdyn}), however, the figure shows that it is a good approximation.
The same conclusion is reached for the $B$-particle (not shown).

\begin{figure}[H]
  \centering
  \includegraphics[width=70mm]{dumbbell_isotherm1_FsA_after}
\end{figure}    
\begin{figure}[H]
  \centering
  \includegraphics[width=70mm]{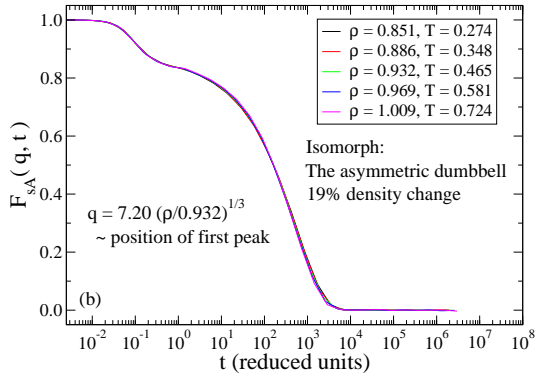}
  \caption{The reduced $A$-particle incoherent intermediate scattering function for the asymmetric dumbbell model keeping the reduced wavevector $q$ constant. 
    (a) Along the isotherm of Figs. \ref{asymrdfAA}-\ref{asymrdfCM} with 12\% density increase. (b) Along the isomorph of Figs. \ref{asymrdfAA}-\ref{asymrdfCM} with 19\% density increase.}
  \label{asymfs}
\end{figure}
\begin{figure}[H]
  \centering
  \includegraphics[width=70mm]{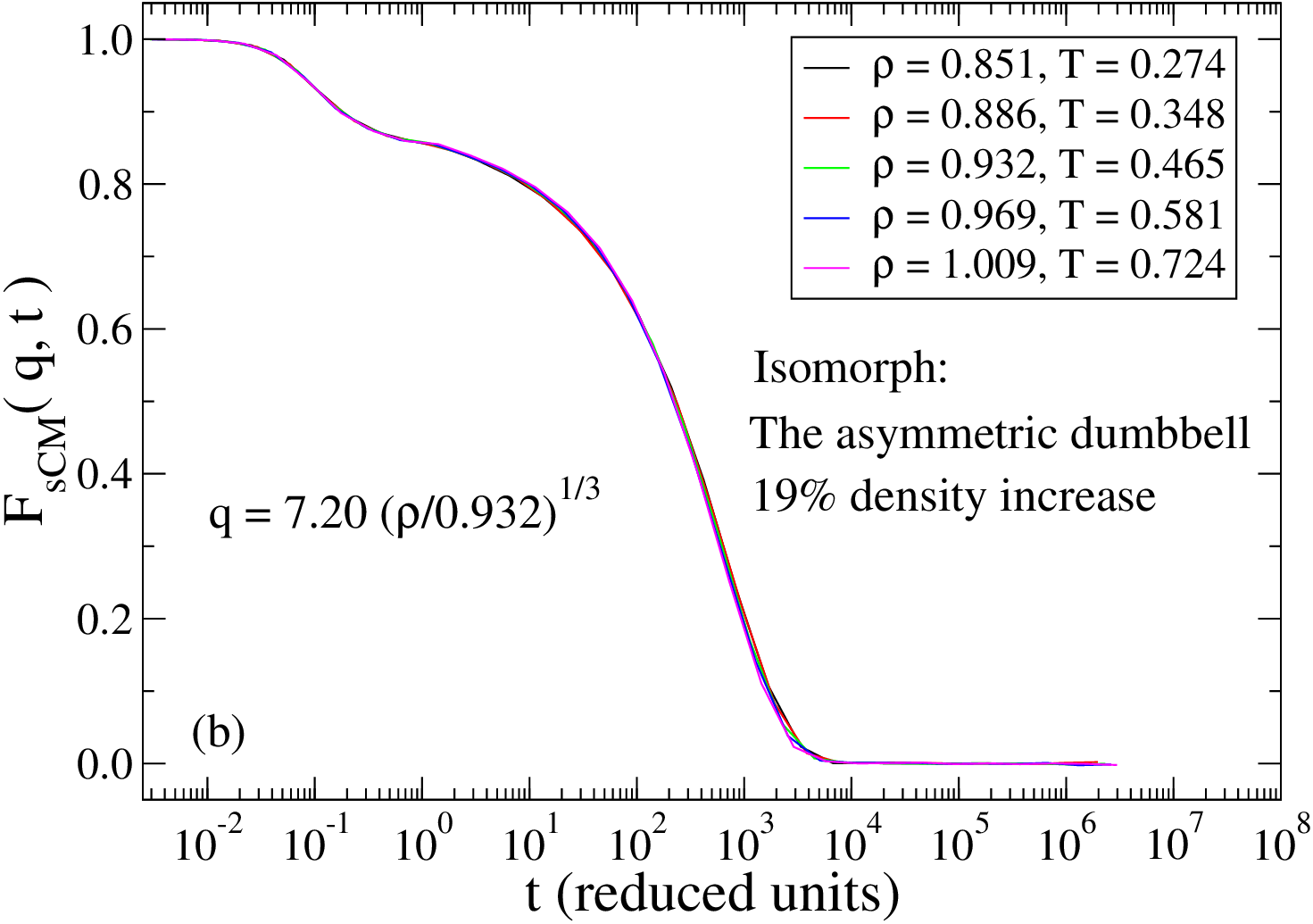}
  \caption{The reduced molecular center-of-mass incoherent intermediate scattering function for the asymmetric dumbbell model keeping the reduced 
    wavevector $q$ constant. (a) Along the isotherm of Figs. \ref{asymrdfAA}-\ref{asymfs} with 12\% 
    density increase. (b) Along the isomorph of Figs. \ref{asymrdfAA}-\ref{asymfs} with 19\% density increase.}
  \label{asymfsCM}
\end{figure}

   
In Figure \ref{asymfsCM} we consider the reduced molecular center-of-mass self part of the intermediate scattering function. This quantity is predicted to be invariant along an 
isomorph (see Appendix \ref{appInvdyn}), and Fig. \ref{asymfsCM} clearly shows this. The dynamics is not invariant along the isotherm.
\newline \newline
\begin{figure}[H]
  \centering
  \includegraphics[width=70mm]{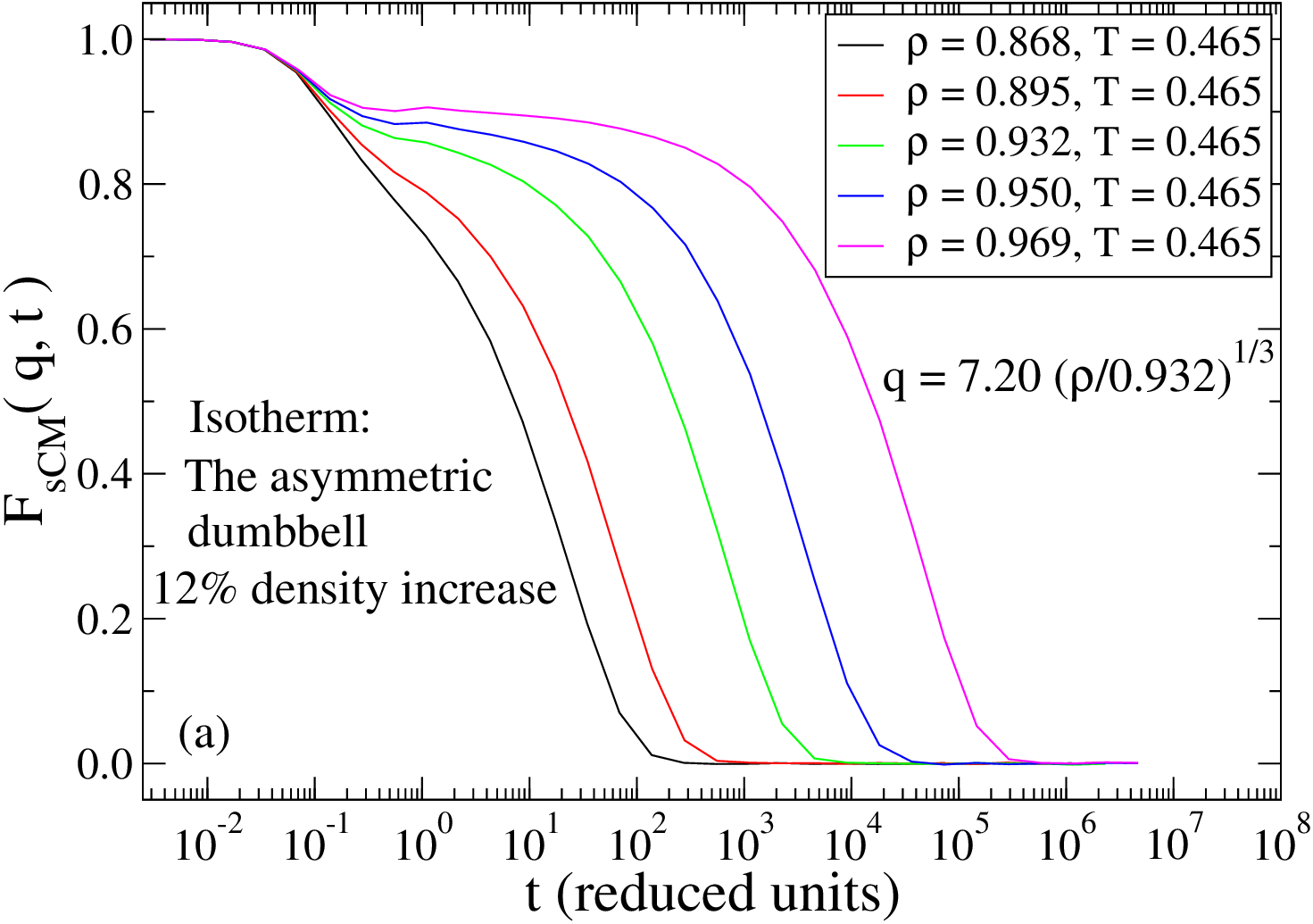}
\end{figure}    
We show the variation of $\gamma$, calculated from the \textit{NVT} fluctuations via Eq. (\ref{iso}), in Fig. \ref{gammadumbbell} along an isochore and 
along the isomorph of Figs. \ref{asymrdfAA}-\ref{asymfsCM} 
in two different versions. The crosses show $\gamma$ calculated from the total virial $W$ while the asterisks show $\gamma$ calculated after subtracting the 
constraint contribution to virial, i.e., replacing $W$ with $W_{LJ} = W - W_ {CON}$. The insets show the corresponding correlation coefficients $R$.
Reference \onlinecite{paper4} predicts that $\gamma$ should be a function of density only $\gamma$ = $\gamma(\rho)$. This is seen in Fig. \ref{gammadumbbell} to be a 
good approximation for both versions of $\gamma$, where the crosses are the $\gamma$ used to keep the excess entropy constant. 

As mentioned in the introduction, density scaling\cite{dscale,dscale1,reviewRoland} is the 
empirical observation that the relaxation time $\tau_{\alpha}$ for many viscous liquids can be written as some 
function $\tau_{\alpha} = f(\rho^{\gamma_{scale}}/T)$ where $\gamma_{scale}$ in experiments is a fitting exponent. In 
performing these fits reduced units are often not used; however, the importance of using reduced units in experimental data has only recently been pointed out\cite{dscale2}. 
If we assume that $\gamma$ is constant along an isomorph, Eq. (\ref{iso}) implies that $\rho^{\gamma}/T = const$ describes the isomorph. In this case density scaling will thus
hold to a good approximation since the reduced relaxation time is an isomorph invariant\cite{hidden}; for the dumbbell system $\gamma$ changes only 
moderately along an isomorph and 
thus density scaling is a good approximation for this system. That $\gamma$ for systems with isomorphs can be identified with the density scaling 
exponent $\gamma_{scale}$ has very recently been verified expementially for a silicone oil\cite{gamma}.
\newline \newline
\begin{figure}[H]
  \centering
  \includegraphics[width=70mm]{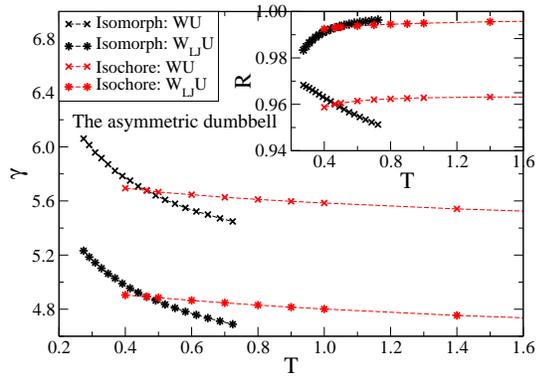}
  \caption{The variation of $\gamma$ [Eq. (\ref{iso})] and the correlation coefficient $R$ [Eq. (\ref{R})] for the asymmetric dumbbell model 
    in two different versions along an isochore (red, $\rho$ = 0.932) and along the isomorph (black) of Figs. \ref{asymrdfAA}-\ref{asymfsCM}. The crosses show $\gamma$ calculated 
    from the total virial $W$ while the asterisks show $\gamma$ calculated after subtracting the 
    constraint contribution to virial, i.e., $W_{LJ} = W - W_ {CON}$. The corresponding $R$'s are shown in the insets. $\gamma$ is predicted in 
    Ref. \onlinecite{paper4} to be a function only of density which is seen to apply to a good approximation for both versions.}
  \label{gammadumbbell}
\end{figure}
Starting from an equilibrated sample at some state point, changing either temperature or density alters the equilibrium Boltzmann distribution of states. 
Two isomorphic state points have identical Boltzmann canonical probabilities (Eq. (\ref{test})). A sudden change of state from one state point to another isomorphic state point should thus not lead 
to any relaxation. This is called an isomorphic jump, and the prediction of no relaxation was shown in Ref. \onlinecite{paper4} to 
work well for the \textit{KABLJ} liquid. 

\begin{figure}[H]
  \centering
  \includegraphics[width=70mm]{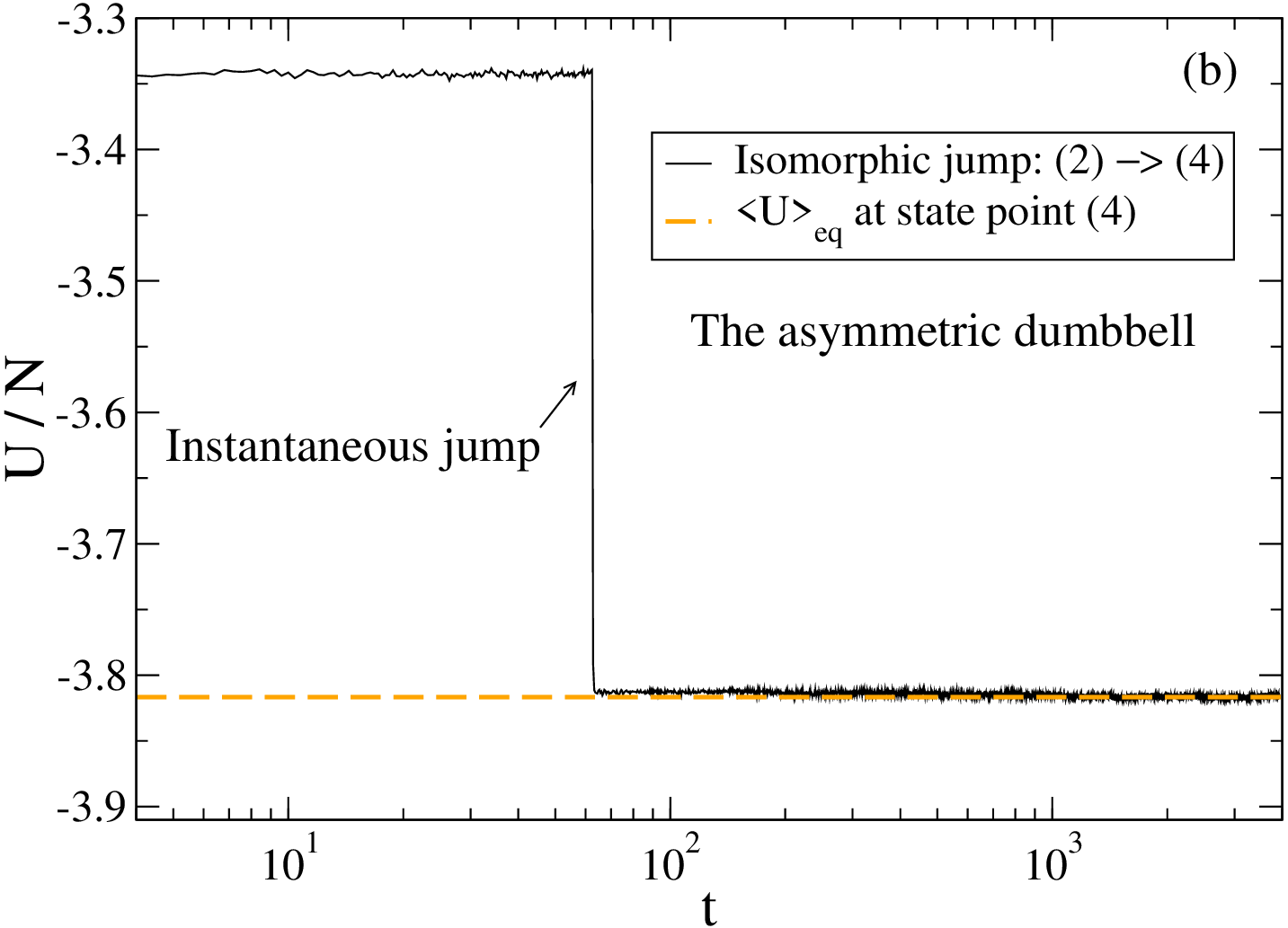}
  \caption{Four state points ($1$), ($2$), ($3$), and ($4$) corresponding to, respectively, ($\rho$, $T$) = ($0.932$, $0.400$), ($0.932$, $0.465$), ($0.932$, $2.000$), and ($0.851$, $0.274$) are given where the 
    first three state points are on the same isochore. State points ($2$) and ($4$) are isomorphic, 
  while ($1$) and ($3$) are not isomorphic to ($4$).  After equilibrating at state points ($1$), ($2$), and ($3$), respectively, the temperature and density are 
  instantaneously changed to that of state point ($4$) 
  via a scaling of the coordinates keeping bond distances and Eulerian angles of the molecules fixed.
  An average has been performed over 100 samples. (a) The relaxational behaviour of all state points quantified by the potential energy $U$.
  The jump ($2$) $\to$ ($4$) shows no relaxation while the other state points do.
  (b) Close up of the potential energy of state point ($2$) before and after the jump, where the jump takes place at $t \approx 60$. }
  \label{isojumpd}
\end{figure}

A similar numerical experiment is carried out for the asymmetric dumbbell model in Fig. \ref{isojumpd}. 
Considering three equilibrated, isochoric state points ($1$), ($2)$, and ($3$), the density and temperature are instantaneously changed to a state point ($4$). State point ($4$) is isomorphic to 
state point ($2$). The isomorph prediction is that jumps from state points ($1$) and ($3$) show relaxation, but not jumps from state point ($2$). This is indeed the case (Fig. \ref{isojumpd}(a)). 
State point ($1$) ages from below  since 
the aging scheme ($1$) $\to$ ($4$) can be described as first an instantaneous isomorphic jump to the correct density, but a lower temperature, and subsequently relaxation from this state point along the  
isochore of state point ($4$).

\begin{figure}[H]
  \centering
  \includegraphics[width=70mm]{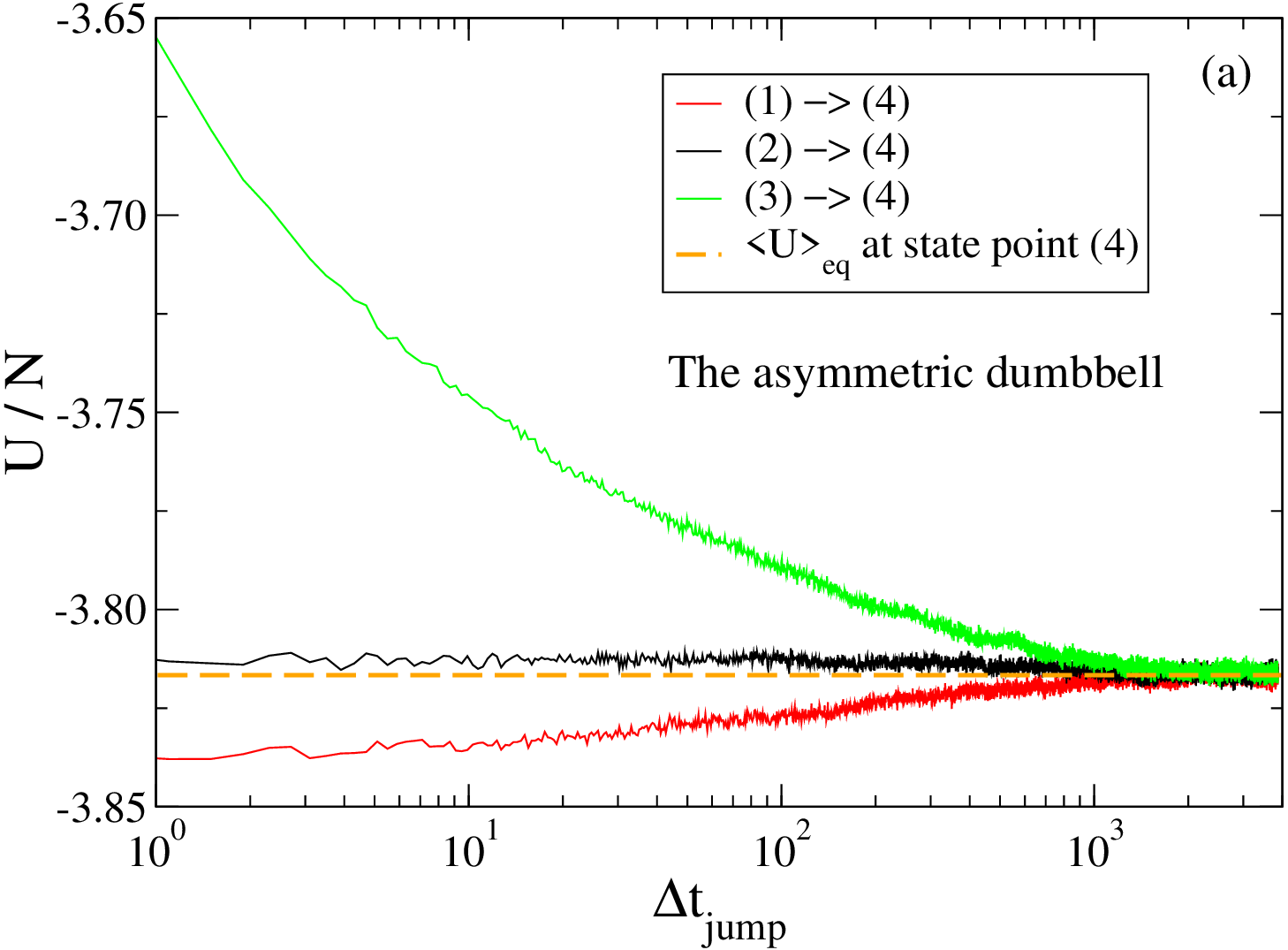}
\end{figure}
\begin{figure}[H]
  \centering
  \includegraphics[width=70mm]{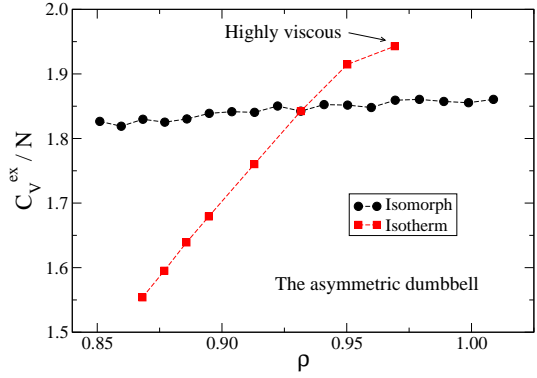}
  \caption{The excess heat capacity per particle $C_{V}^{ex} / N$ for the asymmetric dumbbell model as a function of density along the isomorph (black) and isotherm (red, $T$ = $0.465$)  
    of Figs. \ref{asymrdfAA}-\ref{asymfsCM}. 
    The density increase is 19\% and 12\%, respectively. The excess heat capacity increases less than 2\% along the isomorph, while the isotherm shows a 25\% increase. For the isotherm the dynamics 
    becomes very slow for densities higher than $\rho = 0.950$ and the system becomes difficult to equilibrate properly. }
  \label{heatd}
\end{figure}

\begin{figure}[H]
  \centering
  \includegraphics[width=70mm]{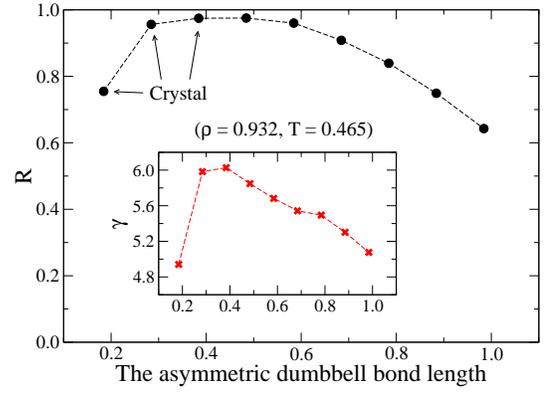}
  \caption{The correlation coefficient $R$ as a function of the bond length in the asymmetric dumbbell model at ($\rho$, $T$) = (0.932, 0.465). The investigated model has bond length $0.584$ with a correlation 
    coefficient $R \approx 0.97$; however, as the bond length increases, the correlation coefficient decreases to $R \approx 0.65$ at unity bond length. The insets 
    shows the corresponding values of $\gamma$ as defined in Eq. (\ref{iso}). As the bond length increase the system becomes very viscous and the statistics is poor at high bond lengths.}
  \label{Rlength}
\end{figure}

We finally consider the excess heat capacity per particle $C_{V}^{ex}/N$ in Fig. \ref{heatd} along 
the isomorph and isotherm of Figs. \ref{asymrdfAA}-\ref{asymfsCM}. The excess heat capacity increases less than 2\% along the isomorph, while the 
$12\%$ density increase on the isotherm results in a 25\% increase in the excess heat capacity. This is consistent with the prediction in Sec. \ref{isoinv} that $C_{V}^{ex}/N$ is an isomorph invariant.
\newline \newline
The previous figures show that isomorphs exists to a good approximation for the asymmetric dumbbell model. 
An important question is whether the specific molecular geometry determines whether or not a particular \textit{LJ} model has good isomorphs. 
In Fig. \ref{Rlength} the correlation coefficient $R$ is given as a function of the bond length. 
The correlation coefficient decreases to $R \approx 0.65$ at unity bond length, and thus one might be tempted to conclude that \textit{LJ} models with large bonds lengths in general do not have good isomorphs. 
In Sec. \ref{secotp} we investigate the Lewis-Wahnstr{\"o}m \textit{OTP} model which have unity bond lengths and show that this model actually has good isomorphs.
A theory connecting the variation of $R$ to the molecular geometry and/or bond lengths remains to be developed.

\subsection{Isomorphs of a symmetric \textit{IPL} dumbbell model}\label{symsec}

In this section we briefly consider the isomorphs of a symmetric inverse power-law (\textit{IPL}) dumbbell with exponent $n = 18$ and bond length $0.584$. 
In Figs. \ref{symall}(a) and (b) we show the particle radial distribution functions along an isomorph before and after scaling the distance $r$ into reduced units. Also shown is 
the reduced particle incoherent intermediate scattering function in Fig. \ref{symall}(c). 

\begin{figure}[H]
  \centering
  \includegraphics[width=70mm]{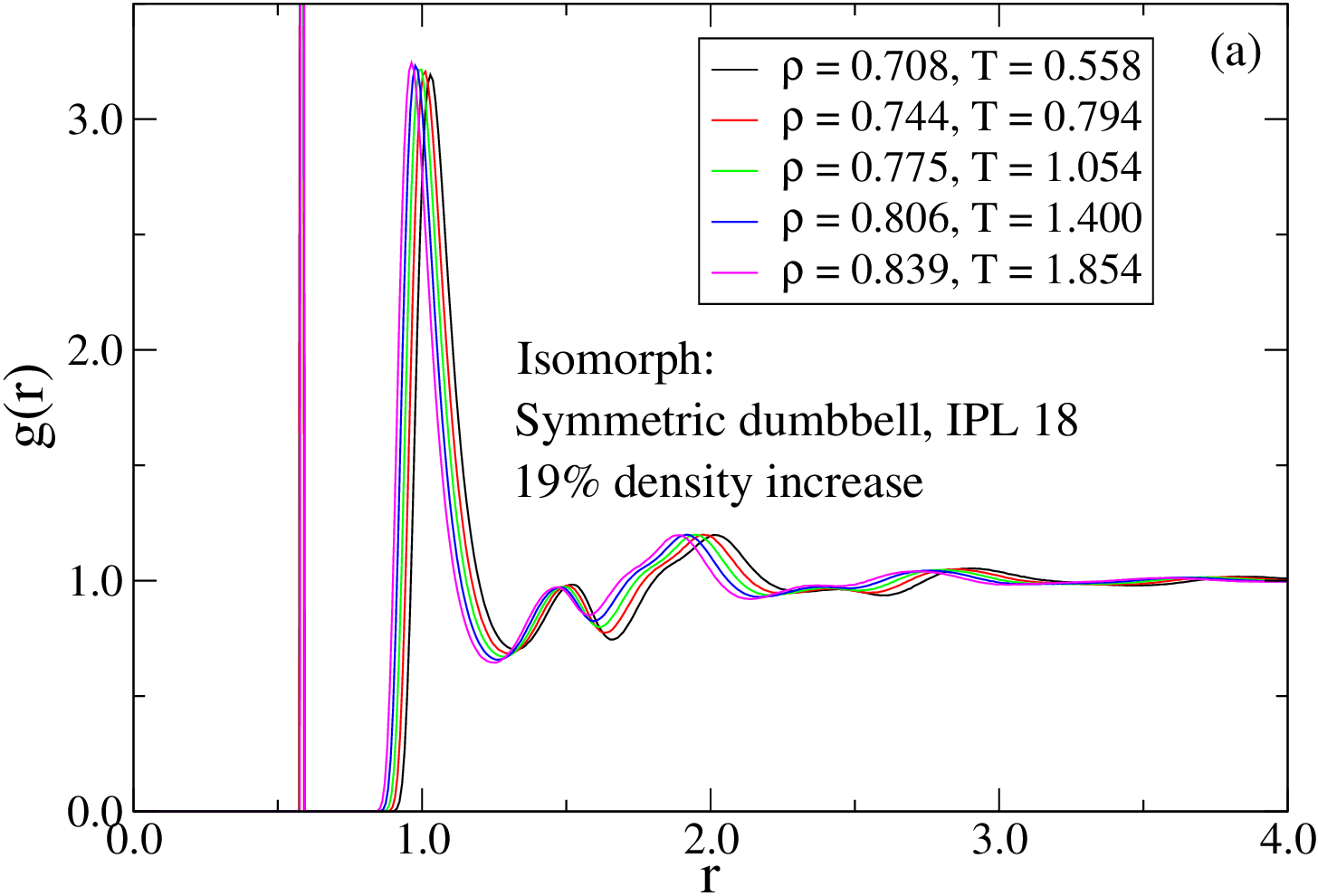}
\end{figure}
\begin{figure}[H]
  \centering
  \includegraphics[width=70mm]{sym_isomorph1_rdfA_after}
\end{figure}    
\begin{figure}[H]
  \centering
  \includegraphics[width=70mm]{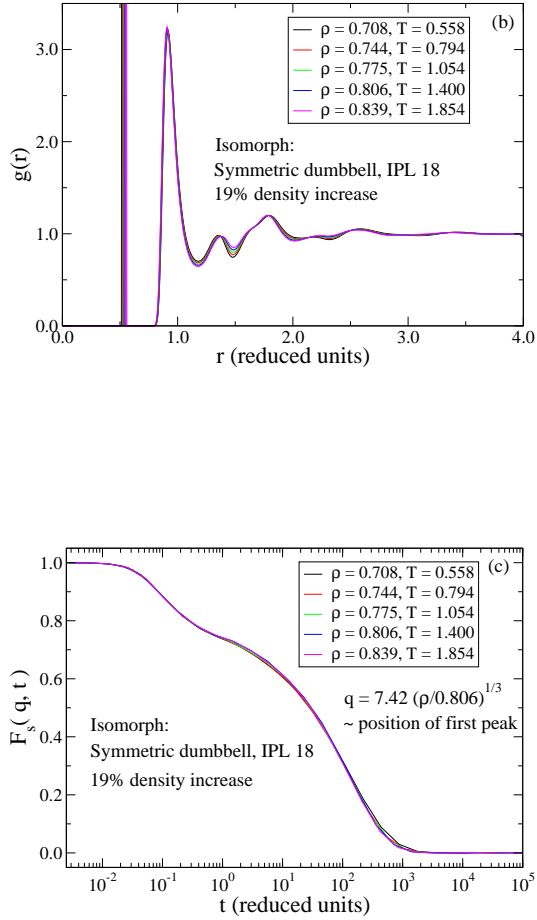}
  \caption{Structure and dynamics along an isomorph with 19\% density increase for the symmetric \textit{IPL} dumbbell model ($n = 18$). (a) Particle pair-correlation function
    before scaling the distance $r$ into reduced units. (b) Particle pair-correlation function  after scaling the distance $r$ into reduced units. 
    (c) The reduced particle incoherent intermediate scattering function at constant reduced wavevector $q$. }
  \label{symall1}
\end{figure}   
The corresponding molecular center-of-mass quantities are shown in Fig. \ref{symall}. Interestingly, the atomic dynamics appears more invariant 
than for the reduced molecular dynamics. The latter is predicted to be invariant along an isomorph while the former is not; however, we have not tried to quantify this observation any further.

Atomic systems with \textit{IPL} interactions have exact isomorphs. This reflects the scale invariance of the \textit{IPL} potential, i.e., that it 
preserves its shape under a scaling of the argument. Since molecules by their fixed geometry define a length scale in the system, isomorphs will always be approximate. However, the previous 
figures show that rigid molecules with \textit{IPL} intermolecular interactions can also have good isomorphs. 
\newline \newline
\begin{figure}[H]
  \centering
  \includegraphics[width=70mm]{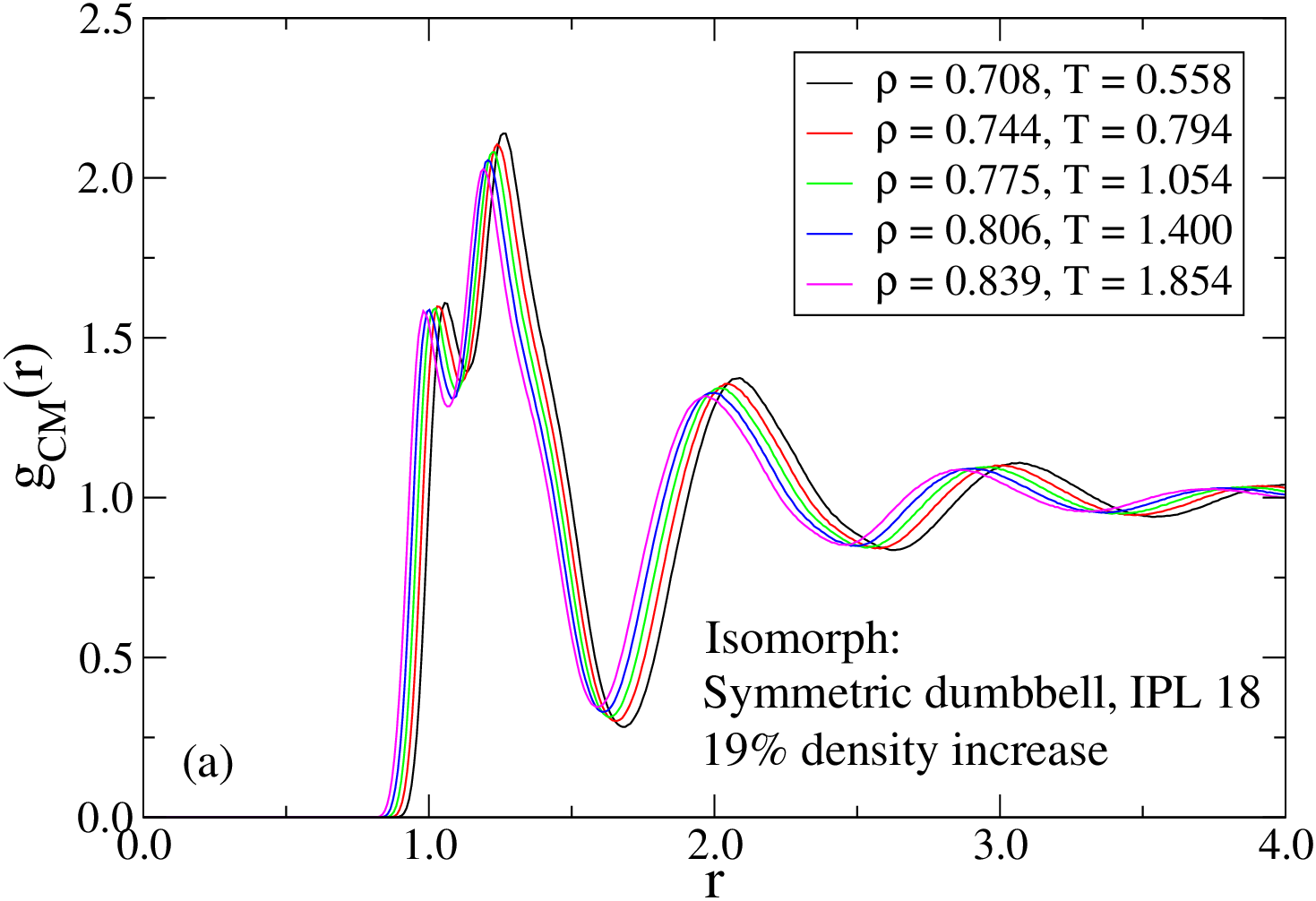}
\end{figure}
\begin{figure}[H]
  \centering
  \includegraphics[width=70mm]{sym_isomorph1_rdfCM_after}
\end{figure}    
\begin{figure}[H]
  \centering
  \includegraphics[width=70mm]{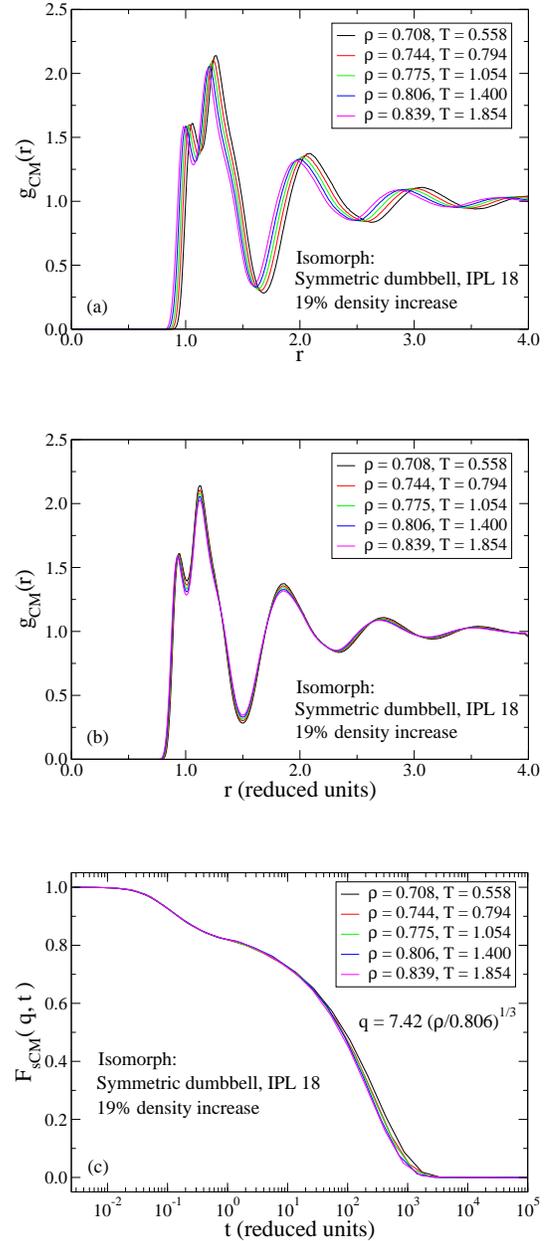}
  \caption{Structure and dynamics along the isomorph of Fig. \ref{symall1} for the symmetric \textit{IPL} dumbbell model ($n = 18$). (a) Molecular center-of-mass pair-correlation function
    before scaling the distance $r$ into reduced units $\tilde{r} = \rho^{1/3}r$. (b) Molecular center-of-mass pair-correlation function  after scaling the distance $r$ into reduced units. 
    (c) The reduced molecular center-of-mass incoherent intermediate scattering function at constant reduced wavevector $q$.}
  \label{symall}
\end{figure}
In Fig. \ref{symgammadumbbell} we consider the variation of $\gamma$ and $R$ along the isomorph, which shows that $R$ decreases 
slightly with increasing temperature and density. The variation of $\gamma$ along the isomorph is 
less than for the asymmetric dumbbell, and is to a good approximation constant. As for the asymmetric dumbbell model (Fig. \ref{gammadumbbell}) the effect 
of the constraints is to increase $\gamma$ and decrease $R$ (these are respectively 6 and 1 for the \textit{IPL} potential used).

\begin{figure}[H]
  \centering
  \includegraphics[width=70mm]{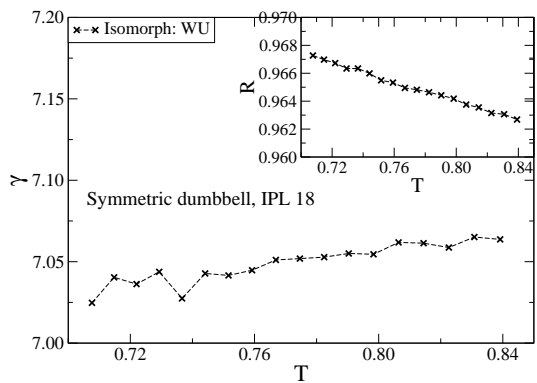}
  \caption{ The variation of $\gamma$ and $R$ (insets) along the isomorph of Figs. \ref{symall1}-\ref{symall} with 19\% density increase 
    for a symmetric dumbbell model ($n$ = $18$). $\gamma$ increases slightly along the isomorph. 
    Excluding the constraints in the virial the correlation coefficient and $\gamma$ are respectively $R$ = $1$ and $\gamma$ = $6$.}
  \label{symgammadumbbell}
\end{figure}

\subsection{Isomorphs of the Lewis$-$Wahnstr{\"o}m \textit{OTP} model}\label{secotp}

\begin{figure}[H]
  \centering
  \includegraphics[width=70mm]{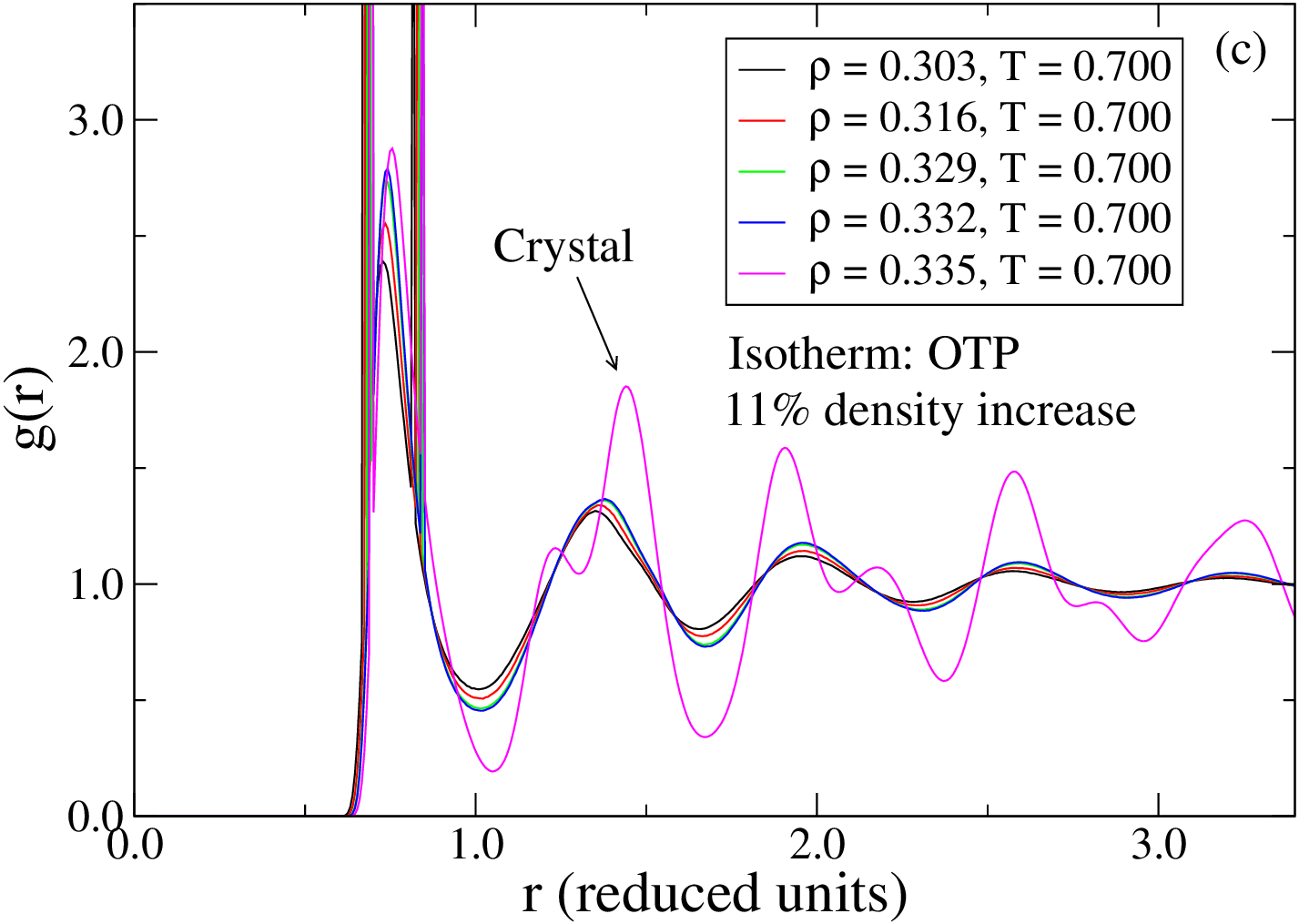}
  \caption{Particle radial distribution functions for the \textit{OTP} model. (a) Along an isomorph with 21\% density increase, shown 
    prior to scaling the distance $r$ into reduced unit $\tilde{r} = \rho^{1/3}r$. (b) Along the same isomorph after scaling the distance $r$ into reduced unit. (c) 
    Along an isotherm with 11\% density increase. At the highest density probed the \textit{OTP} model crystallizes\cite{otpulf} (magenta).}
  \label{otprdf}
\end{figure}

We proceed to investigate the Lewis$-$Wahnstr{\"o}m \textit{OTP} model\cite{otp1,otp2}. Figure \ref{otprdf} shows the 
particle radial distribution functions along an isomorph with 21\% density increase 
before and after scaling the distance $r$ into reduced units. We treat the particles as identical in the quantities probed in 
simulations (i.e., the radial distribution function, etc.) even though the \textit{OTP} model is an isosceles triangle. 
Also shown for reference is an isotherm with 11\% density increase in Fig. \ref{otprdf}(c).

\begin{figure}[H]
  \centering
  \includegraphics[width=70mm]{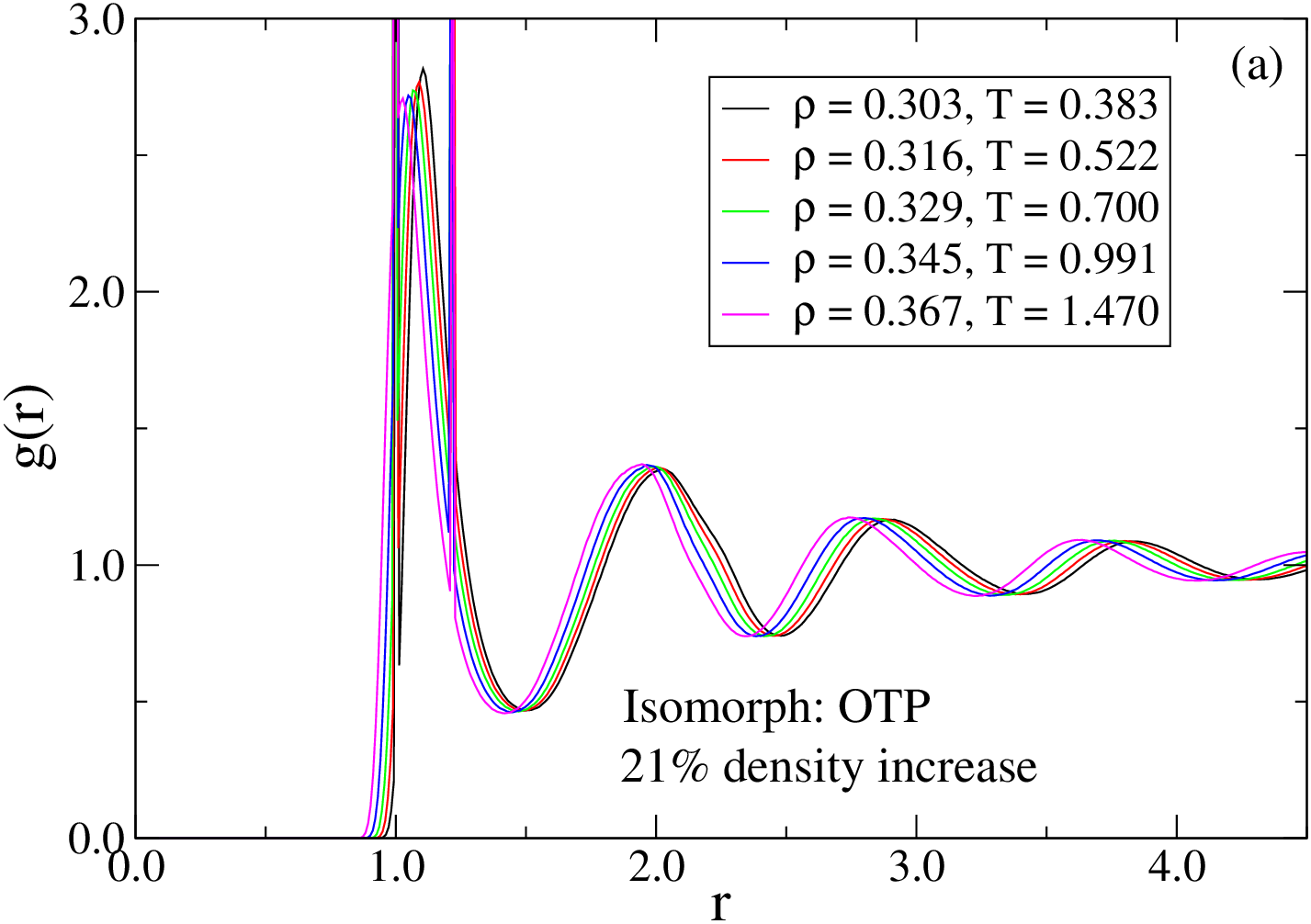}
\end{figure}    
\begin{figure}[H]
  \centering
  \includegraphics[width=70mm]{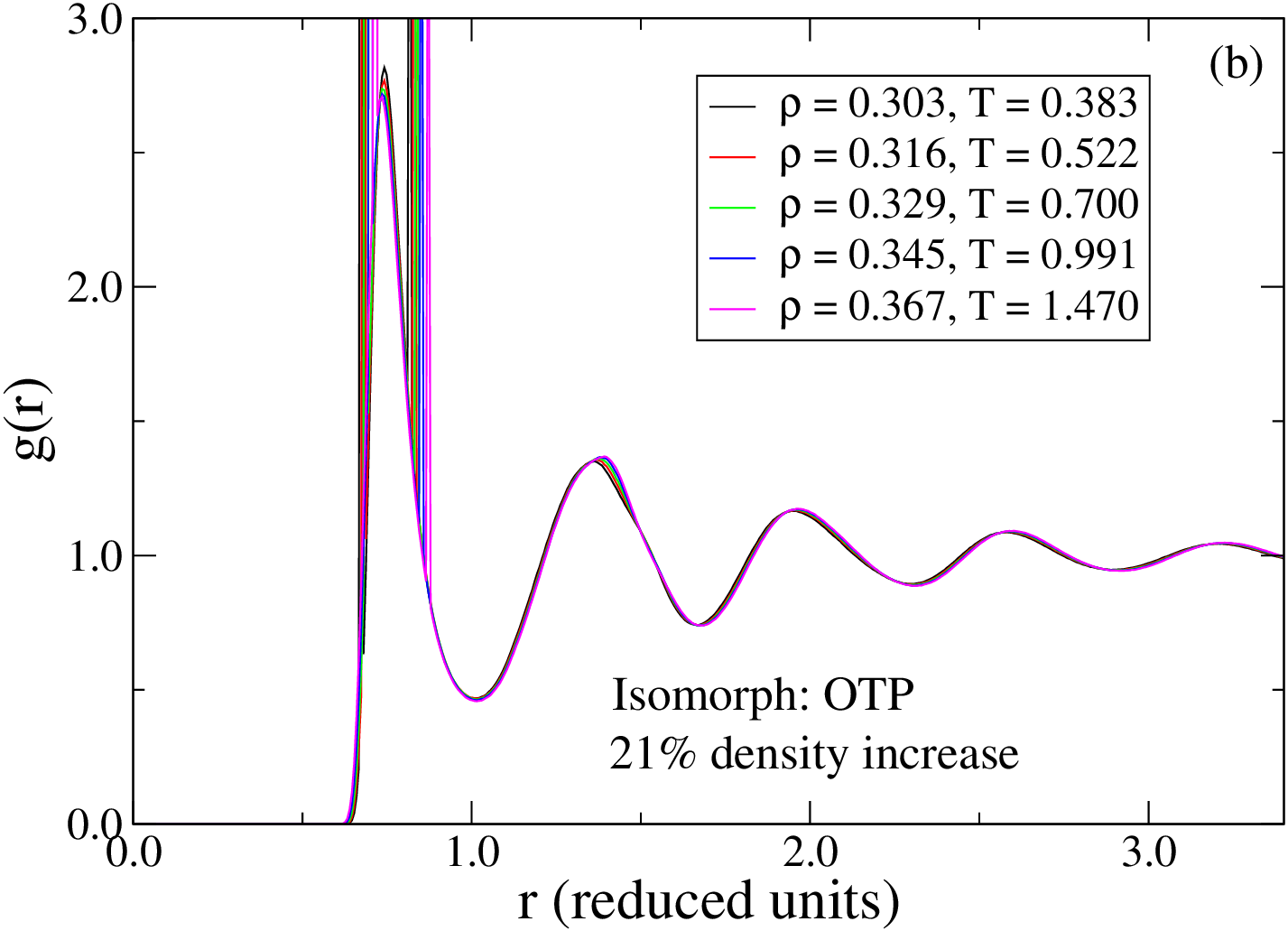}
\end{figure}    
\begin{figure}[H]
  \centering
  \includegraphics[width=70mm]{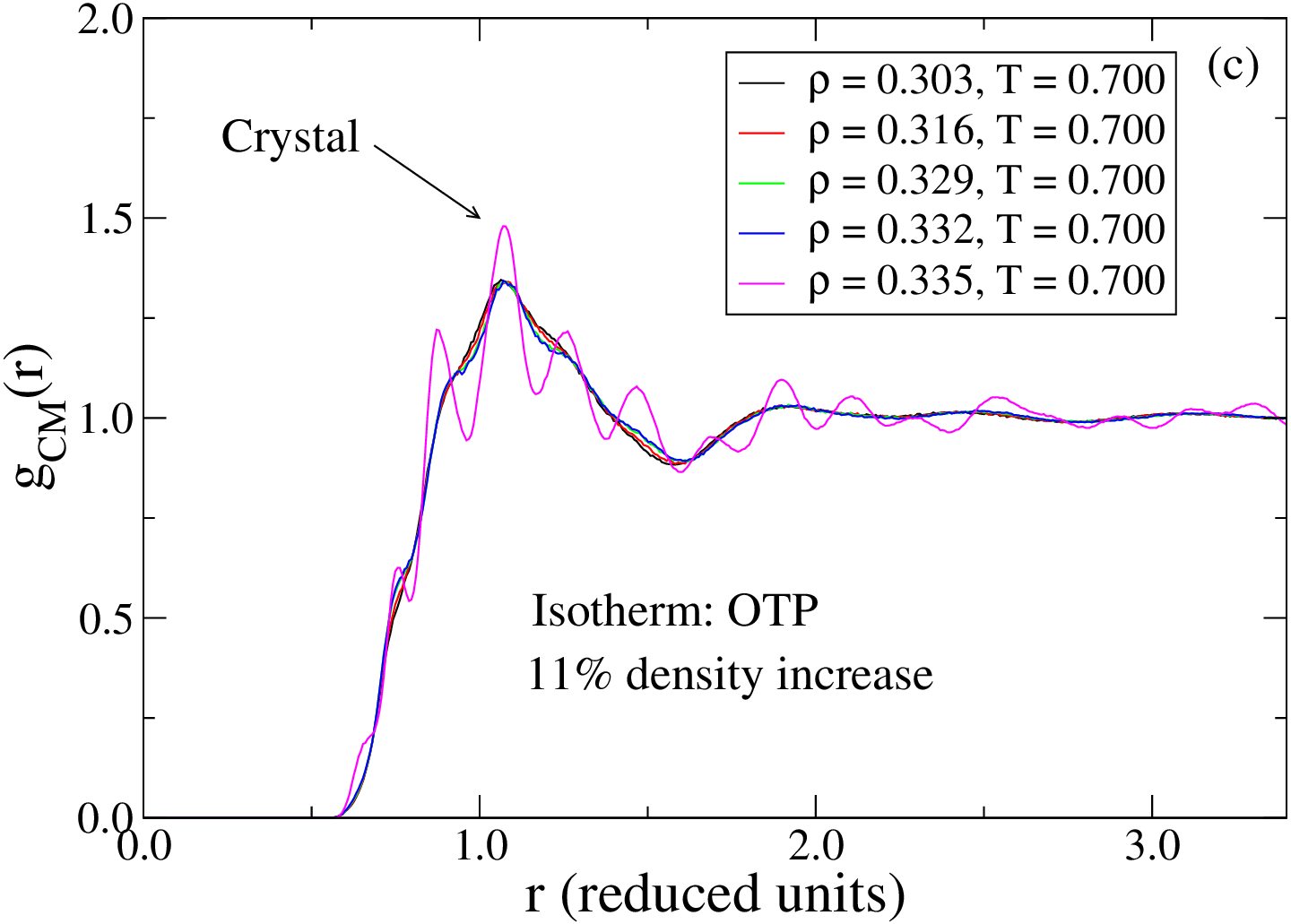}
  \caption{Molecular center-of-mass radial distribution functions for the \textit{OTP} model. (a) Along the isomorph of Fig. \ref{otprdf} with 21\% density increase, shown 
    prior to scaling the distance $r$ into reduced units $\tilde{r} = \rho^{1/3}r$. (b) Along the same isomorph after scaling the distance $r$ into reduced units. (c) 
    Along the isotherm of Fig. \ref{otprdf} with 11\% density increase. At the highest density probed the \textit{OTP} model here crystallizes (magenta).}
  \label{otprdfCM}
\end{figure}

Figure \ref{otprdfCM} shows the corresponding reduced molecular center-of-mass radial distribution functions. The reduced molecular center-of-mass structure seems less invariant along 
the isomorph than for the asymmetric dumbbell (Fig. \ref{asymrdfCM}), consistent with the \textit{OTP} model being less strongly 
correlating ($R \approx 0.90$). However, comparing with the isotherm in Fig. \ref{otprdfCM}(c) the \textit{OTP} model crystallizes at the highest density 
probed\cite{otpulf}, even though the density increase is just 11\% compared with the 21\% density increase along the isomorph. Comparing now with the particle quantities of 
Fig. \ref{otprdf} the latter seems more invariant, even though the reduced molecular center-of-mass structure is predicted to be invariant. We currently have no explanation for this observation.

\begin{figure}[H]
  \centering
  \includegraphics[width=70mm]{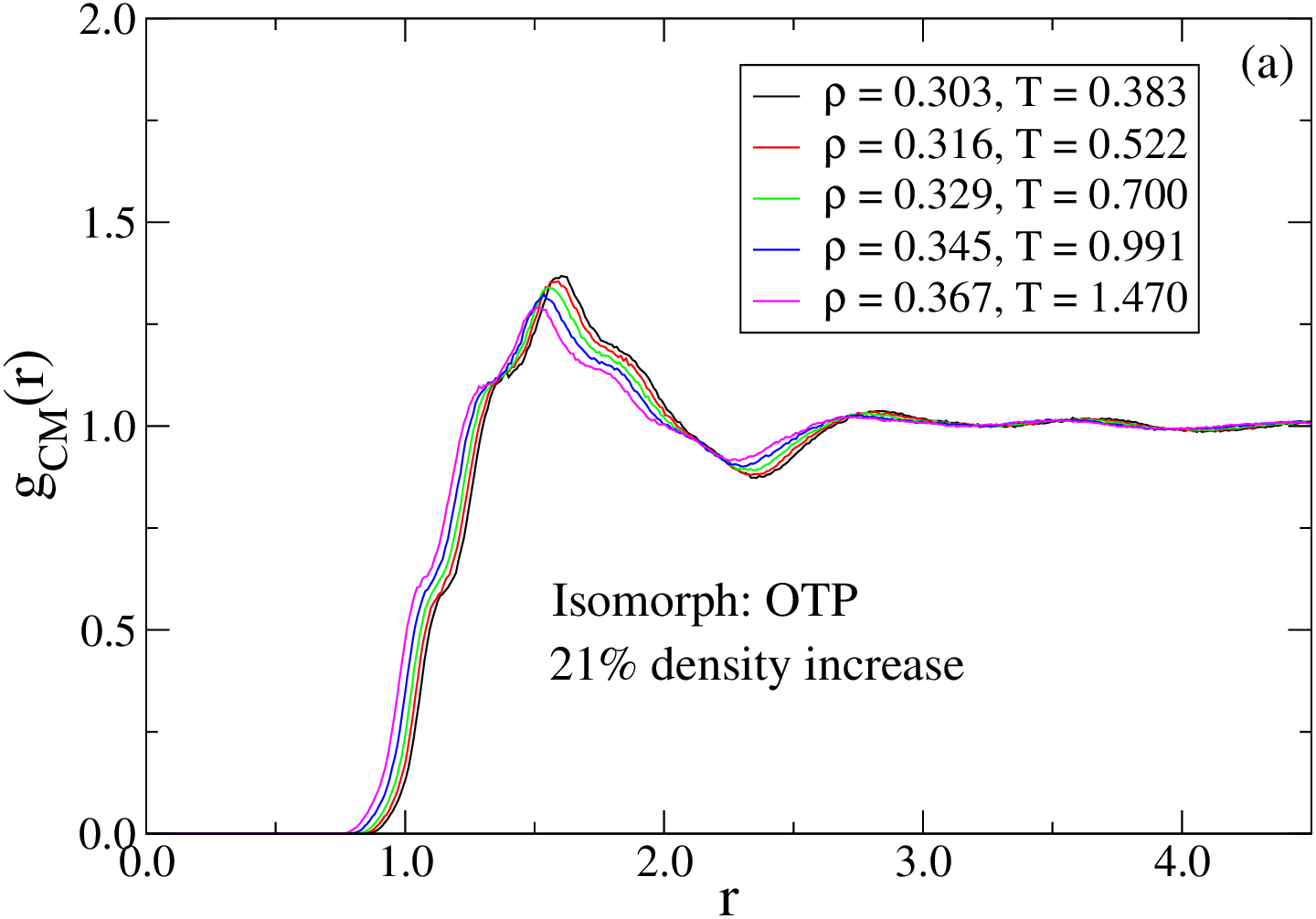}
\end{figure}
\begin{figure}[H]
  \centering
  \includegraphics[width=70mm]{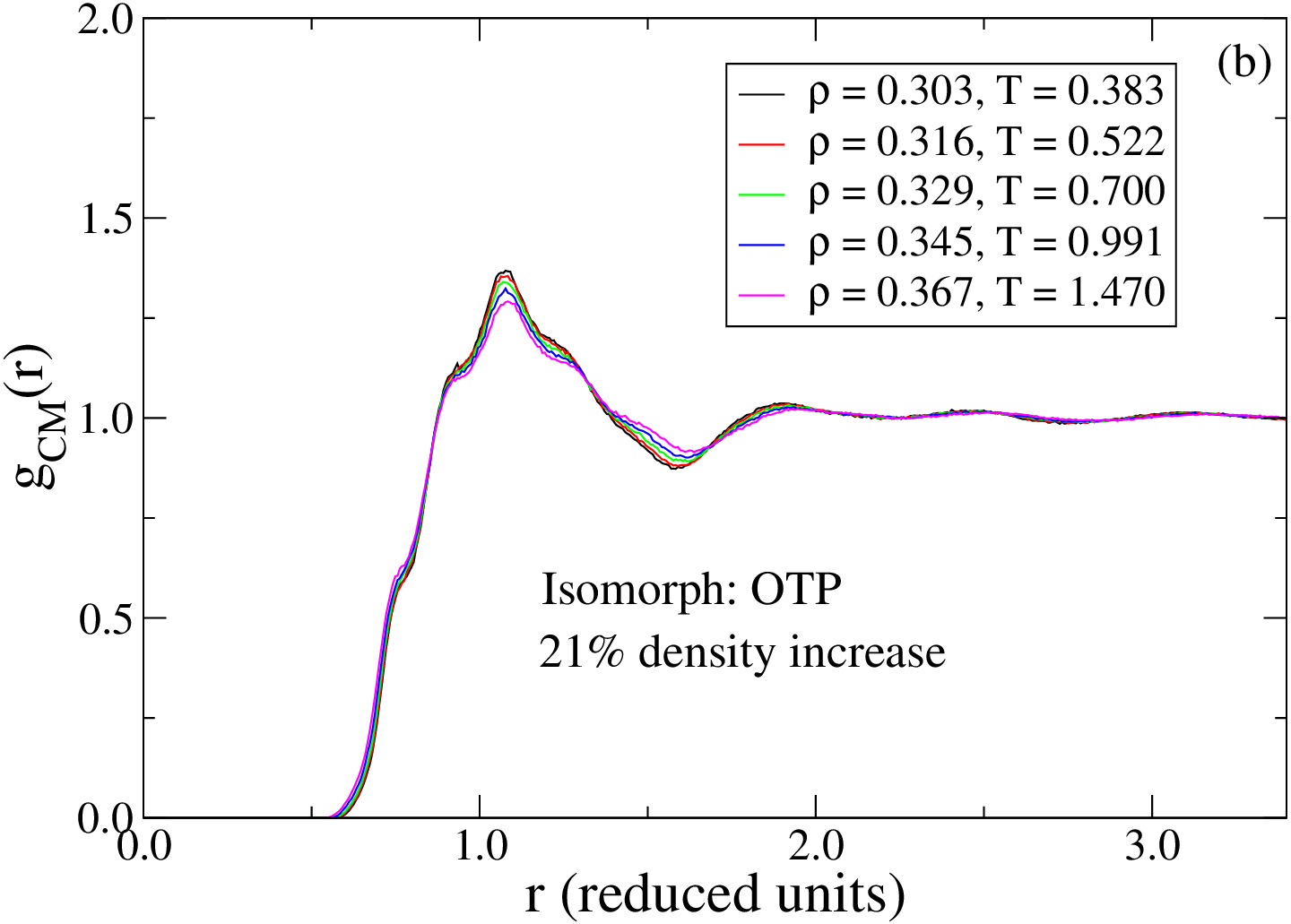}
\end{figure}    
\begin{figure}[H]
  \centering
  \includegraphics[width=70mm]{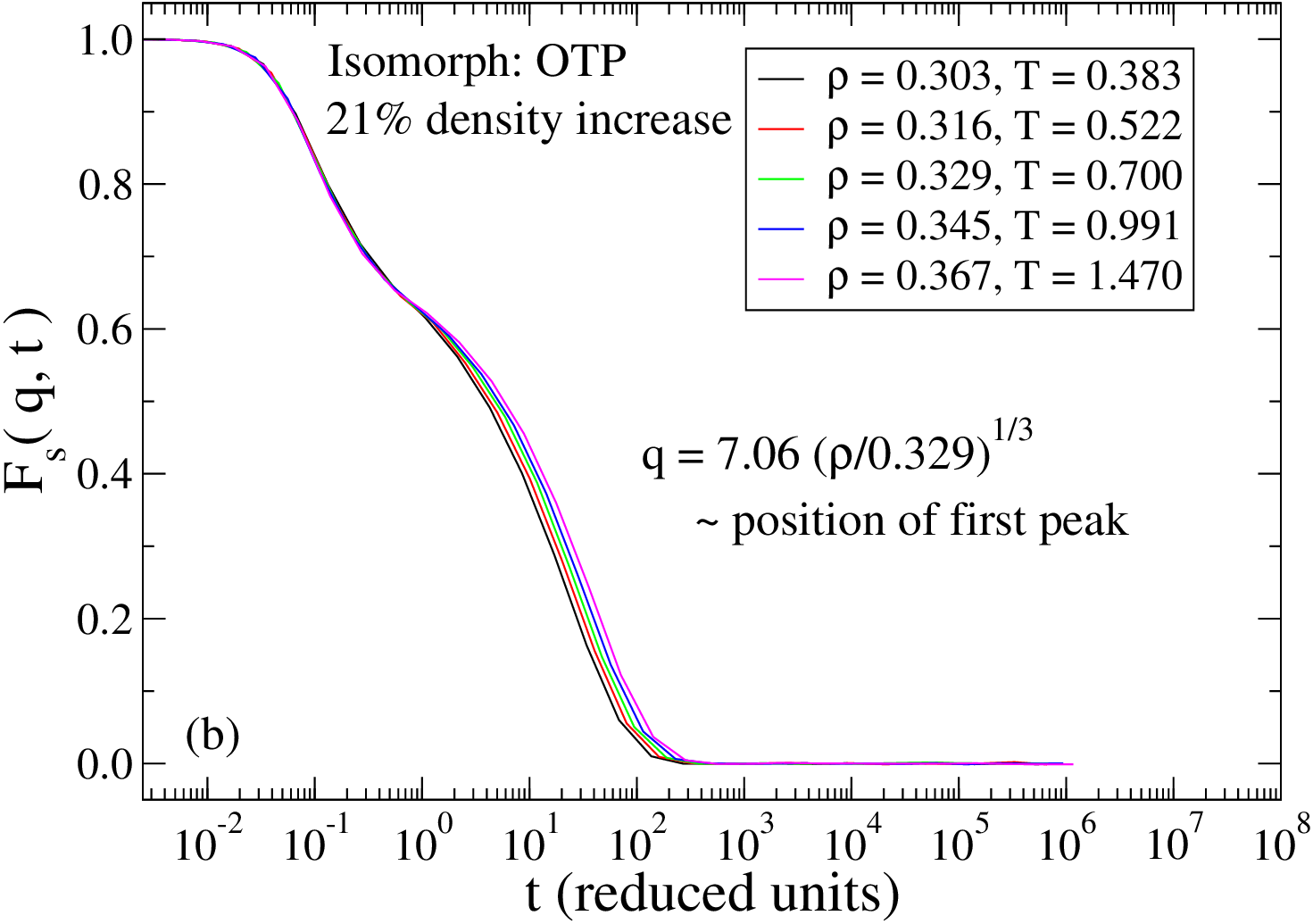}
  \caption{The reduced particle incoherent intermediate scattering functions for the \textit{OTP} model keeping the reduced wavevector q constant. (a) Along the 
    isotherm of Figs. \ref{otprdf}-\ref{otprdfCM} with 11\% density increase. 
    (b) Along the isomorph of Figs. \ref{otprdf}-\ref{otprdfCM} with 21\% density increase and almost a factor of four increase in temperature. The dynamics is roughly invariant 
  along the isomorph, but not along the isotherm.}
  \label{otpfs}
\end{figure}

Figure \ref{otpfs} shows the reduced particle incoherent intermediate scattering functions along the isotherm and isomorph of Figs. \ref{otprdf}-\ref{otprdfCM} while Fig. \ref{otpfsCM} shows the 
reduced molecular center-of-mass incoherent intermediate scattering functions.

\begin{figure}[H]
  \centering
  \includegraphics[width=70mm]{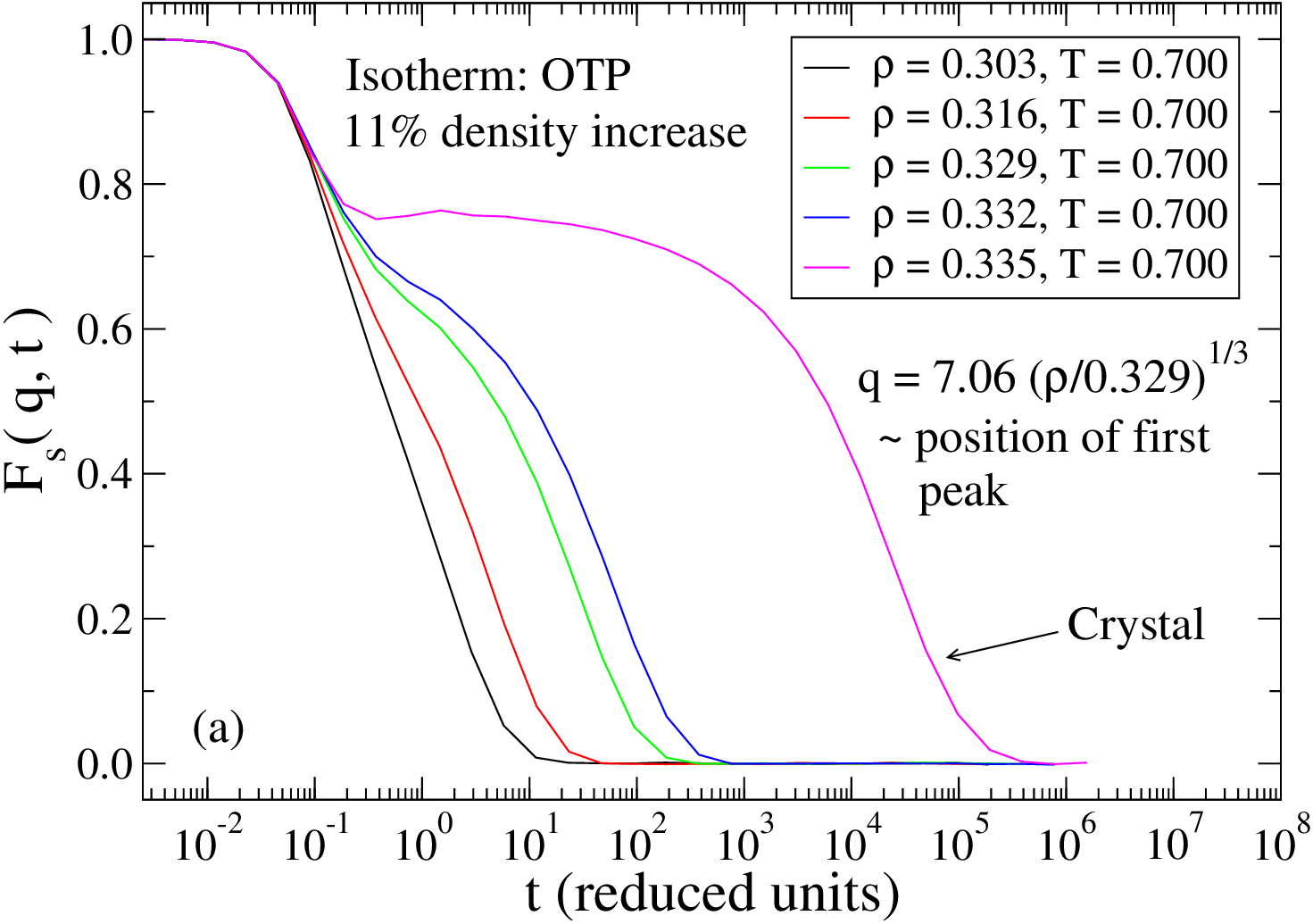}
\end{figure}    
\begin{figure}[H]
  \centering
  \includegraphics[width=70mm]{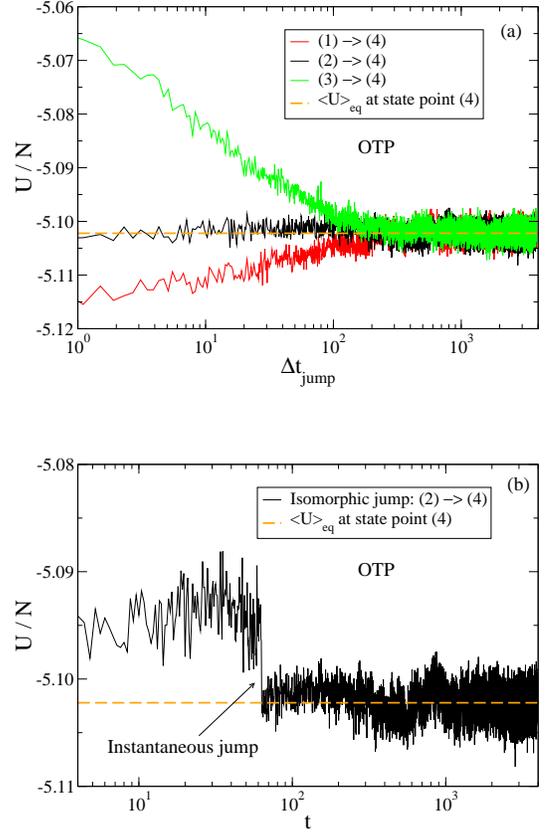}
  \caption{Four state points ($1$), ($2$), ($3$), and ($4$) corresponding to, respectively, ($\rho$, $T$) = ($0.329$, $0.650$), ($0.329$, $0.700$), ($0.329$, $1.000$), and ($0.303$, $0.383$) 
    are given where the first three state points are on the same isochore. State points ($2$) and ($4$) are isomorphic 
  while ($1$) and ($3$) are not isomorphic to ($4$). After equilibrating at state points ($1$), ($2$), and ($3$), respectively, the temperature and density are 
  instantaneously changed to that of state point ($4$) via a scaling of the coordinates keeping bond distances and Eulerian angles of the molecules fixed.
  An average has been performed over 100 samples.
 (a) The relaxational behaviour of all state points quantified by the potential energy $U$.
  The jump ($2$) $\to$ ($4$) shows no relaxation while the other jumps do. (b) Close up of the potential energy of state point ($2$) before and after the jump, where the jump takes place at $t \approx 60$. }
  \label{isojumpo}
\end{figure}

For the molecular quantities, the dynamics is roughly invariant along the 
isomorph but not on the isotherm, even though the density increase is 21\% for the isomorph and only 11\% for the isotherm. In contrast to the reduced molecular 
center-of-mass structure; the molecular dynamics does not seem less invariant than the particle dynamics, consistent with the prediction of Appendix \ref{appInvdyn}.
\newline \newline
\begin{figure}[H]
  \centering
  \includegraphics[width=70mm]{otp_isotherm1_FsCM_after}
\end{figure}    
\begin{figure}[H]
  \centering
  \includegraphics[width=70mm]{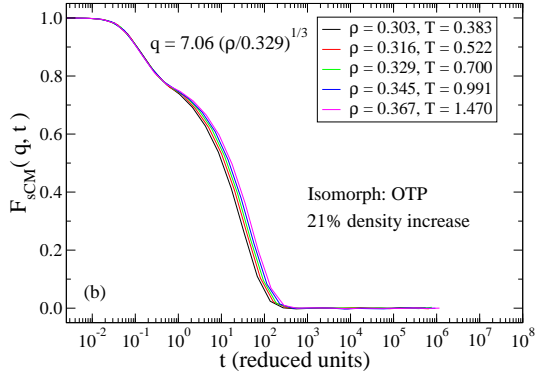}
  \caption{The reduced molecular center-of-mass incoherent intermediate scattering functions 
    for the \textit{OTP} model keeping the reduced wavevector q constant. (a) Along the isotherm of Figs. \ref{otprdf}-\ref{otpfs} with 11\% density increase. 
    (b) Along the isomorph of Figs. \ref{otprdf}-\ref{otpfs} with 21\% density increase and almost a factor of four increase in temperature. The dynamics is roughly invariant 
  along the isomorph, but not along the isotherm.}
  \label{otpfsCM}
\end{figure}    
We consider in Fig. \ref{gammaotp} the variation of $\gamma$ as defined by Eq. (\ref{iso}). 
The large variation in $\gamma$ indicates that density scaling, with a fixed exponent, is of more approximate nature for the \textit{OTP} model\cite{hidden} than for the asymmetric dumbbell, 
even though isomorphs exists to a  good approximation. The isomorphs of the \textit{OTP} model are, however, more approximative than for the asymmetric dumbbell, which is consistent with \textit{OTP} model being less strongly correlating.
\newline \newline
\begin{figure}[H]
  \centering
  \includegraphics[width=70mm]{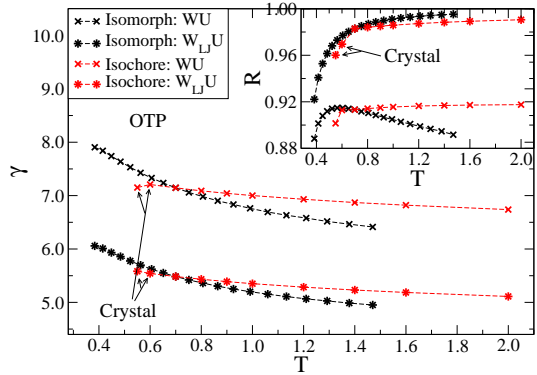}
  \caption{The variation of $\gamma$ (Eq. (\ref{iso})) and the correlation coefficient $R$ (inset, Eq. (\ref{R})) for the \textit{OTP} model 
    in two different versions along an isochore (red, $\rho$ = $0.329$) and the isomorph (black) of Figs. \ref{otprdf}-\ref{otpfsCM}. The crosses show $\gamma$ calculated 
    from the total virial $W$, the asterisks show $\gamma$ calculated after subtracting the 
    constraint contribution to virial, i.e., replacing $W$ by $W_{LJ} = W - W_ {CON}$. $\gamma$ is predicted in 
    Ref. \onlinecite{paper4} to be a function only of density as is seen to be the case for both versions, although the variation is larger than for the asymmetric dumbbell.}
  \label{gammaotp}
\end{figure}
Next we consider isomorph jumps for the \textit{OTP} model. The setup is analogous to that of the asymmetric dumbbell model described in Sec. \ref{isoasymsec}.
It is seen from Fig. \ref{isojumpo} that an isomorph jump shows no relaxation.
\newline \newline
\begin{figure}[H]
  \centering
  \includegraphics[width=70mm]{otp_isomorph1_jump2}
\end{figure}
We close the investigation of the \textit{OTP} model by considering in Fig. \ref{heato} the excess heat capacity per particle $C_{V}^{ex}/N$. 
This quantity increases 7\% over the $21\%$ density increase along the isomorph, while the 
$11\%$ density increase on the isotherm results in a 34\% increase before crystallizing. These results are consistent with the prediction that $C_{V}^{ex}$ is an 
isomorph invariant (see Sec. \ref{isoinv}), although less so than for the asymmetric dumbbell model.

\begin{figure}[H]
  \centering
  \includegraphics[width=70mm]{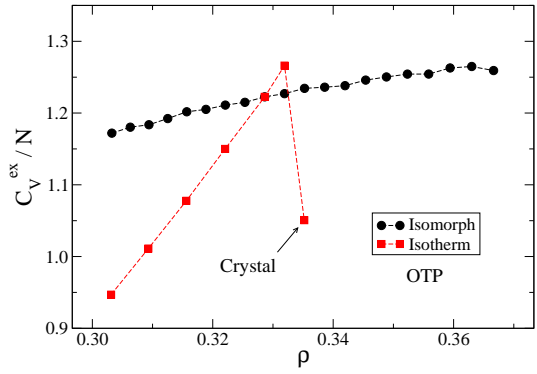}
  \caption{The excess heat capacity per particle $C_{V}^{ex} / N$ for the \textit{OTP} model as a function of density along the isomorph (black) and isotherm (red) 
    of Figs. \ref{otprdf}-\ref{otpfsCM}. The density increase is 21\% and 11\%, respectively. At high densities for the isotherm the \textit{OTP} model crystallizes. The excess heat capacity is to a 
  good approximation invariant along the isomorph, while this is not the case for the isotherm.}
  \label{heato}
\end{figure}

\section{Master isomorphs}\label{master}

The previous section detailed the existence of isomorphs in the phase diagram of liquids of small rigid molecules. We now investigate whether the generated 
isomorphs have the same shape in the $WU$ phase diagram, i.e., whether a so-called master isomorph  exists\cite{paper5}. 
It is also interesting to compare the isomorphs of the dumbbell and the \textit{OTP} models, since both systems have intermolecular 
($12$, $6$)-\textit{LJ} interactions, but different constraint contributions to the virial (one versus three constrained distances per molecule).

\begin{figure}[H]
  \centering
  \includegraphics[width=70mm]{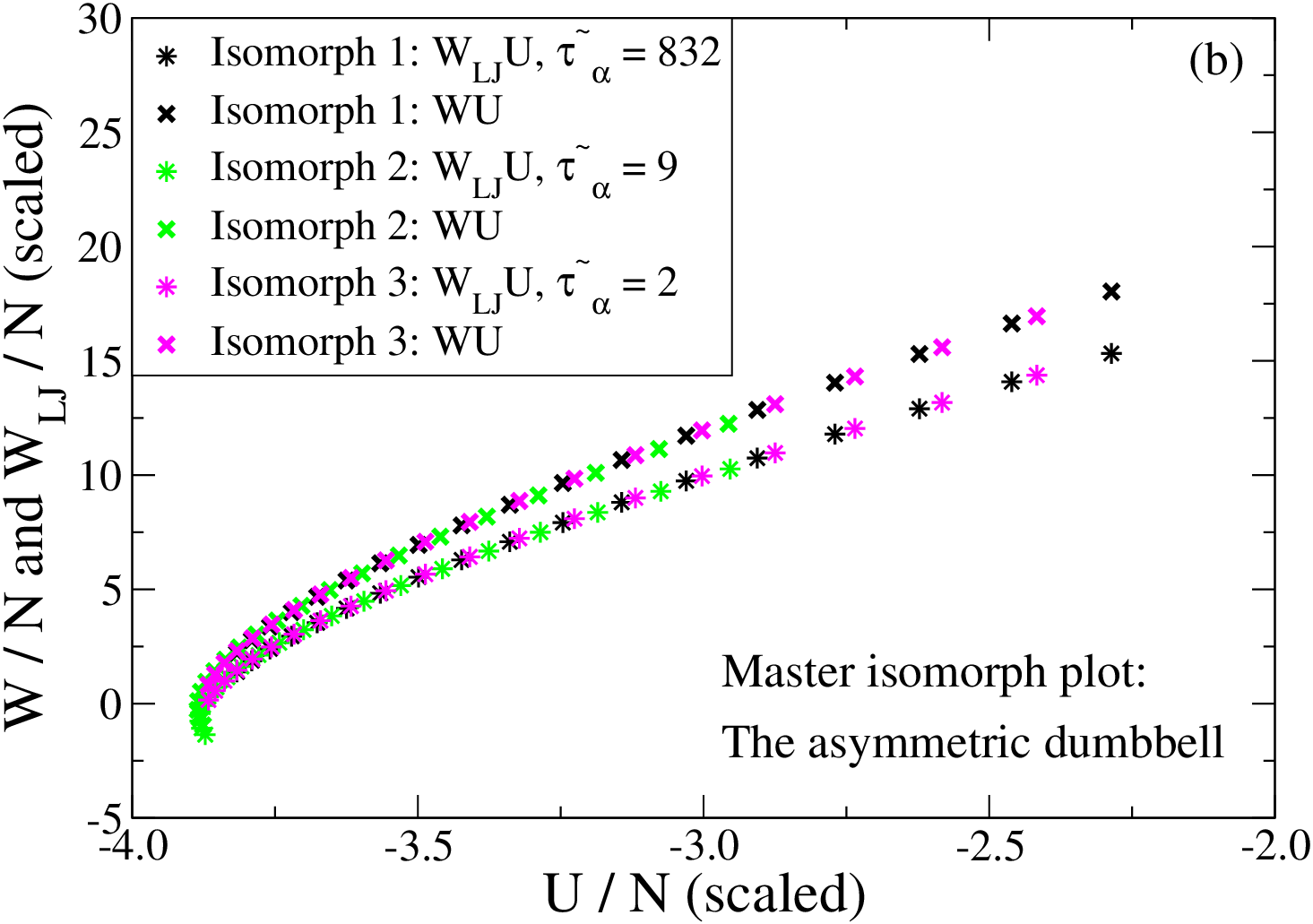}
  \caption{(a) Three different isomorphs for the asymmetric dumbbell model in two different versions with 19\%, 21\% and 22\% density increase, respectively
    (black, magenta and green). The crosses give 
    the total virial $W$, the asterisks give $W_{LJ} = W - W_{CON}$. $\tilde{\tau}_{\alpha}$ is reduced relaxation time 
    extracted from the self part of the intermediate scattering function. (b) The same isomorphs as in (a) where $WU$ and $W_{LJ}U$ are 
    scaled to superpose with a factor identified by trial and error. The black points have unity scaling factor.}
  \label{misodumbb}
\end{figure}

\begin{figure}[H]
  \centering
  \includegraphics[width=70mm]{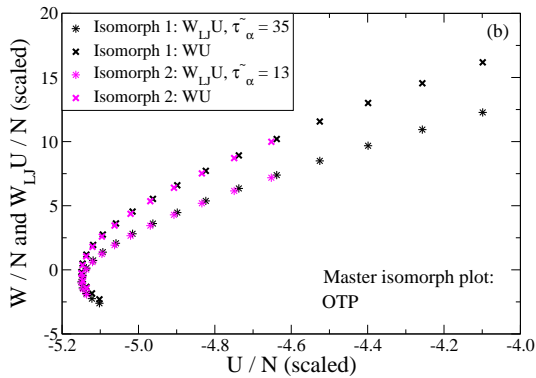}
  \caption{(a) Two different isomorphs for the \textit{OTP} model in two different versions with 21\% and 14\% density increase (black and magenta). The crosses give the 
    total virial $W$, the asterisks give $W_{LJ} = W - W_{CON}$. $\tilde{\tau}_{\alpha}$ is reduced relaxation time 
    extracted from the self part of the intermediate scattering function.
    (b) The same isomorphs as in (a) where $WU$ and $W_{LJ}U$ are scaled to superpose with a factor identified by trial and error. 
    The black points have unity scaling factor.}
  \label{misootp}
\end{figure}

Figure \ref{misodumbb}(a) shows three different isomorphs in the $WU$ phase diagram for the asymmetric dumbbell model, in two different versions: one for the total virial $W$
and one for the ''\textit{LJ}'' virial, i.e., replacing $W$ with $W_{LJ} = W - W_{CON}$. In order to investigate 
whether a master isomorph exists Fig. \ref{misodumbb}(b) shows the same isomorphs 
after scaling of the potential energy and the virial with the \textit{same} factor (depending on the isomorph). The best scaling factor was identified by 
trial and error.
\newline \newline
\begin{figure}[H]
  \centering
  \includegraphics[width=70mm]{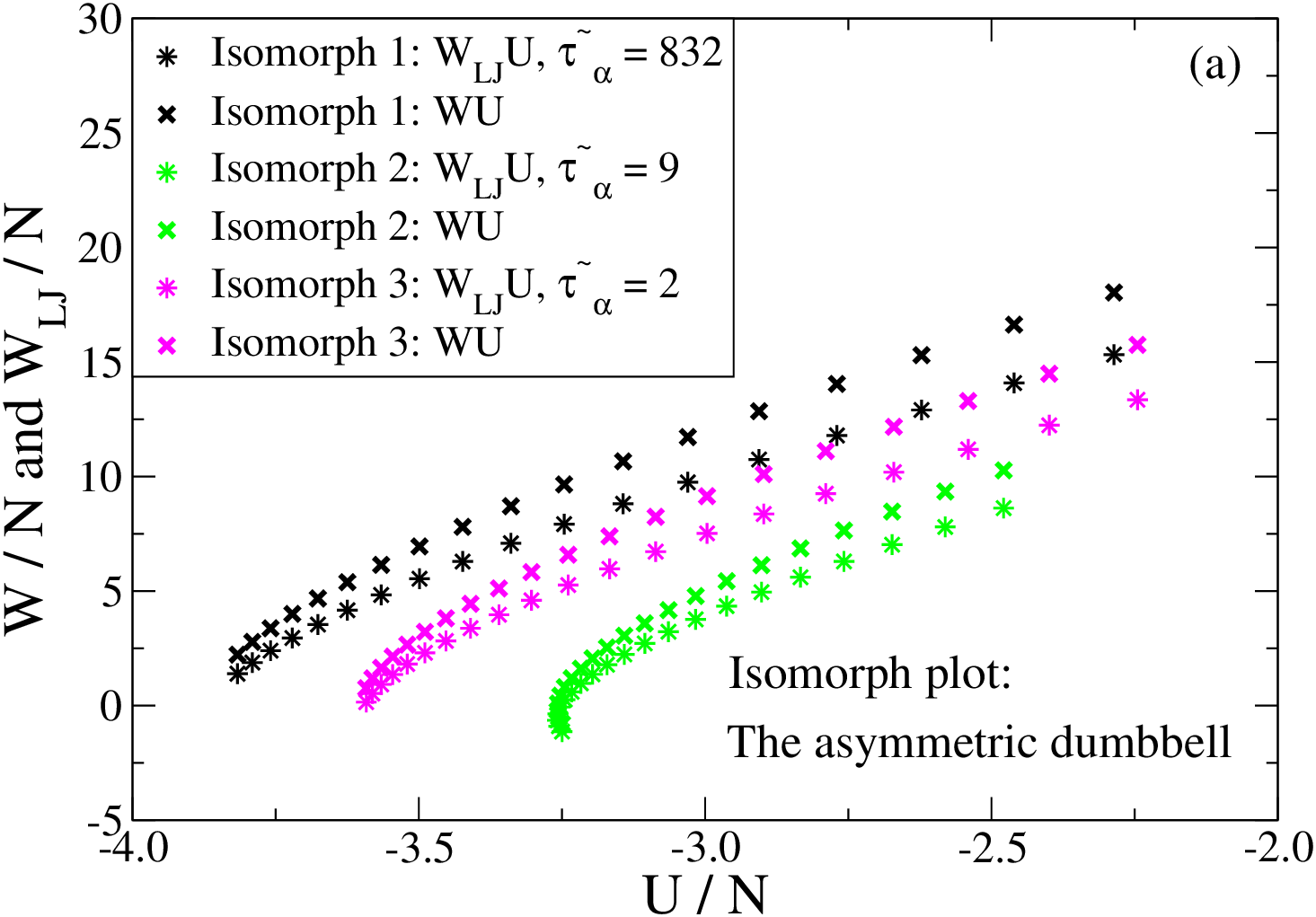}
\end{figure}
Corresponding figures for the \textit{OTP} system are given in Fig. \ref{misootp}. The figures show that for both models
 a master isomorph exists to a good approximation both with and without the constraint contribution to the virial.

\begin{figure}[H]
  \centering
  \includegraphics[width=70mm]{otp_masterIsomorph}
\end{figure}
\begin{figure}[H]
  \centering
  \includegraphics[width=70mm]{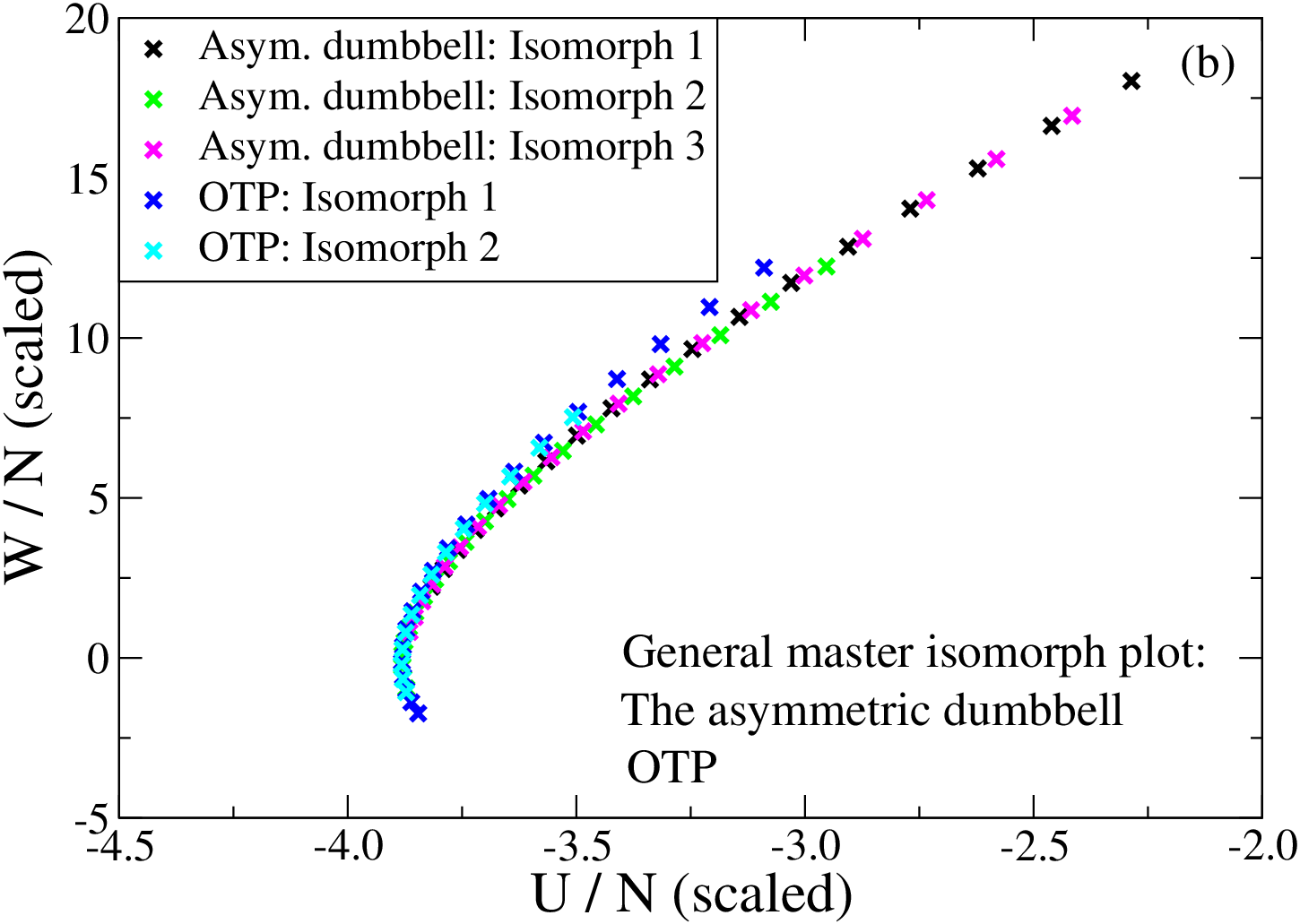}
  \caption{(a) Scaled $W_{LJ}U$ isomorphs for the asymmetric dumbbell and \textit{OTP} models. The black points have unity scaling factor. 
    Both systems have intermolecular ($12$, $6$)-\textit{LJ} interactions, and 
    the dashed curves are the isomorph prediction from Ref. \onlinecite{paper5} for an atomic system, where the \textit{LJ} reference coefficients have respectively 
    been calculated from the dumbbell state points ($\rho$, $T$)  = ($0.932$, $0.465$) and ($0.851$, $0.274$). (b) Scaled $WU$ isomorphs for the systems in (a). 
    The total virial does not show exact scaling between the asymmetric dumbbell and \textit{OTP} models.}
  \label{symmiso}
\end{figure}

As mentioned in the introduction, Ref. \onlinecite{paper5} derived predictions concerning the shape of isomorphs for \textit{atomic} systems with  
pair potential given by a sum 
of two \textit{IPL}s (the generalized \textit{LJ} potential).
The question arises whether $W_{LJ}U$ follows that shape? 
This is studied in Fig. \ref{symmiso}(a) where the $W_{LJ}U$ isomorphs for the asymmetric dumbbell and \textit{OTP} models are shown scaled using the previously 
mentioned procedure. The two dashed curves are the isomorph prediction for an \textit{atomic} ($12$, $6$)-\textit{LJ} system \cite{paper5} (where $\tilde{\rho} = \rho/\rho^{*}$ and $\rho^{*}$ is the density of a chosen reference 
state point)
\begin{align}
  U & = U_{m}^{*}\tilde{\rho}^{4} + U_{n}^{*}\tilde{\rho}^{2}, \label{predu}\\
  W_{LJ} & = 4U_{m}^{*}\tilde{\rho}^{4} + 2U_{n}^{*}\tilde{\rho}^{2} \label{predw},
\end{align}
where the reference coefficients ($U_{m}^{*}$, $U_{n}^{*}$) have been calculated from two different reference state points
along ''Isomorph 1'' of the asymmetric dumbbell (see Ref. \onlinecite{paper5} for details on calculating these coefficients). 
 The only assumption 
used in Ref. \onlinecite{paper5} to derive these formulas is the invariance of the reduced atomic structure along an isomorph, however, this is not predicted to be the case for molecular systems with isomorphs.

It is clear that the atomic isomorph shape is not followed exactly. Nevertheless, 
there seems to exist not only a master isomorph in the \textit{LJ} and total virial for the individual systems, but also for the 
\textit{LJ} virial between these two different model systems.
The same does not hold for the total virial as can been seen in Fig. \ref{symmiso}(b) since the constraint contributions are different.

\begin{figure}[H]
  \centering
  \includegraphics[width=70mm]{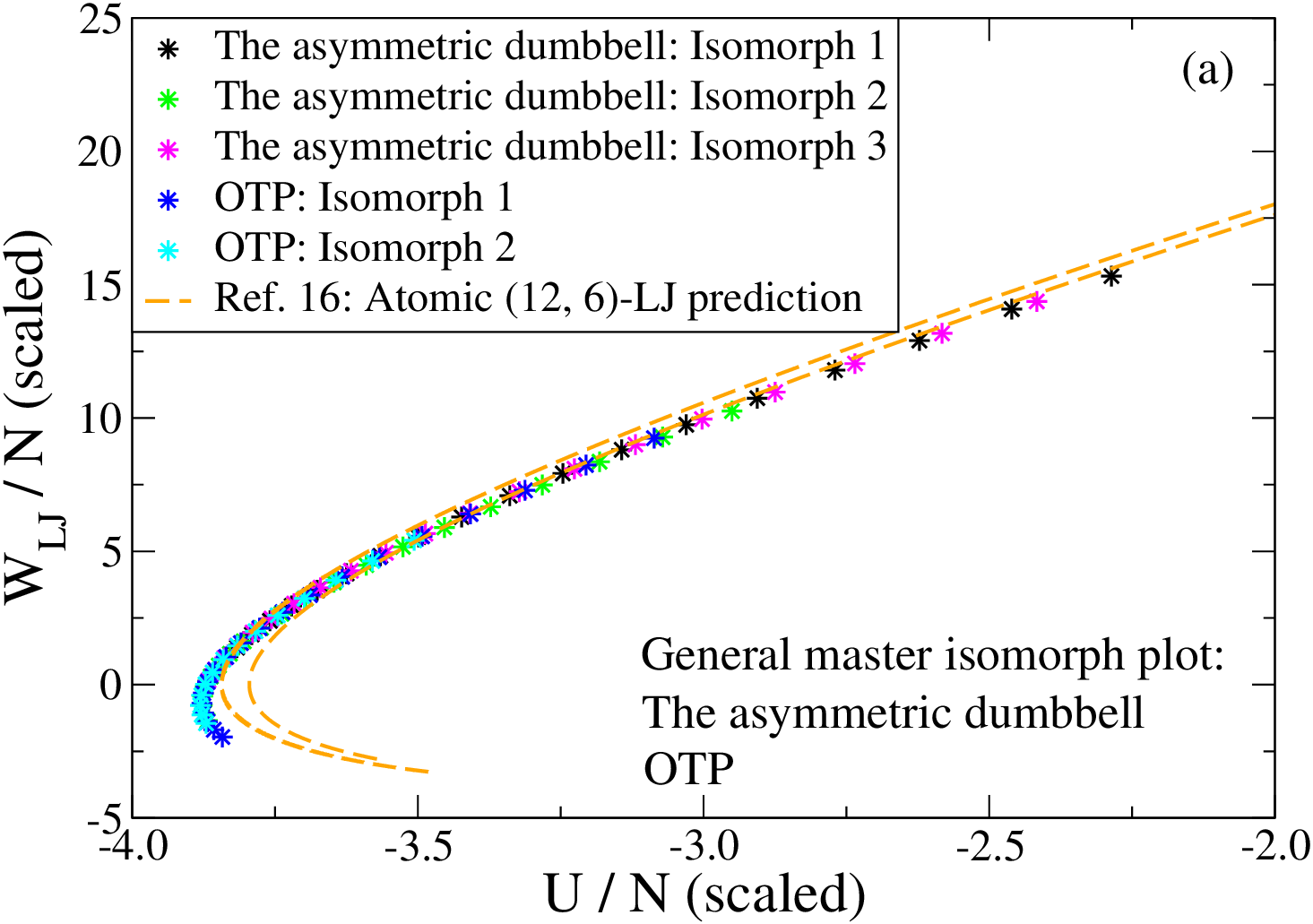}
\end{figure}
\begin{figure}[H]
  \centering
  \includegraphics[width=70mm]{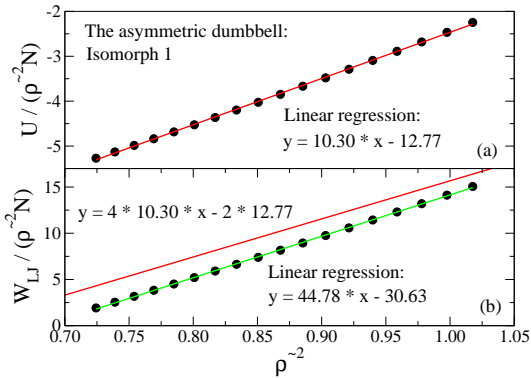}
  \caption{ The potential energy and \textit{LJ} virial as a function of $\tilde{\rho}^{2}$ for ''Isomorph 1'' of the asymmetric dumbbell. (a) A linear regression fit of the potential energy 
    has been performed to calculate the reference coefficients ($U_{m}^{*}$, $U_{n}^{*}$). (b) These coefficients are then used to plot the red straight line, which according to the atomic prediction [Eqs. (\ref{predu}) and (\ref{predw})] should coincide with the black data points. The green line is a linear regression fit to the same data points.  }
  \label{predev}
\end{figure}

To examine the extent of ''deviation'' from Eqs. (\ref{predu}) and (\ref{predw}) we show in Fig. \ref{predev} for the asymmetric dumbbell $U$ and $W_{LJ}$ as a function of the reduced density $\tilde{\rho}^{2}$ 
($\rho^{*} = 1$). 
The reference coefficients can be calculated from a linear regression fit of the potential energy and the estimated coefficients can be used to plot a straight line in the \textit{LJ} virial plot. This is performed in 
Fig. \ref{predev} where it is clear that even though both plots follow a near straight-line, the coefficients are not given by Eqs. (\ref{predu}) and (\ref{predw}). 
It is worth mentioning again that the prediction of Ref. \onlinecite{paper5} is for an atomic system, and is as such not excepted to hold for rigid molecular systems. 

\begin{figure}[H]
  \centering
  \includegraphics[width=70mm]{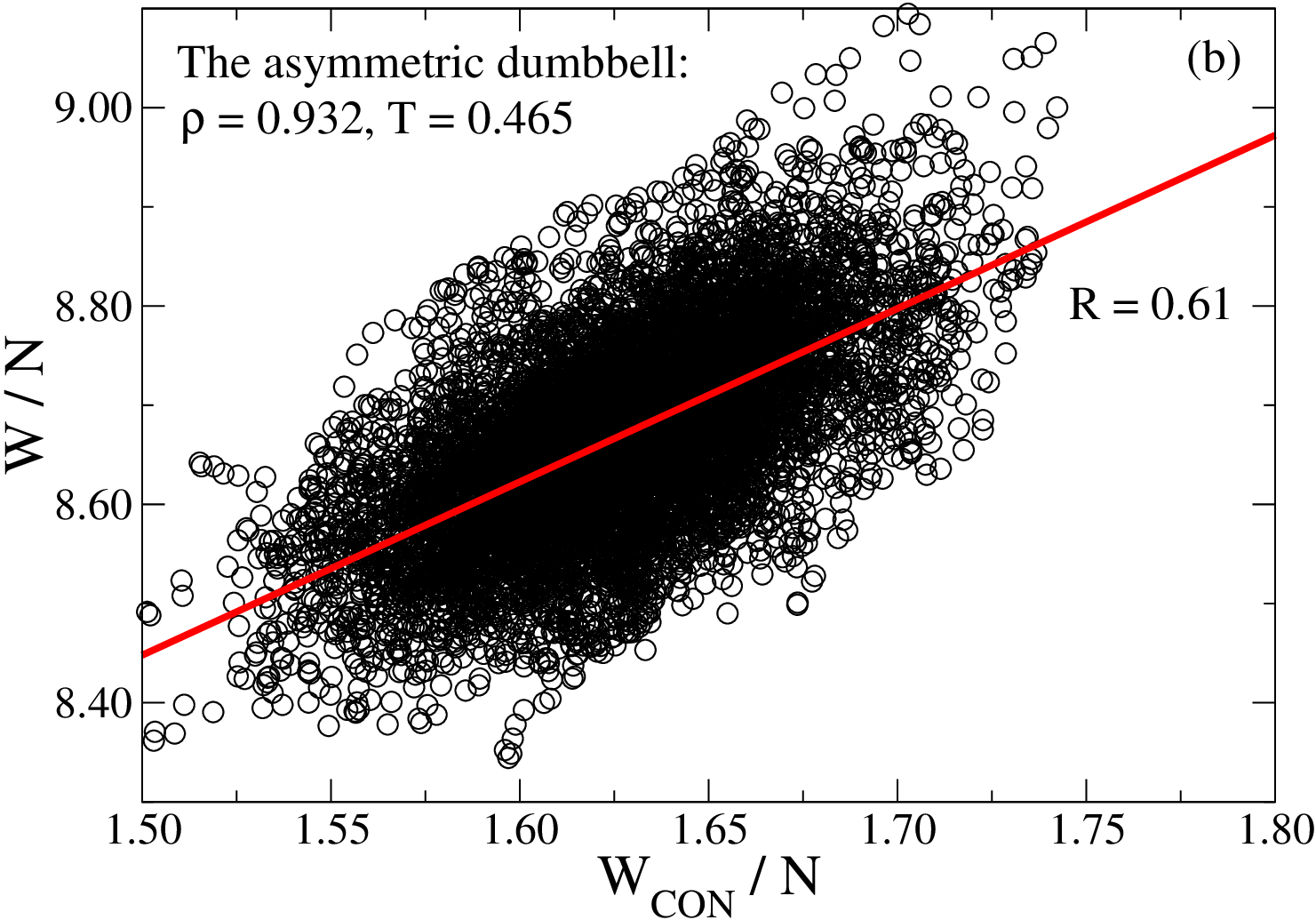}
  \caption{(a) The correlation of the instantaneous fluctuations of $W_{LJ}$ and $W_{CON}$. The correlation coefficient $R$ is 0.31. (b) The correlation of the instantaneous fluctuations of $W$ and $W_{CON}$. 
    The correlation coefficient $R$ is 0.61. }
  \label{wcor}
\end{figure}

Finally we consider in Fig. \ref{wcor} for the asymmetric dumbbell how the instantaneous fluctuations of $W_{CON}$ correlate with $W_{LJ}$ and $W$ respectively. The constraint contribution to the virial at this state point 
does not correlate well with the contribution to the virial coming from the \textit{LJ} interactions ($R$ = $0.31$). Obviously, the correlation is higher when the total virial is considered ($R$ = $0.61$).
The main contribution to the virial, for the asymmetric dumbbell model, comes from the \textit{LJ} interactions, however, the \textit{LJ} virial does not 
correlate well with the constraint virial. The latter observation may indicate a break-down of master isomorph scaling (for the total virial) at high pressures, however, this remains to be 
confirmed.

\begin{figure}[H]
  \centering
  \includegraphics[width=70mm]{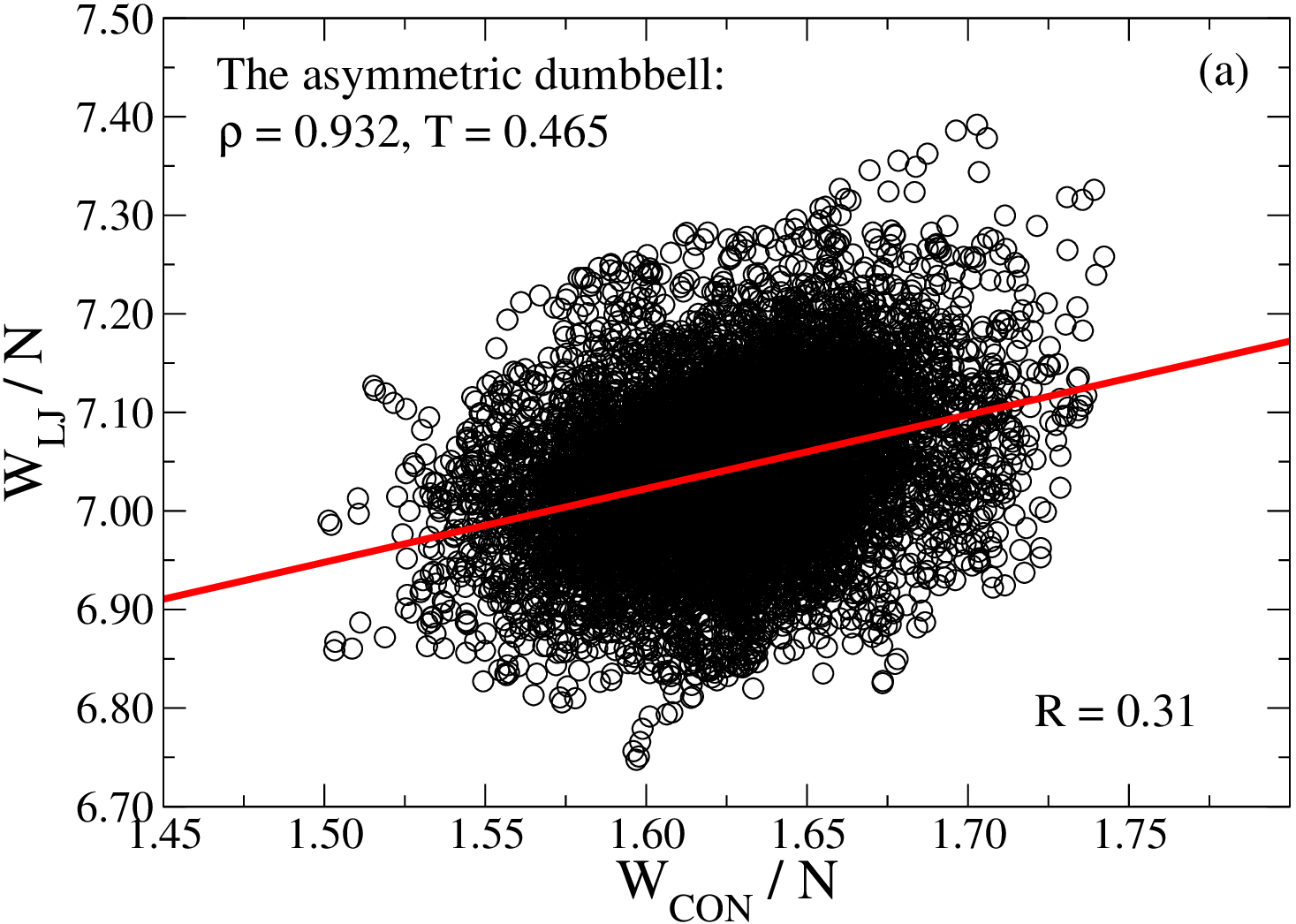}
\end{figure}

\section{Summary and outlook}\label{sum}

Isomorphs are curves in the phase diagram of a strongly correlating liquid along which a number of static and dynamic 
quantities are invariant in reduced units. References \onlinecite{paper4} and \onlinecite{paper5} focused on understanding isomorphs
in  atomic systems. In this paper we generalized the isomorph concept to deal with systems of rigid molecules [Eq. (\ref{defiso1})] and investigated several isomorph 
invariants for the asymmetric dumbbell model and the Lewis-Wahnstr{\"o}m \textit{OTP} model.
We find that these rigid molecular systems also have isomorphs to a good approximation; however, the isomorphs of the \textit{OTP} model were more 
approximative than those of the asymmetric dumbbell, consistent with the \textit{OTP} model being less strongly correlating. 
Moreover, it was found that these systems to a good approximation have master isomorphs, i.e., that all isomorphs have the same shape in the virial/potential energy 
phase diagram. This applies for the total virial, but also after subtracting the constraint contribution.  
A general master isomorph was identified between the investigated model systems after this subtraction. We do not at the present have an explanation for 
this observation.

A full theoretical understanding of the implications of rigid bonds remains to be arrived at. For instance, the shape of molecular isomorphs is different from the 
shape of Ref. \onlinecite{paper5} for atomic isomorphs. The rigid bonds seem in general to increase $\gamma$ and decrease the correlation 
coefficient $R$ with respect to the unconstrained system. More specific; $R$ decreases significantly with increasing asymmetric dumbbell bond length 
($R \approx 0.65$ around unity bond length, see Sec. \ref{isoasymsec}). 
This is consistent with the results of Chopra \textit{et al.}\cite{truskdumb}, who note a worse scaling of the reduced relaxation time and diffusion constant with (excess) 
entropy based quantities when increasing the bond length of a rigid symmetric \textit{LJ} dumbbell model. On the other hand, it is noteworthy that strong correlation is observed 
for the \textit{OTP} model even though it has unity bond lengths.
The molecular center-of-mass structure in reduced units is predicted to be invariant along an isomorph, however, for the \textit{OTP} model the reduced particle structure seems 
more invariant along an isomorph than the reduced molecular center-of-mass structure. The former is not predicted to be invariant along an isomorph, and thus the observed difference should be investigated in 
more detail to clarify this issue.

\acknowledgments 

The centre for viscous liquid dynamics ``Glass and Time'' is sponsored by the Danish National Research Foundation (DNRF).

\appendix

\section{Constrained dynamics and the virial}\label{appConstrained}

Constrained dynamics is discussed in many different places, see for instance Refs.  \onlinecite{tox1}, \onlinecite{edberg}, and \onlinecite{ryck}. We give here 
a brief introduction to constrained 
dynamics and the connection to the virial expression used in this article.

Gauss' principle of least constraint\cite{gauss} states that a classical mechanical system of $N$ particles with constraints deviates 
instantaneously in a least possible sense from Newton's 2nd law, i.e., that 

\begin{equation}
  \sum_{i=1}^{N}m_{i}\Big[\ddot{\textbf{r}}_{i} - \frac{\textbf{F}_{i}}{m_{i}}\Big]^{2},
\end{equation}
is a minimum. Here $\textbf{r}_{i}$ and $\textbf{F}_{i}$ are the position and interaction force of particle $i$. 
In the case of no constraints, setting the partial derivative $\partial / \partial \ddot{\textbf{r}}_{i}$ to zero implies 
$\ddot{\textbf{r}}_{i} - \textbf{F}_{i}/m_{i} = 0$, i.e., Newtons's 2nd law.

In the case of holonomic constraints $\psi^{\alpha}(\textbf{r}^{N}) = 0$ where $\alpha = 1, ..., G$, the variation can be carried out by introducing Lagrangian multipliers, i.e.,  

\begin{equation}
  \sum_{i=1}^{N}m_{i}\Big[\ddot{\textbf{r}}_{i} - \frac{\textbf{F}_{i}}{m_{i}}\Big]^{2} - \sum_{\alpha=1}^{G}\lambda^{\alpha}\ddot{\psi}^{\alpha},
\end{equation}
should be a minimum. Setting the partial derivative $\partial / \partial \ddot{\textbf{r}}_{i}$ to zero implies (where the factor of one half has been absorbed in the Lagrangian multiplier)

\begin{equation}\label{eqm}
  m_{i}\cdot \ddot{\textbf{r}}_{i} = \textbf{F}_{i} +  \sum_{\alpha=1}^{G} \lambda^{\alpha}\nabla_{\textbf{r}_{i}}\psi^{\alpha} = \textbf{F}_{i}+\textbf{G}_{i}.
\end{equation}
Newton's 2nd law thus remains valid if an additional force is added (called the constraint force $\textbf{G}_{i}$). At this point $\lambda^{\alpha}$ 
is undetermined; however, an explicit 
expression\cite{edberg} for $\lambda^{\alpha}$ can be determined from the condition $\textbf{r}_{ij}\cdot\ddot{\textbf{r}}_{ij} + \dot{\textbf{r}}_{ij}^{2} = 0$. 
In molecular dynamics simulations it is imperative to calculate $\lambda^{\alpha}$ correctly to achieve a stable numerical algorithm. The reader 
is referred to Refs. \onlinecite{tox1} and \onlinecite{tox2} for details concerning this aspect.

The virial $W$ is defined by $W$ = $1/3\sum_{i=1}^{N}\textbf{r}_{i}\cdot\textbf{F}_{i}$. In an atomic system with \textit{LJ} pair 
potential interactions the virial is given by $W = W_{LJ} = -1/3\sum_{i<j}^{N}r_{ij}u'(r_{ij})$. 
If the system has bond constraints $\psi^{\alpha}$ = $(\textbf{r}_{\alpha,i} - \textbf{r}_{\alpha,j})^{2}/2$ = $\textbf{r}^{2}_{\alpha,ij}$/2 = $c_{\alpha,ij}^{2}/2$ it follows from Eq. (\ref{eqm}) that the 
constraint force contributes to the virial as $W_{CON} = 1/3\sum_{i=1}^{N}\textbf{r}_{i}\cdot \textbf{G}_{i} =1/3\sum_{\alpha=1}^{G}\lambda^{\alpha}\textbf{r}_{\alpha,ij}^{2}$.

\section{Constrained \textit{NVE} and Nos$\acute{e}$-Hoover \textit{NVT} dynamics in reduced units along an isomorph}\label{appInvdyn}

We start our considerations of this section from the constrained equations of motion derived in Appendix \ref{appConstrained}, Eq. (\ref{eqm}):

\begin{equation}\label{eqm1}
  m_{i} \cdot \ddot{\textbf{r}}_{i} = \textbf{F}_{i} +  \sum_{\alpha=1}^{G}\lambda^{\alpha}\nabla_{\textbf{r}_{i}}\psi^{\alpha} = \textbf{F}_{i}+\textbf{G}_{i}.
\end{equation}
Here $\textbf{r}_{i}$ and $\textbf{F}_{i}$ are, respectively, the position and interaction force of particle $i$, and $\lambda^{\alpha}$ a Lagrangian multiplier for the $\alpha$'te constraint $\psi^{\alpha}$.
For simulating rigid molecules\cite{rigid} the constraints would in general be a combination of constrained bond lengths $\psi^{\alpha}$ = $(\textbf{r}_{\alpha,i} - \textbf{r}_{\alpha,j})^{2}/2$ 
= $\textbf{r}^{2}_{\alpha,ij}$/2 = $c_{\alpha,ij}^{2}/2$ and linear constraints $\psi^{\beta} = \sum_{i=1}^{n_{b}}C_{\beta i}\textbf{r}_{i} - \textbf{r}_{\beta} = 0$, where $C_{\beta i}$ is a factor that 
depends on the geometry of the molecule (see Ref. \onlinecite{rigid} for more details). For simplicity we consider only bond constraints in the following.

A general expression for the Lagrange multiplier $\lambda^{\alpha}$ can be derived and is given by\cite{edberg,mel} 

\begin{align}
  \lambda^{\alpha} & = - \sum_{\beta=1}^{G} \big ( \textbf{Z}^{-1} \big )_{\alpha \beta} \Big[\sum_{i,j=1}^{N} \nabla_{\textbf{r}_{i}} \nabla_{\textbf{r}_{j}}\psi^{\beta}\dot{\textbf{r}}_{j}\dot{\textbf{r}}_{i} +
  \sum_{i=1}^{N}\frac{\nabla_{\textbf{r}_{i}} \psi^{\beta} \cdot \textbf{F}_{i}}{m_{i}} \Big], \\
  Z_{\alpha \beta} & = \sum_{i=1}^{N} \frac{\nabla_{\textbf{r}_{i}} \psi^{\alpha} \cdot \nabla_{\textbf{r}_{i}} \psi^{\beta}}{m_{i}}.\label{eqm2}
\end{align}
Defining reduced units for length, energy, and mass as follows

\begin{align}
  \tilde{\textbf{r}}_{i} & = \rho^{1/3}\textbf{r}_{i}, \\ 
  \tilde{U} & = U / k_{B}T, \\
  \tilde{m}_{i} & = m_{i}/\langle m \rangle,
\end{align}
reduced units for time and force follow as

\begin{align}
  \tilde{t} & = t/\big(\rho^{-1/3} \sqrt{\langle m \rangle/k_{B}T}\big), \\
  \tilde{\textbf{F}}_{i} & = \rho^{-1/3}\textbf{F}_{i}/k_{B}T = -\nabla_{\tilde{\textbf{r}}_{i}}U/k_{B}T.
\end{align}
Inserting the above definitions in Eqs. (\ref{eqm1}) - (\ref{eqm2}) and using that $\nabla_{\textbf{r}_{i}}\psi^{\alpha} = \textbf{r}_{\alpha,ij}$, 
we arrive at the constrained \textit{NVE} equations of motion in reduced units

\begin{equation}
  \tilde{m}_{i} \cdot \ddot{\tilde{\textbf{r}}}_{i} = \tilde{\textbf{F}}_{i} + \sum_{\alpha=1}^{G}\tilde{\lambda}^{\alpha}\tilde{\textbf{r}}_{\alpha,ij} 
  = \tilde{\textbf{F}}_{i} + \tilde{\textbf{G}}_{i},
\end{equation}
where

\begin{align}
  \tilde{\lambda}^{\alpha} & = - \sum_{\beta=1}^{G} \big ( \tilde{\textbf{Z}}^{-1} \big )_{\alpha \beta} \Big[\sum_{i,j=1}^{N}\dot{\tilde{\textbf{r}}}_{j}\dot{\tilde{\textbf{r}}}_{i} +
  \sum_{i=1}^{N}\frac{\tilde{\textbf{r}}_{\beta,ij} \cdot \tilde{\textbf{F}}_{i}}{\tilde{m}_{i}}\Big], \\
  \tilde{Z}_{\alpha \beta} & = \sum_{i=1}^{N}\frac{\tilde{\textbf{r}}_{\alpha,ij}\tilde{\textbf{r}}_{\beta,ij}}{\tilde{m}_{i}}.
\end{align}
Since in general $\tilde{\textbf{r}}_{\alpha,ij}^{2} = \rho^{2/3}c_{\alpha,ij}^{2}$ the reduced constrained equations of motion are \textit{not} invariant along an isomorph.   

Considering instead the molecular center-of-mass motion in reduced units

\begin{equation}\label{redcon}
  \tilde{M}_{i} \cdot \ddot{\tilde{\textbf{r}}}_{CM,i} = \tilde{\textbf{F}}_{CM,i},
\end{equation}
where $\tilde{\textbf{F}}_{CM,i}$ and $\tilde{M}_{i}$ are, respectively, the reduced force and mass of molecule $i$.
Since the reduced force $\tilde{\textbf{F}}_{CM,i}$ is invariant along an isomorph, it follows that the molecular \textit{NVE} 
equations of motion are invariant along an isomorph. 
The invariance of $\tilde{\textbf{F}}_{CM,i}$ can be seen as follows. The isomorph definition Eq. (\ref{defiso1}) implies for a 
fixed state point ($1$) and arbitrary state point ($x$), both along the same isomorph [where $\tilde{\textbf{R}}$ $\equiv$ ($\rho^{-1/3}\tilde{\textbf{r}}_{CM,1}$, $\theta_{1}$, $\phi_{1}$, $\psi_{1}$, ..., 
$\rho^{-1/3}\tilde{\textbf{r}}_{CM,N}$, $\theta_{N}$, $\phi_{N}$, $\psi_{N}$)]

\begin{equation}
  -U(\tilde{\textbf{R}}^{(x)})/k_{B}T_{x} = -U(\tilde{\textbf{R}}^{(1)})/k_{B}T_{1} - \ln\, C_{1x}.
\end{equation}
Taking the gradient $\nabla_{\tilde{\textbf{r}}_{CM,i}}$ it follows that

\begin{equation}
  \tilde{\textbf{F}}_{CM,i}^{(x)} = \tilde{\textbf{F}}_{CM,i}^{(1)}.
\end{equation}
This concludes the proof of the isomorph invariance of the reduced molecular center-of-mass equations of motion. 
A similar situation is given for the constrained \textit{NVT} equations of motion, however; considering 
the molecular motion, the constraint force disappears, and the proof in analogous to the above and shown for atomic systems in Ref. \onlinecite{paper4}. In this case 
the time-constant of the Nos$\acute{e}$-Hoover algorithm needs to be adjusted along the isomorph, otherwise the dynamics is not invariant.


\begin{mcitethebibliography}{46}
\providecommand*\natexlab[1]{#1}
\providecommand*\mciteSetBstSublistMode[1]{}
\providecommand*\mciteSetBstMaxWidthForm[2]{}
\providecommand*\mciteBstWouldAddEndPuncttrue
  {\def\EndOfBibitem{\unskip.}}
\providecommand*\mciteBstWouldAddEndPunctfalse
  {\let\EndOfBibitem\relax}
\providecommand*\mciteSetBstMidEndSepPunct[3]{}
\providecommand*\mciteSetBstSublistLabelBeginEnd[3]{}
\providecommand*\EndOfBibitem{}
\mciteSetBstSublistMode{f}
\mciteSetBstMaxWidthForm{subitem}{(\alph{mcitesubitemcount})}
\mciteSetBstSublistLabelBeginEnd
  {\mcitemaxwidthsubitemform\space}
  {\relax}
  {\relax}

\bibitem[Alba-Simionesco et~al.(2004)Alba-Simionesco, Cailliaux, Alegr{\'i}a,
  and Tarjus]{dscale}
Alba-Simionesco,~C.; Cailliaux,~A.; Alegr{\'i}a,~A.; Tarjus,~G. \emph{Europhys.
  Lett.} \textbf{2004}, \emph{68}, 58\relax
\mciteBstWouldAddEndPuncttrue
\mciteSetBstMidEndSepPunct{\mcitedefaultmidpunct}
{\mcitedefaultendpunct}{\mcitedefaultseppunct}\relax
\EndOfBibitem
\bibitem[Dreyfus et~al.(2004)Dreyfus, Grand, Gapinski, Steffen, and
  Patkowski]{dscale1}
Dreyfus,~C.; Grand,~A.~L.; Gapinski,~J.; Steffen,~W.; Patkowski,~A. \emph{Eur.
  Phys. J. B} \textbf{2004}, \emph{42}, 309\relax
\mciteBstWouldAddEndPuncttrue
\mciteSetBstMidEndSepPunct{\mcitedefaultmidpunct}
{\mcitedefaultendpunct}{\mcitedefaultseppunct}\relax
\EndOfBibitem
\bibitem[Roland(2010)]{reviewRoland}
Roland,~C.~M. \emph{Macromolecules} \textbf{2010}, \emph{43}, 7875\relax
\mciteBstWouldAddEndPuncttrue
\mciteSetBstMidEndSepPunct{\mcitedefaultmidpunct}
{\mcitedefaultendpunct}{\mcitedefaultseppunct}\relax
\EndOfBibitem
\bibitem[Roland et~al.(2003)Roland, Casalini, and Paluch]{isochone1}
Roland,~C.~M.; Casalini,~R.; Paluch,~M. \emph{Chem. Phys. Lett.} \textbf{2003},
  \emph{367}, 259\relax
\mciteBstWouldAddEndPuncttrue
\mciteSetBstMidEndSepPunct{\mcitedefaultmidpunct}
{\mcitedefaultendpunct}{\mcitedefaultseppunct}\relax
\EndOfBibitem
\bibitem[Ngai et~al.(2005)Ngai, Casalini, Capaccioli, Paluch, and
  Roland]{isochone2}
Ngai,~K.~L.; Casalini,~R.; Capaccioli,~S.; Paluch,~M.; Roland,~C.~M. \emph{J.
  Phys. Chem. B} \textbf{2005}, \emph{109}, 17356\relax
\mciteBstWouldAddEndPuncttrue
\mciteSetBstMidEndSepPunct{\mcitedefaultmidpunct}
{\mcitedefaultendpunct}{\mcitedefaultseppunct}\relax
\EndOfBibitem
\bibitem[Rosenfeld(1977)]{rosenfeld1}
Rosenfeld,~Y. \emph{Phys. Rev. A} \textbf{1977}, \emph{15}, 2545\relax
\mciteBstWouldAddEndPuncttrue
\mciteSetBstMidEndSepPunct{\mcitedefaultmidpunct}
{\mcitedefaultendpunct}{\mcitedefaultseppunct}\relax
\EndOfBibitem
\bibitem[Rosenfeld(1999)]{rosenfeld2}
Rosenfeld,~Y. \emph{J. Phys.: Condens. Matter} \textbf{1999}, \emph{11},
  5415\relax
\mciteBstWouldAddEndPuncttrue
\mciteSetBstMidEndSepPunct{\mcitedefaultmidpunct}
{\mcitedefaultendpunct}{\mcitedefaultseppunct}\relax
\EndOfBibitem
\bibitem[Chopra et~al.(2010)Chopra, Truskett, and Errington]{truskwater}
Chopra,~R.; Truskett,~T.~M.; Errington,~J.~R. \emph{J. Phys. Chem. B}
  \textbf{2010}, \emph{114}, 10558\relax
\mciteBstWouldAddEndPuncttrue
\mciteSetBstMidEndSepPunct{\mcitedefaultmidpunct}
{\mcitedefaultendpunct}{\mcitedefaultseppunct}\relax
\EndOfBibitem
\bibitem[Chopra et~al.(2010)Chopra, Truskett, and Errington]{truskhydro}
Chopra,~R.; Truskett,~T.~M.; Errington,~J.~R. \emph{J. Phys. Chem. B}
  \textbf{2010}, \emph{114}, 16487\relax
\mciteBstWouldAddEndPuncttrue
\mciteSetBstMidEndSepPunct{\mcitedefaultmidpunct}
{\mcitedefaultendpunct}{\mcitedefaultseppunct}\relax
\EndOfBibitem
\bibitem[Chopra et~al.(2010)Chopra, Truskett, and Errington]{truskdumb}
Chopra,~R.; Truskett,~T.~M.; Errington,~J.~R. \emph{J. Chem. Phys.}
  \textbf{2010}, \emph{133}, 104506\relax
\mciteBstWouldAddEndPuncttrue
\mciteSetBstMidEndSepPunct{\mcitedefaultmidpunct}
{\mcitedefaultendpunct}{\mcitedefaultseppunct}\relax
\EndOfBibitem
\bibitem[Galliero et~al.(2011)Galliero, Boned, and
  Fern$\acute{a}$ndez]{exviscous}
Galliero,~G.; Boned,~C.; Fern$\acute{a}$ndez,~J. \emph{J. Phys. Chem.}
  \textbf{2011}, \emph{134}, 064505\relax
\mciteBstWouldAddEndPuncttrue
\mciteSetBstMidEndSepPunct{\mcitedefaultmidpunct}
{\mcitedefaultendpunct}{\mcitedefaultseppunct}\relax
\EndOfBibitem
\bibitem[Bailey et~al.(2008)Bailey, Pedersen, Gnan, Schr{\o}der, and
  Dyre]{paper1}
Bailey,~N.~P.; Pedersen,~U.~R.; Gnan,~N.; Schr{\o}der,~T.~B.; Dyre,~J.~C.
  \emph{J. Chem. Phys.} \textbf{2008}, \emph{128}, 184507\relax
\mciteBstWouldAddEndPuncttrue
\mciteSetBstMidEndSepPunct{\mcitedefaultmidpunct}
{\mcitedefaultendpunct}{\mcitedefaultseppunct}\relax
\EndOfBibitem
\bibitem[Bailey et~al.(2008)Bailey, Pedersen, Gnan, Schr{\o}der, and
  Dyre]{paper2}
Bailey,~N.~P.; Pedersen,~U.~R.; Gnan,~N.; Schr{\o}der,~T.~B.; Dyre,~J.~C.
  \emph{J. Chem. Phys.} \textbf{2008}, \emph{129}, 184508\relax
\mciteBstWouldAddEndPuncttrue
\mciteSetBstMidEndSepPunct{\mcitedefaultmidpunct}
{\mcitedefaultendpunct}{\mcitedefaultseppunct}\relax
\EndOfBibitem
\bibitem[Schr{\o}der et~al.(2009)Schr{\o}der, Bailey, Pedersen, Gnan, and
  Dyre]{paper3}
Schr{\o}der,~T.~B.; Bailey,~N.~P.; Pedersen,~U.~R.; Gnan,~N.; Dyre,~J.~C.
  \emph{J. Chem. Phys.} \textbf{2009}, \emph{131}, 234503\relax
\mciteBstWouldAddEndPuncttrue
\mciteSetBstMidEndSepPunct{\mcitedefaultmidpunct}
{\mcitedefaultendpunct}{\mcitedefaultseppunct}\relax
\EndOfBibitem
\bibitem[Gnan et~al.(2009)Gnan, Schr{\o}der, Pedersen, Bailey, and
  Dyre]{paper4}
Gnan,~N.; Schr{\o}der,~T. .~B.; Pedersen,~U.~R.; Bailey,~N.~P.; Dyre,~J.~C.
  \emph{J. Chem. Phys.} \textbf{2009}, \emph{131}, 234504\relax
\mciteBstWouldAddEndPuncttrue
\mciteSetBstMidEndSepPunct{\mcitedefaultmidpunct}
{\mcitedefaultendpunct}{\mcitedefaultseppunct}\relax
\EndOfBibitem
\bibitem[Schr{\o}der et~al.(2011)Schr{\o}der, Gnan, Pedersen, Bailey, and
  Dyre]{paper5}
Schr{\o}der,~T.~B.; Gnan,~N.; Pedersen,~U.~R.; Bailey,~N.~P.; Dyre,~J.~C.
  \emph{J. Chem. Phys.} \textbf{2011}, \emph{134}, 164505\relax
\mciteBstWouldAddEndPuncttrue
\mciteSetBstMidEndSepPunct{\mcitedefaultmidpunct}
{\mcitedefaultendpunct}{\mcitedefaultseppunct}\relax
\EndOfBibitem
\bibitem[Pedersen et~al.(2008)Pedersen, Bailey, Schr{\o}der, and Dyre]{first}
Pedersen,~U.~R.; Bailey,~N.~P.; Schr{\o}der,~T.~B.; Dyre,~J.~C. \emph{Phys.
  Rev. Lett.} \textbf{2008}, \emph{100}, 015701\relax
\mciteBstWouldAddEndPuncttrue
\mciteSetBstMidEndSepPunct{\mcitedefaultmidpunct}
{\mcitedefaultendpunct}{\mcitedefaultseppunct}\relax
\EndOfBibitem
\bibitem[Coslovich and Roland(2009)Coslovich, and Roland]{coslovich1}
Coslovich,~D.; Roland,~C.~M. \emph{J. Chem. Phys.} \textbf{2009}, \emph{130},
  014508\relax
\mciteBstWouldAddEndPuncttrue
\mciteSetBstMidEndSepPunct{\mcitedefaultmidpunct}
{\mcitedefaultendpunct}{\mcitedefaultseppunct}\relax
\EndOfBibitem
\bibitem[Coslovich and Roland(2008)Coslovich, and Roland]{coslovich2}
Coslovich,~D.; Roland,~C.~M. \emph{J. Phys. Chem. B} \textbf{2008}, \emph{112},
  1329\relax
\mciteBstWouldAddEndPuncttrue
\mciteSetBstMidEndSepPunct{\mcitedefaultmidpunct}
{\mcitedefaultendpunct}{\mcitedefaultseppunct}\relax
\EndOfBibitem
\bibitem[Schr{\o}der et~al.(2009)Schr{\o}der, Pedersen, Bailey, Toxvaerd, and
  Dyre]{hidden}
Schr{\o}der,~T.~B.; Pedersen,~U.~R.; Bailey,~N.~P.; Toxvaerd,~S.; Dyre,~J.~C.
  \emph{Phys. Rev. E} \textbf{2009}, \emph{80}, 041502\relax
\mciteBstWouldAddEndPuncttrue
\mciteSetBstMidEndSepPunct{\mcitedefaultmidpunct}
{\mcitedefaultendpunct}{\mcitedefaultseppunct}\relax
\EndOfBibitem
\bibitem[Kob and Andersen(1995)Kob, and Andersen]{ka1}
Kob,~W.; Andersen,~H.~C. \emph{Phys. Rev. E} \textbf{1995}, \emph{51},
  4626\relax
\mciteBstWouldAddEndPuncttrue
\mciteSetBstMidEndSepPunct{\mcitedefaultmidpunct}
{\mcitedefaultendpunct}{\mcitedefaultseppunct}\relax
\EndOfBibitem
\bibitem[Kob and Andersen(1995)Kob, and Andersen]{ka2}
Kob,~W.; Andersen,~H.~C. \emph{Phys. Rev. E} \textbf{1995}, \emph{52},
  4134\relax
\mciteBstWouldAddEndPuncttrue
\mciteSetBstMidEndSepPunct{\mcitedefaultmidpunct}
{\mcitedefaultendpunct}{\mcitedefaultseppunct}\relax
\EndOfBibitem
\bibitem[Wahnstr{\"o}m and Lewis(1993)Wahnstr{\"o}m, and Lewis]{otp1}
Wahnstr{\"o}m,~G.; Lewis,~L.~J. \emph{Physica A} \textbf{1993}, \emph{201},
  150\relax
\mciteBstWouldAddEndPuncttrue
\mciteSetBstMidEndSepPunct{\mcitedefaultmidpunct}
{\mcitedefaultendpunct}{\mcitedefaultseppunct}\relax
\EndOfBibitem
\bibitem[Lewis and Wahnstr{\"o}m(1994)Lewis, and Wahnstr{\"o}m]{otp2}
Lewis,~L.~J.; Wahnstr{\"o}m,~G. \emph{Phys. Rev. E} \textbf{1994}, \emph{50},
  3865\relax
\mciteBstWouldAddEndPuncttrue
\mciteSetBstMidEndSepPunct{\mcitedefaultmidpunct}
{\mcitedefaultendpunct}{\mcitedefaultseppunct}\relax
\EndOfBibitem
\bibitem[Gundermann et~al.(2011)Gundermann, Pedersen, Hecksher, Bailey,
  Jakobsen, Christensen, Olsen, Schr{\o}der, Fragiadakis, Casalini, Roland,
  Dyre, and Niss]{gamma}
Gundermann,~D.; Pedersen,~U.~R.; Hecksher,~T.; Bailey,~N.~P.; Jakobsen,~B.;
  Christensen,~T.; Olsen,~N.~B.; Schr{\o}der,~T.~B.; Fragiadakis,~D.;
  Casalini,~R.; Roland,~C.~M.; Dyre,~J.~C.; Niss,~K. \emph{Nature Physics}
  \textbf{2011}, DOI: 10.1038\relax
\mciteBstWouldAddEndPuncttrue
\mciteSetBstMidEndSepPunct{\mcitedefaultmidpunct}
{\mcitedefaultendpunct}{\mcitedefaultseppunct}\relax
\EndOfBibitem
\bibitem[Pedersen et~al.(2011)Pedersen, Schr{\o}der, Bailey, Toxvaerd, and
  Dyre]{overviewscl}
Pedersen,~U.~R.; Schr{\o}der,~T.~B.; Bailey,~N.~P.; Toxvaerd,~S.; Dyre,~J.~C.
  \emph{J. of Non-Crystalline Solids} \textbf{2011}, \emph{357}, 320\relax
\mciteBstWouldAddEndPuncttrue
\mciteSetBstMidEndSepPunct{\mcitedefaultmidpunct}
{\mcitedefaultendpunct}{\mcitedefaultseppunct}\relax
\EndOfBibitem
\bibitem[pra()]{prac}
In practice it's only required that the physically relevant configurations obey
  this scaling, i.e., at least those that contribute significantly to the
  partition function.\relax
\mciteBstWouldAddEndPunctfalse
\mciteSetBstMidEndSepPunct{\mcitedefaultmidpunct}
{}{\mcitedefaultseppunct}\relax
\EndOfBibitem
\bibitem[Goldstein et~al.(2002)Goldstein, Poole, and Safko]{goldstein}
Goldstein,~H.; Poole,~C.; Safko,~J. \emph{Classical Mechanics}, 3rd ed.;
  Addison Wesley, 2002\relax
\mciteBstWouldAddEndPuncttrue
\mciteSetBstMidEndSepPunct{\mcitedefaultmidpunct}
{\mcitedefaultendpunct}{\mcitedefaultseppunct}\relax
\EndOfBibitem
\bibitem[Gray and Gubbins(1984)Gray, and Gubbins]{gubbins}
Gray,~C.~G.; Gubbins,~K.~E. \emph{Theory of Molecular Fluids}; Oxford
  University Press, 1984\relax
\mciteBstWouldAddEndPuncttrue
\mciteSetBstMidEndSepPunct{\mcitedefaultmidpunct}
{\mcitedefaultendpunct}{\mcitedefaultseppunct}\relax
\EndOfBibitem
\bibitem[Lazaridis and Paulaitis(1992)Lazaridis, and Paulaitis]{excessexp}
Lazaridis,~T.; Paulaitis,~M.~E. \emph{J. Phys. Chem.} \textbf{1992}, \emph{96},
  3847\relax
\mciteBstWouldAddEndPuncttrue
\mciteSetBstMidEndSepPunct{\mcitedefaultmidpunct}
{\mcitedefaultendpunct}{\mcitedefaultseppunct}\relax
\EndOfBibitem
\bibitem[Lazaridis and Karplus(1996)Lazaridis, and Karplus]{ex}
Lazaridis,~T.; Karplus,~M. \emph{J. Chem. Phys.} \textbf{1996}, \emph{105},
  4294\relax
\mciteBstWouldAddEndPuncttrue
\mciteSetBstMidEndSepPunct{\mcitedefaultmidpunct}
{\mcitedefaultendpunct}{\mcitedefaultseppunct}\relax
\EndOfBibitem
\bibitem[Pedersen et~al.(2011)Pedersen, Hudson, and Harrowell]{otpulf}
Pedersen,~U.~R.; Hudson,~T.~S.; Harrowell,~P. \emph{J. Chem. Phys.}
  \textbf{2011}, \emph{134}, 114501\relax
\mciteBstWouldAddEndPuncttrue
\mciteSetBstMidEndSepPunct{\mcitedefaultmidpunct}
{\mcitedefaultendpunct}{\mcitedefaultseppunct}\relax
\EndOfBibitem
\bibitem[Hess et~al.(2008)Hess, Kutzner, van~der Spoel, and Lindahl]{gromacs}
Hess,~B.; Kutzner,~C.; van~der Spoel,~D.; Lindahl,~E. \emph{J. Chem. Theory
  Comput.} \textbf{2008}, \emph{4}, 435\relax
\mciteBstWouldAddEndPuncttrue
\mciteSetBstMidEndSepPunct{\mcitedefaultmidpunct}
{\mcitedefaultendpunct}{\mcitedefaultseppunct}\relax
\EndOfBibitem
\bibitem[Toxvaerd et~al.(2009)Toxvaerd, Heilmann, Ingebrigtsen, Schr{\o}der,
  and Dyre]{tox1}
Toxvaerd,~S.; Heilmann,~O.~J.; Ingebrigtsen,~T.; Schr{\o}der,~T.~B.;
  Dyre,~J.~C. \emph{J. Chem. Phys.} \textbf{2009}, \emph{131}, 064102\relax
\mciteBstWouldAddEndPuncttrue
\mciteSetBstMidEndSepPunct{\mcitedefaultmidpunct}
{\mcitedefaultendpunct}{\mcitedefaultseppunct}\relax
\EndOfBibitem
\bibitem[Ingebrigtsen et~al.(2010)Ingebrigtsen, Heilmann, Toxvaerd, and
  Dyre]{tox2}
Ingebrigtsen,~T.; Heilmann,~O.~J.; Toxvaerd,~S.; Dyre,~J.~C. \emph{J. Chem.
  Phys} \textbf{2010}, \emph{132}, 154106\relax
\mciteBstWouldAddEndPuncttrue
\mciteSetBstMidEndSepPunct{\mcitedefaultmidpunct}
{\mcitedefaultendpunct}{\mcitedefaultseppunct}\relax
\EndOfBibitem
\bibitem[Nos$\acute{e}$(1984)]{nose}
Nos$\acute{e}$,~S. \emph{J. Chem. Phys.} \textbf{1984}, \emph{81}, 511\relax
\mciteBstWouldAddEndPuncttrue
\mciteSetBstMidEndSepPunct{\mcitedefaultmidpunct}
{\mcitedefaultendpunct}{\mcitedefaultseppunct}\relax
\EndOfBibitem
\bibitem[Hoover(1985)]{hoover}
Hoover,~W.~G. \emph{Phys. Rev. A} \textbf{1985}, \emph{31}, 1695\relax
\mciteBstWouldAddEndPuncttrue
\mciteSetBstMidEndSepPunct{\mcitedefaultmidpunct}
{\mcitedefaultendpunct}{\mcitedefaultseppunct}\relax
\EndOfBibitem
\bibitem[Toxvaerd(1991)]{canotox}
Toxvaerd,~S. \emph{Mol. Phys.} \textbf{1991}, \emph{72}, 159\relax
\mciteBstWouldAddEndPuncttrue
\mciteSetBstMidEndSepPunct{\mcitedefaultmidpunct}
{\mcitedefaultendpunct}{\mcitedefaultseppunct}\relax
\EndOfBibitem
\bibitem[rum()]{rumd}
All simulations were performed using a molecular dynamics code optimized for
  \textit{NVIDIA} graphics cards, which is available as open source code at
  www.rumd.org.\relax
\mciteBstWouldAddEndPunctfalse
\mciteSetBstMidEndSepPunct{\mcitedefaultmidpunct}
{}{\mcitedefaultseppunct}\relax
\EndOfBibitem
\bibitem[Frenkel and Smit(2002)Frenkel, and Smit]{frenkel}
Frenkel,~D.; Smit,~B. \emph{Understanding Molecular Simulation}; Academic
  Press, 2002\relax
\mciteBstWouldAddEndPuncttrue
\mciteSetBstMidEndSepPunct{\mcitedefaultmidpunct}
{\mcitedefaultendpunct}{\mcitedefaultseppunct}\relax
\EndOfBibitem
\bibitem[Edberg et~al.(1986)Edberg, Evans, and Morriss]{edberg}
Edberg,~R.; Evans,~D.~J.; Morriss,~G.~P. \emph{J. Chem. Phys.} \textbf{1986},
  \emph{84}, 6933\relax
\mciteBstWouldAddEndPuncttrue
\mciteSetBstMidEndSepPunct{\mcitedefaultmidpunct}
{\mcitedefaultendpunct}{\mcitedefaultseppunct}\relax
\EndOfBibitem
\bibitem[Ryckaert et~al.(1977)Ryckaert, Ciccotti, and Berendsen]{ryck}
Ryckaert,~J.~P.; Ciccotti,~G.; Berendsen,~H. J.~C. \emph{J. Comput. Phys.}
  \textbf{1977}, \emph{23}, 327\relax
\mciteBstWouldAddEndPuncttrue
\mciteSetBstMidEndSepPunct{\mcitedefaultmidpunct}
{\mcitedefaultendpunct}{\mcitedefaultseppunct}\relax
\EndOfBibitem
\bibitem[Fragiadakis and Roland(2011)Fragiadakis, and Roland]{dscale2}
Fragiadakis,~D.; Roland,~C.~M. \emph{J. Chem. Phys.} \textbf{2011}, \emph{134},
  044504\relax
\mciteBstWouldAddEndPuncttrue
\mciteSetBstMidEndSepPunct{\mcitedefaultmidpunct}
{\mcitedefaultendpunct}{\mcitedefaultseppunct}\relax
\EndOfBibitem
\bibitem[Gauss(1829)]{gauss}
Gauss,~K.~F. \emph{J. Reine Angew. Math.} \textbf{1829}, \emph{4}, 232\relax
\mciteBstWouldAddEndPuncttrue
\mciteSetBstMidEndSepPunct{\mcitedefaultmidpunct}
{\mcitedefaultendpunct}{\mcitedefaultseppunct}\relax
\EndOfBibitem
\bibitem[Ciccotti et~al.(1982)Ciccotti, Ferrario, and Ryckaert]{rigid}
Ciccotti,~G.; Ferrario,~M.; Ryckaert,~J.~P. \emph{Mol. Phys.} \textbf{1982},
  \emph{47}, 1253\relax
\mciteBstWouldAddEndPuncttrue
\mciteSetBstMidEndSepPunct{\mcitedefaultmidpunct}
{\mcitedefaultendpunct}{\mcitedefaultseppunct}\relax
\EndOfBibitem
\bibitem[Melchionna(2000)]{mel}
Melchionna,~S. \emph{Phys. Rev. E} \textbf{2000}, \emph{61}, 6165\relax
\mciteBstWouldAddEndPuncttrue
\mciteSetBstMidEndSepPunct{\mcitedefaultmidpunct}
{\mcitedefaultendpunct}{\mcitedefaultseppunct}\relax
\EndOfBibitem
\end{mcitethebibliography}

\providecommand*\mcitethebibliography{\thebibliography}
\csname @ifundefined\endcsname{endmcitethebibliography}
  {\let\endmcitethebibliography\endthebibliography}{}

\end{document}